\documentclass{article}
\usepackage{amsmath}
\usepackage{amssymb}
\usepackage{amsxtra}
\usepackage{amsthm}
\numberwithin{equation}{section}
\allowdisplaybreaks[1]
\newcommand{\comment}[1]{}
\newcommand\hs[3]{\left(
\begin{array}{c}{#1}\\[6pt]{#2}\end{array}\,;\,{#3}\,\right)
}

\title{The connection problem 
associated with a Selberg type
integral and the $q$-Racah polynomials}
\author{Katsuhisa Mimachi}
\date{}
\begin{document}
\maketitle

\begin{abstract}
The connection problem associated with a
Selberg type integral is solved.  The connection
coefficients are given in terms of the $q$-Racah polynomials.
As an application of the explicit expression of 
the connection coefficients, examples of the monodromy-invariant 
Hermitian form of non-diagonal type are presented. It is noteworthy
that such Hermitian forms are intimately related with the 
correlation functions of non-diagonal type in 
$\widehat{sl_2}$-confromal field theory.

\end{abstract}

\bigskip
\begin{quote}
{\small
{\bf Key Words and Phrases}. A Selberg type integral, 
connection problem, connection coefficient, $q$-Racah polynomial,
twisted homology,  monodromy-invariant Hermitian form,
conformal field theory, correlation functions of non-diagonal type.}
\end{quote}
\medskip 
\bigskip


\begin{center}
{\bf Introduction}
\end{center}

A Selberg type integral 

\begin{equation}
\int_{\gamma}
\prod_{1\le i<j\le m}(t_j-t_i)^g
\prod_{\substack{1\le i\le m\\1\le j\le n}}
(t_i-z_j)^{\lambda_j}
dt_1\cdots dt_m,
\end{equation}
where $g$ and $\lambda_j$ are complex numbers and
$\gamma$ is a suitable cycle, is a natural generalization
of the Gauss hypergeometric function and the
Selberg integral.  It is used to express  conformal blocks
in conformal field theory 
\cite{DJMM}\cite{DF1}\cite{DF2}\cite{MY1}\cite{SV}\cite{V}
and to represent the hypergeometric function associated with
a root system due to Heckman and Opdam \cite{Kan}\cite{M1}. 
The integral (0.1) can be thought of
as an element of the pairing between the de Rham cohomology group
and the twisted homology group (the homology group with coefficients in
local system). Such a viewpoint to the integral represetation 
of special functions was introduced by Aomoto around 1970, and have
been developed after the name of the
{\it twisted de Rham theory} \cite{Ao1}\cite{Ao2}\cite{Ao3}.

The main purpose of this article is
to solve the {\it connection problem} associated with a special
case of (0.1):
\begin{equation}
\int_{\gamma}
\prod_{1\le i<j\le m}(t_j-t_i)^g
\prod_{1\le i\le m}
t_i^a(1-t_i)^b(t_i-z)^c\;
dt_1\cdots dt_m,
\end{equation}
which satisfies an ordinary differential
equation of order $m+1$ with 
three regular singular points $0, 1$ and $\infty.$ 
The connection problem we mean here is to give 
linear relations between the fundamental sets of solutions
around the singularities, moreover, to write down such
coefficients explicitly.
  
Generally speaking, the connection problem is important 
to know the global property of the solution space of a
given differential equation, but only rare cases are known to be solved.

In the case of  the Gauss hypergeometric function, 
Kummer discovered the relations between the fundamental sets
of solutions around three singularities in 1836 \cite{Ku}\;(Afterwards, 
it was found in his nachlass that Gauss had also discovered such 
relations in 1812). In the case of the generalized  
hypergeometric function ${}_nF_{n-1}$ for $n\ge 3$, 
Thomae obtained the coefficients between the solutions around $0$ 
and those around $\infty$ in 1870 
(See also \cite {W} \cite{OTY} and \cite{M3}).
These are all the cases of regular singular type
in which the connection problem is solved.

Contents in this article is the following. 
The sets of solutions around $0$ and $1$ are given in Proposition 2.1, 
and those around $0$ and $\infty$ in Proposition 2.6. 
The connection formula which connects the solutions in Proposition 2.1
is given in Theorem 2.3-4, and the formula 
which connects the solutions in Proposition 2.6 is given in Theorem 2.7-8.
In particilar, the connection coefficients in Theorem 2.4 and Theorem 2.8
are represented by the $q$-Racah polynomials. The $q$-Racah polynomials
are essentially the same as the $q$-6j symbol and
are known to be ingredients to construct some kinds of link invariants
including the Jones polynomial \cite{KR}. It is also noteworthy
that the formulas in Theorem 2.3-4 correspond to the braiding matrices
and the formulas in Theorem 2.7-8  to the fusion matrices in the
context of conformal field theory.
 
The connection formulas in Section 2 are actually derived in
Section 3 in a unified form;this section is separated into two parts.  
The first part is devoted to the manipulation
of twisted cycles to obtain the connection formula in 
Proposition 3.3.  We note that Theorem 2.3 and Theorem 2.7
are two special cases of Proposition 3.3.
Next, the second part is devoted to the change of the expression
of Proposition 3.3 into several forms by means of the transformation 
formulas of the basic hypergeometric series. 
The expression in Proposition 3.4 is in terms of ${}_8\varphi_7$,
the expressions in Proposition 3.5 are in terms of ${}_4\varphi_3$,
and the expression in Proposition 3.5 is in terms of the
$q$-Racah polynomial.

Finally, in Section 4, examples of the monodromy-invariant
Hermitian form of non-diagonal type are presented in
Theorem 4.3-4. These are intimately related with the 
correlation functions of non-diagonal type in conformal
field theory classified by
Kato \cite{Kat} and Cappelli-Itzykson-Zuber \cite{CIZ}.

It is also noteworthy that
the intersection number of twisted cycles
associated with the function
$$u(t)=
\prod_{1\le i<j\le m}(t_j-t_i)^g
\prod_{\substack{1\le i\le m\\1\le j\le n}}
(t_i-z_j)^{-{g}/{2}}
$$
is used to construct the Jones polynomial
of link invariant \cite{M4}. On the other hand,
$q$-6j symbols, or the $q$-Racah polynomials are 
used in \cite{KR} to construct the Jones polynomial.
To clarify the linkage of the $q$-Racah polynomials
with the intersection number of twisted cycles
is our future problem. 

\bigskip

\noindent
{\bf Acknowledgments.} The author  would like to thank Professor
Yasuhiko Yamada for valuable comments mainly on the correlation
functions of non-diagonal type in conformal field theory.

\tableofcontents

\newpage

\section{Preliminaries}

\subsection{Twisted homology groups}

Let $u(t)=\prod_i f_i(t)^{\alpha_i}$ be a multivalued
function on $T\subset{\Bbb C}^m$, where $\alpha_i\in{\Bbb C}$
and $T$ is the complement of the singular locus 
$\cup_i\{t=(t_1,\ldots,t_m)\in {\Bbb C}^m \mid f_i(t)=0\}$ in ${\Bbb C}^m$.
Let ${\mathcal L}$ be the local system
(locally constant sheaf) defined by $u$:
the sheaf consisting of the local solutions of $dL=L\omega$ for 
$\omega=du(t)/u(t).$  

Let $H_m(T,{\mathcal L})$ be the $m$-th homology
group with  coefficients in ${\mathcal L},$
$H_m^{\rm lf}(T,{\mathcal L})$ the
$m$-th locally finite homology group with coefficients in ${\mathcal L}.$
Elements of these twisted homology groups, called
{\it twisted cycles} or  {\it loaded cycles}, 
are represented by $\partial$-closed
twisted (finite  or locally finite) chains
\smallskip

$$C=\sum_\rho\,a_\rho\rho\otimes 
v_\rho,\quad(a_\rho\in{\Bbb C}),$$
\smallskip

\noindent
where each $\rho$ is an $m$-simplex
and $v_\rho$ a section of ${\mathcal L}$ on $\rho$.
The boundary operator $\partial$
is defined to be a ${\Bbb C}$-linear mapping satisfying
$\partial(\rho\otimes\;v)=\sum_{i=0}^m(-1)^i\rho^i
\otimes\;v|_{\rho^i},$ where $\rho$ is an $m$-simplex,
$\rho^i$ denotes the $i$-th face of $\rho$, and
$v|_{\rho^i}$ is the restriction of $v$ on $\rho^i$. 
\smallskip

If each factor $f_i(t)$ of $u(t)$ is defined 
over ${\Bbb R},$ and $D$ is a domain of the real manifold 
$T_{\Bbb R}$ (the real locus of $T$), 
then it is convenient to load $D$ with a section 
$$u_D(t)=\prod_i(\epsilon_i\,f_i(t))^{\alpha_i}$$ 
of ${\mathcal L}$ on $D$,
and to make a loaded cycle $D\otimes u_D(t)$,
where $\epsilon_i=\pm$ is so determined that 
$\epsilon_i\,f_i(t)$ is positive on $D$, and the argument of
 $\epsilon_i\,f_i(t)$ is assigned to be zero.
This choice of a section is said to be {\it standard}.

In this paper, we adopt mainly the standard loading. 
Thus, we frequently omit the assignment of loading and 
denote just the topological cycles for simplicity.
For example, in case  $T={\Bbb C}\backslash\{0,1\}$
and $u(t)=t^{\alpha}(1-t)^{\beta}$,
we denote by $\overrightarrow{(0,1)}$ to express 
$\overrightarrow{(0,1)}\otimes u(t)$, and $\overrightarrow{(1,\infty)}$ 
for
$\overrightarrow{(1,\infty)}\otimes t^{\alpha}(t-1)^{\beta}$.
\medskip

Under some genericity condition on the exponents $\alpha_i$, 
we  have the isomorphism, called the {\it regularization},

$$\;\,{\rm reg}\;:\;
H_m^{\rm lf}(T,{\mathcal L})\;\longrightarrow\;
H_m(T,{\mathcal L}),$$

\noindent
which is the inverse of the
natural map $\iota\,:\,H_m(T,{\mathcal L})\,\rightarrow
\,H_m^{\rm lf}(T,{\mathcal L})$. 

For example, in case  
$T={\Bbb C}\backslash\{0,1\}$
and $u(t)=t^{\alpha}(t-1)^{\beta}$, where  
$\alpha,\beta,\alpha+\beta\in {\Bbb R}\backslash{\Bbb Z},$
a regularization (regularized cycle)
$\;\,{\rm reg}\,C \in H_1(T,{\mathcal L})$ of
$C=\overrightarrow{(0,1)}\in H_1^{\rm lf}(T,{\mathcal L})$ can be given by
\begin{equation*}
\;\,{\rm reg}\,C=\left\{
\frac{1}{d_\alpha}S(\epsilon\,;0)
+\overrightarrow{[\epsilon, 1-\epsilon]}
-\frac{1}{d_\beta}S(1-\epsilon\,;1)\right\}\otimes u(t).
\end{equation*}

\noindent
Here $d_a=e(2a)-1$ with $e(A)=\exp(\pi\sqrt{-1} A)$, 
$\epsilon$ is a small positive number,
the symbol $S(a\,;z)$ stands for the positively oriented circle
centered at the point $z$ with starting and ending at the point $a$, 
and the argument of each factor of $u(t)$ on  
$S(\epsilon\,;0)$ or $S(1-\epsilon\,;1)$ 
is defined so that $\arg t$ takes values 
from $0$ to $2\pi$ on $S(\epsilon\,;0),$ 
and $\arg (1-t)$ from $0$ to $2\pi$.
\smallskip

We refer the reader to \cite{Ki} for
the construction of regularized cycles in higher dimensional cases.
\bigskip

The intersection form 
$$\bullet:H_m^{\rm lf}(T,{\mathcal L})\times {H_m^{\rm lf}(T,{\mathcal L})}\longrightarrow {\Bbb C}$$
is the Hermitian form defined by 
$$(C, C')\longmapsto 
C\bullet C'=\sum_{\rho,\,\sigma}a_\rho\,\overline{a'_\sigma}
\sum_{t\in\rho\cap\sigma}
I_{t}(\rho,\sigma)v_\rho(t)\overline{v'_\sigma(t)}/|u|^2
$$
for $C,\, C'\in H_m^{\rm lf}(T,{\mathcal L})$, if 
$\mbox{reg}\, C$ and $C'$ 
are represented by
$$\mbox{reg}\ C=\sum_\rho a_\rho\,\rho\otimes v_\rho,\quad 
C'=\sum_\sigma a'_\sigma\,\sigma\otimes{v'_\sigma},$$
where $a_\rho, a'_\sigma\in {\Bbb C}$,
each $\rho$ or $\sigma$ is an $m$-simplex,
$v_\rho$ or $v'_\sigma$ a section of ${\mathcal L}$ on $\rho$
or $\sigma$,
${}^-$ the complex conjugation, 
and $I_{x}(\rho,\sigma)$ the topological intersection
number of $\rho$ and $\sigma$ at $x$. 
The value $C\bullet C'$ of the intersection form 
for $C,\ C'\in H_m^{\rm lf}(T,{\mathcal L})$ 
is called the  intersection number
of $C$ and $C'$. 
\smallskip

For example,  if the local system ${\mathcal L}$
is defined by
$$u(t)=\prod_{i=1}^m\ t_i^\alpha\ (1-t_i)^\beta\ 
\prod_{1\le i<j\le m}(t_j-t_i)^{{2\gamma}},$$
where $\alpha, \beta, \gamma \in {\Bbb R}\backslash
{\Bbb Z}$ with some genericity condition on $\alpha, \beta, \gamma$,
and 
$$C:=\sum_{\sigma\in {\frak S}_m} C_\sigma,$$
where
$C_\sigma=D_\sigma\otimes u_{D_\sigma}(t)$ and
$D_\sigma$ is a bounded domain 
$$\{(t_1,\dots,t_m)\in {\Bbb R}^m\mid 
0<t_{\sigma(1)}<\cdots<t_{\sigma(m)}<1\}$$
with the standard orientation, then we have the self-intersection
number 
\begin{align}
&J_m(\alpha,\beta,\gamma):=C^2=C\bullet C\nonumber\\[3mm]
&=m!\left(\frac{\sqrt{-1}}{2}\right)^m
\prod_{j=1}^m
\frac{s(\alpha+\beta+(m+j-2)\gamma)s(\gamma)}
{s(\alpha+(j-1)\gamma)s(\beta+(j-1)\gamma)s(j\gamma)},
\end{align}
where  $s(A)=\sin(\pi A).$ We refer the reader to \cite{KY1}\cite{KY2}
\cite{MOY}\cite{MY1}\cite{MY2}
for more details of the intersection numbers of twisted cycles.

\subsection{A Selberg type integral}

Let  ${\mathcal L}_z$ be the local system determined by
a function
\smallskip

\begin{equation*}
u(t)=
\prod_{1\le i<j\le m}(t_j-t_i)^g
\prod_{1\le i\le m}t_i^a(1-t_i)^b(t_i-z)^c
\end{equation*}

\noindent
on the domain 

$$T_z=\{t=(t_1,\ldots, t_m)\in {\Bbb C}^m\mid
t_i\neq t_j\; (i\neq j),\; t_i\neq 0,\,1,\,z\;\}.$$

\noindent
for $z\in {\Bbb C}\backslash\{0,1\}$. Let 
$H_m^{\rm lf}(T_z, {\mathcal L}_z)^{{\frak S}_m}_{-}$ 
and $H_m(T_z, {\mathcal L}_z)^{{\frak S}_m}_{-}$
stand for the anti-symmetric part of
$H_m^{\rm lf}(T_z, {\mathcal L}_z)$ 
and $H_m(T_z, {\mathcal L}_z)$
with respect to the action of the symmetric group
${\frak S}_m$ on the coordinate $t=(t_1,\ldots, t_m)$ of $T_{z}$. 

\noindent
Under the {\it genericity} condition on the exponents $a, b, c$ and $g$, 
it follows  that  
$H_j(T_z, {\mathcal L}_z)=H_j^{\rm lf}(T_z, {\mathcal L}_z)=0$  
for $j\neq m$
and 
\begin{equation}
{\rm dim}\, H_m(T_z, {\mathcal L}_z)^{{\frak S}_m}_{-}=
{\rm dim}\,H_m^{\rm lf}(T_z, {\mathcal L}_z)^{{\frak S}_m}_{-}=m+1,
\end{equation}
and that the natural map
$\iota\,:\,H_m(T_z, {\mathcal L}_z)^{{\frak S}_m}_{-}
\,\rightarrow
{\rm dim}\,H_m^{\rm lf}(T_z, {\mathcal L}_z)^{{\frak S}_m}_{-}$
is an isomorphism (See \cite{Ao3}\cite{C}\cite{KN}\cite{MOY}).
Here the genericity condition on the exponents is that 
none of the following is an integer: 

\begin{equation*}
ia+{i \choose 2}g, \quad ib+{i \choose 2}g, \quad
\quad ic+{i \choose 2}g, \quad
i\lambda_\infty+{i \choose 2}g, \quad {i \choose 2}g,
\qquad (1\le i\le m)
\end{equation*}
\smallskip

\noindent
where
$$\lambda_\infty=-a-b-c-(m-1)g\quad{\rm and}\quad
{1\choose 2}=0.$$
\medskip

In this paper, the genericity condition is assumed and 
the inverse map 
$$\;\,{\rm reg}\;:\;
H_m^{\rm lf}(T_z, {\mathcal L}_z)^{{\frak S}_m}_{-}
\;\longrightarrow\;
H_m(T_z, {\mathcal L}_z)^{{\frak S}_m}_{-}$$

\noindent
is freely used.
\medskip

It is noteworthy that $(1.2)$,  or more directly
$${\rm dim}\,H^m(T_z, {\mathcal L}_z^\vee)^{{\frak S}_m}_{-}=m+1,$$
where $H^m(T_z, {\mathcal L}_z^\vee)^{{\frak S}_m}_{-}$ is the
anti-symmetric part of the twisted de Rham cohomology
$H^m(T_z, {\mathcal L}_z^\vee)$ 
( ${\mathcal L}_z^\vee$ is the sheaf of the local solutions of
$dL=-\omega L, \omega=du(t)/u(t)$)
guarantees the existence of the
ordinary differential equation of order $m+1$ 
which is satisfied by the Selberg type integral
\begin{equation}
\int_{\gamma}
\prod_{1\le i<j\le m}(t_j-t_i)^g
\prod_{1\le i\le m}
t_i^a(1-t_i)^b(t_i-z)^c\;
dt_1\cdots dt_m,
\end{equation}

\noindent
where $\gamma$ is a suitable cycle.
Indeed, we have the following differential
equation of the first order with  matrix coefficients: 
(\cite{M1}): For $0\le i\le m$, set
$$\tilde{\varphi}_i=\sum_{\sigma\in{\frak S}_m}
\left\{
\prod_{1\le s\le i}t_{\sigma(s)}^{-1}
\prod_{i< s\le m}(t_{\sigma(s)}-1)^{-1}\right\},$$
which corresponds to an element of the basis of
the twisted de Rham cohomology 
$H^m(T,{\mathcal L}^\vee)^{{\frak S}_m}_{-}.$
For the fixed cycle $\gamma$, set  
\begin{equation*}
\langle \varphi\rangle=
\int_\gamma \varphi\,u(t)\,dt_1\cdots dt_m.
\end{equation*}

\noindent
Then we have 
\begin{align*}
\frac{d}{dz}\langle \tilde{\varphi}_0\rangle
&=\frac{m}{z-1}\left\{\bigl(b+c+(m-1)\frac{g}{2}\bigr)
\langle \tilde{\varphi}_0\rangle
+a\langle \tilde{\varphi}_{1}\rangle\right\},\\[6pt]
\frac{d}{dz}\langle \tilde{\varphi}_m\rangle
&=\frac{m}{z}\left\{\bigl(a+c+(m-1)\frac{g}{2}\bigr)
\langle \tilde{\varphi}_m\rangle
+b\langle \tilde{\varphi}_{m-1}\rangle\right\}
\end{align*}
and

\begin{align*}
\frac{d}{dz}\langle \tilde{\varphi}_i\rangle
&=\frac{i}{z}\left\{\bigl(a+c+(i-1)\frac{g}{2}\bigr)
\langle \tilde{\varphi}_i\rangle
+\bigl(b+(m-i)\frac{g}{2}\bigr)
\langle \tilde{\varphi}_{i-1}\rangle\right\}
\\[6pt]
&+\frac{m-i}{z-1}\left\{\bigl(b+c+(m-i-1)\frac{g}{2}\bigr)
\langle \tilde{\varphi}_i\rangle
+\bigl(a+i\frac{g}{2}\bigr)
\langle \tilde{\varphi}_{i+1}\rangle\right\},
\end{align*}
for $1<i<m.$  This system of equations induces the 
the scalar-valued differential equation satisfied by (1.3),
the order of which is $m+1$ and the characteristic exponents
$e_j^{(0)}, e_j^{(1)}, e_j^{(\infty)}$ 
of which at the singularities $0, 1, \infty$ are given by
\begin{align*}
&e_j^{(0)}=(a+c+1)j+{j\choose 2} g, \nonumber\\[6pt] 
&e_j^{(1)}=(b+c+1)j+{j\choose 2} g, \nonumber\\[6pt] 
&e_j^{(\infty)}=-(a+b+1)j-cm-({j\choose 2}+j(m-j)) g
\end{align*}
for $0\le j\le m.$  
\medskip

\noindent
When $m=1$, the differential equation satisfied by (1.3) is 

\begin{equation}
z(z-1)I''+\{a+c-(a+b+2c)z\}I'+c(a+b+c+1)I=0,
\end{equation}

\noindent
which is nothing but the hypergeometric differential
equation.

\noindent
When $m=2$, it is 
\begin{align}
&z^2(z-1)^2 I'''+(K_1z+K_2(z-1))z(z-1)I''
\nonumber\\[6pt]
&+(L_1z^2+L_2(z-1)^2+L_3z(z-1))I'+(M_1z+M_2(z-1))I=0
\end{align}

\noindent
with
\begin{align*}
&K_1=-g-3b-3c, \quad K_2=-g-3a-3c,\\[8pt]
&L_1=(b+c)(2b+2c+g+1), \quad L_2= (a+c)(2a+2c+g+1),\\[8pt]
&L_3=(b+c)(2a+2c+g+1)+(a+c)(2b+2c+g+1)\\[8pt]
&+(c-1)(a+b+c)+(3c+g)(a+b+c+g+1),\\[8pt]
&M_1=-c(2b+2c+g+1)(2a+2b+2c+g+2),\\[8pt]
&M_2=-c(2a+2c+g+1)(2a+2b+2c+g+2),
\end{align*}

\noindent
which was first derived by Dotsenko-Fateev \cite{DF1}.
\smallskip

\noindent
In more general $m$ case, such an explicit expression  
is not known.

\medskip

Let $V=\sum_{1\le i\le l}{\Bbb C} 
C_i\subset H_m^{\rm lf}(T_z, {\mathcal L}_z)^{{\frak S}_m}_{-}$
be an invariant subspace  under the action of the fundamental
group $\pi_1(z, {\Bbb C}\backslash\{0,1\}).$
If the loaded cycle $C_i$ is expressed as  
$\sum_\rho\,a_{i\rho}\,\rho\otimes u_\rho(t),$ where
each $\rho$ is an $m$-simplex in $T_z$, and $u_\rho(t)$
a section of ${\mathcal L}_z$ on $\rho,$
then we define a function $I_i(z)$  by the integral
$$
\sum_\rho a_{i\rho}\int_{\rho}u_\rho(t)
dt_1\cdots dt_m.$$ 
Let
$$I_h=(\,C_i\bullet{C_j}\,)
_{1\le i,j\le l}$$
be the intersection matrix. 
Then, the Hermitian form
\begin{equation}
F(z,\overline{z})=\sum_{1\le i,j\le l}
(I_h^{-1})_{ij}\,\overline{I_i(z)}I_j(z)
\end{equation}
is invariant under the action of
$\pi_1(z, {\Bbb C}\backslash\{0,1\})$.
\medskip

For a while, we assume that $z$ is real and $0<z<1.$
Let us define  $C_k$ to be
\begin{equation}
C_k=\sum_{\sigma\in {\frak S}_m}\sigma
\left(\Delta_k(t)\otimes u_{\Delta_k}(t)\right),
\qquad0\le k\le m\end{equation}
where
$$
\Delta_k(t)=
\{t\mid 0<t_1<\cdots
<t_{k}<z, \,1<t_{k+1}<\cdots<t_{m}\}.
$$
\comment{
and
$$u_{\Delta_k}=\prod_{1\le i<j\le m}(t_j-t_i)^g
\prod_{i=1}^{k}t_i^a(1-t_i)^b(z-t_i)^c
\prod_{i=k+1}^{m}t_i^a(t_i-1)^b(t_i-z)^c.
$$}

\noindent
Then $C_0,\dots,C_m$ form a basis of  
$H^{\rm lf}_m(T_z, {\mathcal L}_z)^{{\frak S}_m}_{-}$. Note that
$C_k$ is equal to the domain
\begin{align*}
&{m\choose k}
\sum_{\sigma\in {\frak S}_k}\{(t_1,\ldots,t_k)\mid
0<t_{\sigma(1)}<\cdots< t_{\sigma(k)}<z\,\}\\[5pt]
&\times
\sum_{\sigma\in {\frak S}_{m-k}}\{(t_{k+1},\ldots,t_m)\mid
1<t_{\sigma(k+1)}<\cdots< t_{\sigma(m)}<\infty\,\}
\end{align*}
standardly loaded with $u(t)$.
Hence, (1.1) implies that
\begin{equation}
C_k\bullet C_k={m\choose k}
J_k(a, c, g/2)J_{m-k}(b, -a-b-c-(m-1)g, g/2).
\end{equation}
\smallskip

Moreover, since $C_0,\dots,C_m$ 
are mutually disjoint, if we set 
\begin{equation*}
I_k(z)=\langle C_k, dt_1\cdots dt_m\rangle
=m!\int_{\Delta_k(t)}u_{\Delta_k}dt_1\cdots dt_m,
\end{equation*}
we obtain the monodromy-invariant Hermitian form

\begin{align}
&F(z,\overline{z})=\frac{1}{m!}
\left(\frac{2}{\sqrt{-1}}\right)^m
\sum_{k=0}^m
\prod_{j=1}^k\frac{s\left(a+(j-1)\frac{g}{2}\right)
s\left(c+(j-1)\frac{g}{2}\right)
s\left(j\frac{g}{2}\right)}
{s\left(a+c+(k+j-2)\frac{g}{2}\right)s\left(\frac{g}{2}\right)}
\nonumber\\[5pt]
&
\prod_{j=1}^{m-k}
\frac{s\left(-a-b-c-(m-1)g+(j-1)\frac{g}{2}\right)
s\left(b+(j-1)\frac{g}{2}\right)s\left(j\frac{g}{2}\right)}
{s\left(-a-c-(m-1)g+(m-k+j-2)\frac{g}{2}\right)s\left(\frac{g}{2}\right)}
\times|I_k(z)|^2.
\end{align}

\noindent
We refer the reader to \cite{MY2} for more details of (1.9).

\medskip

\comment{
\noindent
This coincides with $(3.19)$ in $\cite{DF2}$, the
four-point correlation function in the conformal
field theory, 
up to a constant multiplicative factor.}
\par\bigskip


\subsection{$q$-Racah polynomials}

Let  ${}_m\varphi_{m-1}$ be the basic hypergeometric series 

\begin{equation*}
{}_m\varphi_{m-1}\hs{a_1, \cdots, a_m}
{b_1, \cdots, b_{m-1}}
{q, z}
=\sum_{n\ge 0}
\frac{(a_1, \cdots\cdots, a_m;q)_n}{(b_1, \cdots, b_{m-1}, q;q)_n}z^n,
\end{equation*}
where 
$$(a_1, \cdots\cdots, a_m;q)_n
=(a_1;q)_n\cdots(a_m;q)_n$$
and
$$(a;q)_n=
\prod_{0\le i\le n-1}(1-aq^i).$$

\comment{
\noindent
The basic hypergeometric series ${}_m\varphi_{m-1}$
is called {\it $k$-balanced} if $z=q$ and
$$b_1b_2\cdots b_{m-1}=q^ka_1a_2\cdots a_m.$$
When $k=1$, the series is called {\it balanced}.
When 
$$qa_1=b_1a_2=b_{m-1}a_m,$$
the series is called {\it well poised.}
}

The $q$-Racah polynomials $W_n(x; a,b,c, N;q)$ are
defined by
\begin{equation}
W_n(x;q)=W_n(x; a,b,c, N;q)=
{}_4\varphi_3\hs{q^{-n}, abq^{n+1}, q^{-x}, cq^{x-N}}
{aq, q^{-N}, bcq}
{q, q}
\end{equation}
for $n=0,1,\ldots, N$,
which is of degree $n$ in the variable $\mu(x)=q^{-x}+cq^{x-N}$. 
Their orthogonality relation is
\begin{equation}
\sum_{x=0}^N
\rho(x;q)W_m(x;q)W_n(x;q)=
\frac{\delta_{m,n}}{h_n(q)},
\end{equation}
where
\begin{align}
\rho(x;q)&=\rho(x; a, b, c, N;q)\nonumber\\[6pt]
&=\frac{(1-cq^{2x-N})\,(cq^{-N}, q^{-N}, aq, bcq\,;q)_x}
{(1-cq^{-N})\,(ca^{-1}q^{-N}, b^{-1}q^{-N}, q, cq\,;q)_x}
(abq)^{-x}
\end{align} 
and
\begin{align}
h_n(q)&=h_n(a,b,c, N\,;q)\nonumber\\[6pt]
&=\frac{(bq,aq/c\,;q)_N}{(abq^2, 1/c\,;q)_N}
\frac{ (1-abq^{2n+1})\,(aq, abq, bcq,q^{-N}\,;q)_n}
{(1-abq)\,(q, bq, aq/c, abq^{N+2}\,;q)_n}
(q^N/c)^n.
\end{align} 

\noindent
We refer the reader to \cite{GR} for more detail on the 
$q$-Racah polynomials.
\medskip

It is noteworthy  that the  $q$-Racah polynomial $W_n(x;;q)$ 
is essentially the same as the $q$-$6j$ symbol \cite{KR}; actually we have

$$\left\{
\begin{array}{ccc}
a, & b, &e \\[5pt]
d, & c, &f
\end{array}
\right\}_q=\{\rho(x;q)h_n(q)\}^{1/2}
W_n(x; \alpha, \beta, \gamma, N;q),$$
where

\begin{align*}
&n=a+b-e, \quad x=c+d-e, \quad N=a+b+c+d+1,\\[5pt]
&\alpha=q^{-a-d+e+f},\quad \beta=q^{-a-b-c+d-1},\quad
\gamma=q^{a+e+f-d+1}.
\end{align*}

\noindent
The orthogonality relation satisfied by these $q$-$6j$ symbols
is

$$\sum_{0\le j\le N}
\left\{
\begin{array}{ccc}
j_2, & j_1, & j\\[5pt]
j_3, & j_5, & j_4
\end{array}
\right\}_q
\left\{
\begin{array}{ccc}
j_3, & j_1, & j_6\\[5pt]
j_2, & j_5, & j
\end{array}
\right\}_q=\delta_{j_4, j_6}.$$

\noindent
The $q$-6j symbols are used to construct the invariants of
links related with the quantum group $U_q(sl_2)$ \cite{KR}. 

\bigskip

\section{Connection formulas}

Let ${\mathcal L}_z$ be the local system determined by
the function
\smallskip

\begin{equation*}
u(t)=\prod_{1\le i<j\le m}(t_j-t_i)^g
\prod_{1\le i\le m}
t_i^a(1-t_i)^b(t_i-z)^c
\end{equation*}

\noindent
on the domain 

$$T_z=\{t=(t_1,\ldots, t_m)\in {\Bbb C}^m\mid
t_i\neq t_j\; (i\neq j),\; t_i\neq 0,\,1,\,z\;\},$$

\noindent
where $z\in {\Bbb C}\backslash\{0,1\}.$  \medskip

\subsection{The solutions around $0$ in terms of
those around $1$}

In this subsection, we give two formulas that 
connect the fundamental set of solutions around $0$
with that around $1$.

For convenience to our purpose,  
we fix a complex variable $z$ to be real with
$0<z<1$ and assign the domains
$D_{0,j,0, m-j}(t)$ and  $D_{m-j,0,j,0}(t)$ for $0\le j\le m$  
of the real manifold $T_{\Bbb R}$  by
\begin{align*}
&D_{0,j,0, m-j}(t)=\{(t_1,\ldots,t_m)\mid
0<t_1<\cdots<t_j<z,\; 1<t_{j+1}<\cdots<t_m\},\\[6pt]
&D_{m-j,0,j,0}(t)=\{(t_1,\ldots,t_m)\mid
t_1<\cdots<t_{m-j}<0,\; z<t_{m-j+1}<\cdots<t_m<1\},
\end{align*}

\noindent
where each orientation is fixed to be
natural one induced from  $T_{\Bbb R}.$ Moreover,
we define the loaded cycle $C_{0,j,0, m-j}$ and
$C_{m-j,0,j,0}$ to be
$$C_{0,j,0, m-j}=\sum_{\sigma\in S_m}\sigma
\left\{D_{0,j,0, m-j}(t)\otimes u_{D_{0,j,0, m-j}}(t)\right\}$$
and
$$C_{m-j,0,j,0}=\sum_{\sigma\in S_m}\sigma
\left\{D_{m-j,0,j,0}(t)\otimes u_{D_{m-j,0,j,0}}(t)\right\},$$ 
where the action of $\sigma\in S_m$ is on the coordinates
$t_1,\ldots, t_m$  of ${\Bbb C}^m$. The families of cycles 
$\{\;C_{0,j,0, m-j}\mid 0\le j\le m\;\}$ and 
$\{\;C_{m-j,0, j,0}\mid 0\le j\le m\;\}$
give  fundamental sets of solutions around $0$ and 
that around $1$, respectively. Indeed, the integrals 
\begin{align*}
&I_j(a,b,c; g;z)
=\langle\, {\rm reg}\,C_{0,j,0, m-j}, dt_1\ldots dt_m\rangle\\[5pt]
&=m!\int_{D_{0,j,0, m-j}(t)}u_{D_{0,j,0, m-j}}(t)\,dt_1\cdots\,dt_m
\end{align*}
and
\begin{align*}
&J_j(a,b,c; g;z)
=\langle\, {\rm reg}\,C_{m-j, 0, j, 0}, dt_1\ldots dt_m\rangle\\[5pt]
&=m!\int_{D_{m-j, 0, j, 0}(t)}u_{D_{m-j, 0, j, 0}}(t)\,dt_1\cdots\,dt_m
\end{align*}
give fundamental sets of solutions around $0$ and 
that around $1$, respectively.

\bigskip

\noindent
{\bf Proposition 2.1.} (1)\;{\it For $0\le j\le m,$ we have}

\begin{align*}
&I_j(a,b,c; g;z)
\nonumber\\[6pt]
&=m!\,
S_j(a+1, b+1, g/2)\,
S_{m-j}(-a-b-c-(m-1)g-1, b+1, {g}/{2})
\nonumber\\[6pt]
&\times z^{(a+c+1)j+{j\choose 2} g}\,(1+O(z))\quad (z\rightarrow 0).
\end{align*}
\medskip

\noindent
(2)\; {\it For $0\le j\le m$, we have}

\begin{align*}
&J_j(a,b,c;g;z)
\nonumber\\[6pt]
&=m!\,
S_j(b+1, c+1, g/2)\,
S_{m-j}(-a-b-c-(m-1)g-1, a+1, g/2)
\nonumber\\[6pt]
&\times (1-z)^{(b+c+1)j+{j\choose 2} g}\,(1+O(1-z))\quad (z\rightarrow 1).
\end{align*}
\medskip

\noindent
{\it Here the arguments of $z$ of $z^{(a+c+1)j+{j\choose 2} g}$
and $1-z$ of  $(1-z)^{(b+c+1)j+{j\choose 2} g}$ are fixed to be zero
on $0<z<1$, and
$S_m(\alpha,\beta,\gamma)$ denotes the Selberg integral}
$\cite{Se}$:

\begin{align}
&S_m(\alpha,\beta,\gamma)\nonumber\\[6pt]
&=\int_{0<t_1<\cdots<t_m<1} \prod_{i=1}^{m}t_i^{\alpha-1}(1-t_i)^{\beta-1}
\prod_{1\le i< j \le m}(t_j-t_i)^{2\gamma}dt_1\cdots dt_m\nonumber\\[6pt]
&=\frac{1}{m!}\prod_{j=1}^m\frac{\Gamma(\alpha+(j-1)\gamma)
\Gamma(\beta+(j-1)\gamma)\Gamma(j\gamma+1)}
{\Gamma(\alpha+\beta+(m+j-2)\gamma)\Gamma(\gamma+1)}.
\end{align}
\comment{
for
$$
{\rm Re}\,\alpha>0, \;{\rm Re}\,\beta>0,\; 
{\rm Re}\;\gamma>-{\rm mim}
\{1/m,({\rm Re}\,\alpha)/(m-1),({\rm Re}\,\beta)/(m-1)\}.
$$
}
\medskip

\noindent
{\bf Proof.}\;(1)\; 
\comment{Note that
\begin{align*}
&\frac{1}{m!}I_j(a,b,c;g;z)\nonumber\\[6pt]
&=\int_{D_{0,j,0, m-j}}
\prod_{1\le i_1<i_2\le j}(t_{i_2}-t_{i_1})^g
\prod_{1\le i\le j}
t_i^a(z-t_i)^c(1-t_i)^b
\;\nonumber\\[6pt]
&\times
\prod_{j< i_1<i_2\le m}(t_{i_2}-t_{i_1})^g
\prod_{j< i\le m}
t_i^a(t_i-z)^c(t_i-1)^b
\;\nonumber\\[6pt]
&\times
\prod_{\substack{
1\le i_1\le j\\[3pt]
j\le i_2\le m}}
(t_{i_2}-t_{i_1})^g\;
dt_1\cdots dt_m,
\end{align*}

\noindent
where $D_{0,j,0, m-j}$ has the standard orientation.}
The change of the integration variables such as
$t_i\mapsto zt_i\;(1\le i\le j)$ and
$t_i\mapsto 1/t_i\;(j< i\le m)$ leads to
\begin{align*}
\comment{
&z^{(\lambda_{12}+1)j+{j\choose 2}g}
\int
\prod_{1\le i_1<i_2\le j}(t_{i_2}-t_{i_1})^g
\prod_{1\le i\le j}
t_i^{\lambda_1}(1-t_i)^{\lambda_2}(1-zt_i)^{\lambda_3}
\;\nonumber\\[6pt]
&\times
\prod_{j< i_1<i_2\le m}(t_{i_2}^{-1}-t_{i_1}^{-1})^g
\prod_{j< i\le m}
t_i^{-\lambda_1-2}(t_i^{-1}-z)^{\lambda_2}(t_i^{-1}-1)^{\lambda_3}
\;\nonumber\\[6pt]
&\times
\prod_{\substack{
1\le i_1\le j\\[3pt]
j\le i_2\le m}}
(t_{i_2}^{-1}-zt_{i_1})^g\;
dt_1\cdots dt_m\;\nonumber\\[6pt]}
&\frac{1}{m!}I_j(a,b,c;g;z)\\[5pt]
&=z^{(a+c+1)j+{j\choose 2}g}
\int
\prod_{1\le i_1<i_2\le j}(t_{i_2}-t_{i_1})^g
\prod_{1\le i\le j}
t_i^a(1-t_i)^c(1-zt_i)^b
\;\nonumber\\[6pt]
&\times
\prod_{j< i_1<i_2\le m}(t_{i_1}-t_{i_2})^g
\prod_{j< i\le m}
t_i^{-a-b-c-(m-1)g-2}(1-zt_i)^c
(1-t_i)^b
\;\nonumber\\[6pt]
&\times
\prod_{\substack{
1\le i_1\le j\\[3pt]
j\le i_2\le m}}
(1-zt_{i_1}t_{i_2})^g\;
dt_1\cdots dt_m,
\end{align*}
where the domain of integration is
$$0<t_1<\cdots<t_j<1,\quad 0<t_m<\cdots<t_{j+1}<1$$
with the standard orientation.
This implies the required result by
using the binomial theorem
$$(1-zt_{i_1}t_{i_2})^g
=\sum_{n\ge 0}\frac{(-g)_n}{n!}(zt_{i_1}t_{i_2})^n.
$$

\medskip

\noindent
(2)\; The change of the integration variables such as
$t_i\mapsto 1/(1-t_i)\;(1\le i\le m-j)$ and
$t_i\mapsto (1-t_i)/(1-z)\;(m-j< i\le m)$
\comment{in

\begin{align*}
&\frac{1}{m!}J_j(a,b,c;g;z)\nonumber\\[6pt]
&=\int_{D_{m-j, 0,j,0}}
\prod_{1\le i_1<i_2\le m-j}(t_{i_2}-t_{i_1})^g
\prod_{1\le i\le m-j}
(-t_i)^a(z-t_i)^c(1-t_i)^b
\;\nonumber\\[6pt]
&\times
\prod_{m-j< i_1<i_2\le m}(t_{i_2}-t_{i_1})^g
\prod_{m-j< i\le m}
t_i^a(t_i-z)^c(1-t_i)^b
\;\nonumber\\[6pt]
&\times
\prod_{\substack{
1\le i_1\le m-j\\[3pt]
m-j< i_2\le m}}
(t_{i_2}-t_{i_1})^g\;
dt_1\cdots dt_m
\end{align*}

\noindent}
leads to

\begin{align*}
&\frac{1}{m!}J_j(a,b,c;g;z)\nonumber\\[6pt]
&=
(1-z)^{(b+c+1)j+{j\choose 2}g}\nonumber\\[6pt]
&\times\int
\prod_{1\le i_1<i_2\le m-j}(t_{i_2}-t_{i_1})^g
\prod_{1\le i\le m-j}
t_i^{-a-b-c-(m-1)g-2}(1-t_i)^a(1-(1-z)t_i)^c
\;\nonumber\\[6pt]
&\times
\prod_{m-j< i_1<i_2\le m}(t_{i_2}-t_{i_1})^g
\prod_{m-j+1< i\le m}
t_i^b(1-(1-z)t_i)^a(1-t_i)^c
\;\nonumber\\[6pt]
&\times
\prod_{\substack{
1\le i_1\le m-j\\[3pt]
m-j< i_2\le m}}
(1-(1-z)t_{i_1}t_{i_2})^g\;
dt_1\cdots dt_m
\comment{
\nonumber\\[6pt]
&=S_j(b+1, c+1, g/2)
S_{m-j}(-a-b-c-(m-1)g-1, a+1, g/2)\nonumber\\[6pt]
&\times 
(1-z)^{(b+c+1)j+{j\choose 2} g}\;
(1+O(1-z))\quad (z\rightarrow 1)},
\end{align*}
where the domain of integration is
$$0<t_{m-j}<\cdots<t_1<1-z, \quad
0<t_m<\cdots<t_{m-j+1}<1$$
with the standard orientation. This implies the
required result.
\hfill$\square$

\bigskip

\noindent
In case of $m=1$, i.e. Gauss hypergeometric functions,
Proposition 2.1 corresponds to
\begin{align}
&I_0(a,b,c;z)\nonumber\\[6pt]
&=B(b+1,-a-b-c-1)\;{}_2F_1
\hs{-c,-a-b-c-1}{-a-c}{z},\\[6pt]
&I_1(a,b,c;z)\nonumber\\[6pt]
&=B(a+1,c+1)\;z^{a+c+1}\,{}_2F_1
\hs{-b,a+1}{a+c+2}{z}
\end{align}
and
\begin{align}
&J_0(a,b,c;z)\nonumber\\[6pt]
&=B(a+1,-a-b-c-1)\;{}_2F_1
\hs{-c,-a-b-c-1}{-b-c}{1-z},\\[6pt]
&J_1(a,b,c;z)
\nonumber\\[6pt]
&=B(b+1,c+1)\;(1-z)^{b+c+1}\,{}_2F_1
\hs{-a,b+1}{b+c+2}{1-z},
\end{align}
where
$B(a,b)$ denotes the beta function
$$B(a,b)=\frac{\Gamma(a)\Gamma(b)}{\Gamma(a+b)}.$$

\noindent

\medskip

In this case, the linear relations between $\{I_0, I_1\}$ and $\{J_0, J_1\}$
are given by

\begin{align*}
&I_0(a,b,c;z)=\dfrac{s(a)}{s(b+c)}J_0(a,b,c;z)
+\dfrac{-s(c)}{s(b+c)}J_1(a,b,c;z),\\[6pt]
&I_1(a,b,c;z)=\dfrac{-s(a+b+c)}{s(b+c)}J_0(a,b,c;z)
+\dfrac{-s(b)}{s(b+c)}J_1(a,b,c;z)
\end{align*}

\noindent
and

\smallskip
\begin{align*}
&J_0(a,b,c;z)=\frac{s(b)}{s(a+c)}I_0(a,b,c;z)
+\frac{-s(c)}{s(a+c)}I_1(a,b,c;z),\\[6pt]
&J_1(a,b,c;z)=\frac{-s(a+b+c)}{s(a+c)}I_0(a,b,c;z)
+\frac{-s(a)}{s(a+c)}I_1(a,b,c;z),
\end{align*}
where $s(A)=\sin(\pi A).$

These are called the {\it connection formulas} between
the fundamental set of solutions around $0$ and that around $1$
in the case of the Gauss hypergeometric functions.
\medskip

In general $m$ case,
if we define $p_{ij}^{(0,1)}=p_{ij}^{(0,1)}(a,b,c;g)$ 
for $0\le i,j\le m$, called
the {\it connection coefficients}, by 
\begin{equation*}
I_i(a,b,c;g;z)=\sum_{0\le j\le m} p_{ij}^{(0,1)}(a,b,c;g)\, J_j(a,b,c;g;z),
\end{equation*}

\noindent
we have the following:
\comment{two expressions, both of which are usefull,
actually, both of which are used in Section 5.}

\medskip

\noindent
{\bf Theorem 2.2.} {\it For $0\le i,j\le m$, we have
\comment{
\noindent
(1)\;}
\begin{align}
&p_{ij}^{(0,1)}(a,b,c;g)\nonumber\\[5pt]
&=(-)^i\sum_{\substack{
0\le k\le m-i\\[3pt]
0\le l\le i\\[3pt]
k+l=j}}
(-)^{k}
\prod_{r=1}^{m-i-k}
\frac{s\bigl(\,a+\frac{i+r-1}{2}g\,\bigr)}
{s\bigl(\,b+c+\left(k+\frac{i+r-1}{2}\right)g\,\bigr)}
\prod_{r=1}^{k}
\frac{s\bigl(c+\frac{i+r-1}{2}g\,\bigr)}
{s\bigl(b+c+\left(k+\frac{i-r-1}{2}\right)g\,\bigr)}
\nonumber\\[6pt]
&\times
\prod_{r=1}^{i-l}
\frac{s\bigl(a+b+c+\frac{m+i+k-r-1}{2}g\,\bigr)
s\bigl(\frac{m-i-k+r}{2}g\,\bigr)}
{s\bigl(b+c+\left(j+\frac{r-1}{2}\right)g\,\bigr)
s\bigl(\frac{r}{2}g\,\bigr)}
\prod_{r=1}^{l}
\frac{s\bigl(b+\frac{k+r-1}{2}g\,\bigr)
s\bigl(\frac{k+r}{2}g\,\bigr)}
{s\bigl(b+c+\left(j-\frac{r+1}{2}\right)g\,\bigr)
s\bigl(\frac{r}{2}g\,\bigr)}
\end{align}

\noindent
and

\comment{
\noindent
(2)\;}

\begin{align}
&p_{ij}^{(0,1)}(a,b,c;g)\nonumber\\[5pt]
&=(-)^i
\sum_{\substack{
0\le k\le i\\[3pt]
0\le l\le m-i\\[3pt]
k+l=j}}
(-)^{l}
\prod_{r=1}^{i-k}
\frac{s\bigl(\,a+b+c+\frac{m+i-r-1}{2}g\,\bigr)}
{s\bigl(\,b+c+\left(k+\frac{m-i+r-1}{2}\right)g\,\bigr)}
\prod_{r=1}^{k}
\frac{s\bigl(\,b+\frac{m-i+r-1}{2}g\,\bigr)}
{s\bigl(\,b+c+\left(k+\frac{m-i-r-1}{2}\right)g\,\bigr)}
\nonumber\\[6pt]
&\times
\prod_{r=1}^{m-i-l}
\frac{s\bigl(\,a+\frac{i-k+r-1}{2}g\,\bigr)
s\bigl(\,\frac{i-k+r}{2}g\,\bigr)}
{s\bigl(\,b+c+\left(j+\frac{r-1}{2}\right)g\,\bigr)
s\bigl(\,\frac{r}{2}g\,\bigr)}
\prod_{r=1}^{l}
\frac{s\bigl(\,c+\frac{k+r-1}{2}g\,\bigr)
s\bigl(\,\frac{k+r}{2}g\,\bigr)}
{s\bigl(\,b+c+\left(j-\frac{r+1}{2}\right)g\,\bigr)
s\bigl(\,\frac{r}{2}g\,\bigr)},
\end{align}

\noindent
where $s(A)=\sin(\pi\,A).$}

\medskip

\noindent
{\bf Proof.}\; Set $\lambda_1=a, \lambda_2=c, \lambda_3=b$
and $z_1=0, z_2=z, z_3=1$ in Proposition 3.3 in the next section. 
Then we obtain the required result. \hfill$\square$

\bigskip

When $m=2$, Theorem 2.2 implies that

\begin{align*}
&P=(p_{ij}^{(0,1)})_{0\le i,j \le 2}\\[6pt]
&=
\left[
\begin{array}{ccc}
\frac{s(a)s(a+\frac{1}{2}g)}{s(b+c)s(b+c+\frac{1}{2}g)}
&-\frac{s(a)s(c)}{s(b+c)s(b+c+g)}
&\frac{s(c)s(c+\frac{1}{2}g)}{s(b+c+g)s(b+c+\frac{1}{2}g)}\\[12pt]
-\frac{s(a+\frac{1}{2}g)s(a+b+c+\frac{1}{2}g)s(g)}
{s(b+c)s(b+c+\frac{1}{2}g)s(\frac{1}{2}g)}&p_{11}^{(0,1)}&
\frac{s(b+\frac{1}{2}g)s(c+\frac{1}{2}g)s(g)}
{s(b+c+g)s(b+c+\frac{1}{2}g)s(\frac{1}{2}g)}\\[12pt]
\frac{s(a+b+c+\frac{1}{2}g)s(a+b+c+g)}
{s(b+c)s(b+c+\frac{1}{2}g)}&
\frac{s(b)s(a+b+c+g)}
{s(b+c)s(b+c+g)}&
\frac{s(b+\frac{1}{2}g)s(b)}
{s(b+c+g)s(b+c+\frac{1}{2}g)}
\end{array}\right],
\end{align*}
where
\begin{align*}
p_{11}^{(0,1)}&=-\frac{s(b)s(a+\frac{1}{2}g)}{s(b+c)s(b+c+\frac{1}{2}g)}
+\frac{s(a+b+c+g)s(c+\frac{1}{2}g)}{s(b+c+g)s(b+c+\frac{1}{2}g)}\\[6pt]
&=\frac{s(c)s(a+b+c+\frac{1}{2}g)}{s(b+c)s(b+c+\frac{1}{2}g)}
-\frac{s(a)s(b+\frac{1}{2}g)}{s(b+c+g)s(b+c+\frac{1}{2}g)}.
\end{align*}
\medskip

\noindent
The first expression of $p_{11}^{(0,1)}$ is implied by (2.6), 
and the second one is by (2.7).
The second expression coinsides with $(5.11)$ of \cite{DF1}.

\medskip

We have many expressions of the connection coefficient
$p_{ij}^{(0,1)}=p_{ij}^{(0,1)}(a,b,c;g)$ other than those in Theorem 2.2.
As an example of them, we give the following, which is
essentially given by the $q$-Racah polynomial.

\bigskip

\noindent
{\bf Theorem 2.3.}\; {\it For $0\le i, j\le m,$ we have 
\begin{align*}
&p_{ij}^{(0,1)}(a,b,c;g)=(-)^{i+j}
\prod_{r=1}^{i}
\frac{s(\frac{m-i+r}{2}g)}
{s(\frac{r}{2}g)}
\nonumber\\[6pt]
&\times
\frac{s(b+c+(j-\frac{1}{2})g)
\prod_{r=1}^{i}s(a+b+c+\frac{m+r-2}{2}g)
\prod_{r=1}^{j}s(c+\frac{r-1}{2}g)
\prod_{r=1}^{m}s(a+\frac{r-1}{2}g)}
{\prod_{r=1}^{m+1}s(b+c+\frac{j+r-2}{2}g)
\prod_{r=1}^{i}s(a+\frac{r-1}{2}g)
\prod_{r=1}^{j}s(a+\frac{m-j+r-1}{2}g)}
\nonumber\\[6pt]
&\times\sum_{l\ge 0}\; \prod_{r=1}^l\;
\frac{s(\frac{-j+r-1}{2}g)s(\frac{-i+r-1}{2}g)
s(b+c+\frac{j+r-2}{2}g)s(a+c+\frac{i+r-2}{2}g)}
{s(\frac{-m+r-1}{2}g)s(c+\frac{r-1}{2}g)
s(a+b+c+\frac{m+r-2}{2}g)
s(\frac{r}{2}g)},
\end{align*}

\noindent
or, equivalently, 

\begin{align*}
&p_{ij}^{(0,1)}(a,b,c;g)\\[5pt]
&=\frac{1-e(2(b+c))q^{2j-1}}{1-e(2(b+c))q^{j-1}}
\frac{(e(2(a+b+c))q^{m-1}, q^{-m};q)_i}{(e(2a), q;q)_i}\nonumber\\[6pt]
&\times\frac{(e(2(b+c)), e(2c);q)_j}{(e(2(b+c))q^m, e(-2a)q^{1-m};q)_j}
\frac{(e(2a);q)_m}{(e(2(b+c));q)_m}\nonumber\\[6pt]
&\times
e((-m-j)a+(m-i)b+(m-i-j)c)q^i\nonumber\\[6pt]
&\times{}_4\varphi_3\hs{q^{-i},\; e(2(a+c))q^{i-1},\; 
q^{-j},\; e(2(c+b))q^{j-1}}
{e(2c),\; q^{-m},\; e(2(a+b+c))q^{m-1}}
{q,\; q}.
\end{align*}

\noindent
Here 
$$
{}_4\varphi_3\hs{q^{-i},\; e(2(a+c))q^{i-1},\; 
q^{-j},\; e(2(c+b))q^{j-1}}
{e(2c),\; q^{-m},\; e(2(a+b+c))q^{m-1}}
{q,\; q}$$
can be considered as 
$$
W_i(j; e(2c)q^{-1}, e(2a)q^{-1}, e(2(b+c))q^{m-1}, m ;q)
$$
or
$$
W_j(i; e(2c)q^{-1}, e(2b)q^{-1}, e(2(a+c))q^{m-1}, m ;q),
$$

\noindent
where $W_n(x; a', b', c', N ;q)$ denotes the $q$-Racah polynomial
defined by (1.6).

}
\medskip

\noindent
{\bf Proof.} In Proposition 3.6 in the next section, 
set $\lambda_1=a, \lambda_2=c, \lambda_3=b$
and $z_1=0, z_2=z, z_3=1$. Then we obtain
the required result. \hfill$\square$

\bigskip

On the other hand, it is easily seen from (2.2-5)  that
$J_j(a,b,c;g;z)=I_j(b,a,c;g;1-z)$ for $j=0,1$. 
In general $m$ case, we have the following.
\medskip

\noindent
{\bf Proposition 2.4.}\; {\it For $0\le j\le m$, }
\begin{equation*}
J_j(a,b,c;g;z)=I_j(b,a,c;g;1-z).
\end{equation*}

\noindent
{\bf Proof.}\; \comment{Note that
\begin{align*}
&J_j(a,b,c; g;z)\nonumber\\[6pt]
&:=\int_{D_{m-j,0, j,0}(t)}u_{D_{m-j,0, j,0}}(t)\,dt_1\cdots\,dt_m
\nonumber\\[6pt]
&=\int
\prod_{1\le i_1<i_2\le m-j}(t_{i_2}-t_{i_1})^g
\prod_{1\le i\le m-j}
(-t_i)^a(1-t_i)^b(z-t_i)^c
\;\nonumber\\[6pt]
&\times
\prod_{m-j< i_1<i_2\le m}(t_{i_2}-t_{i_1})^g
\prod_{m-j< i\le m}
t_i^a(1-t_i)^b(t_i-z)^c
\;\nonumber\\[6pt]
&\times
\prod_{\substack{
1\le i_1\le m-j\\[3pt]
m-j\le i_2\le m}}
(t_{i_2}-t_{i_1})^g\;
dt_1\cdots dt_m,
\end{align*}
where the domain of integration is

$$t_1<\cdots<t_{m-j}<0,\quad z<t_{j+1}<\cdots<t_m<1$$
\smallskip

\noindent
with the standard orientation.}
The change of the integration variables 
$t_i\mapsto 1-t_i\;(1\le i\le m)$ in
$$J_j(a,b,c; g;z)
=m!\,\int_{D_{m-j,0, j,0}(t)}u_{D_{m-j,0, j,0}}(t)\,dt_1\cdots\,dt_m$$ 
leads to 
\comment{
\begin{align*}
&\int
\prod_{1\le i_1<i_2\le m-j}(t_{i_1}-t_{i_2})^g
\prod_{1\le i\le m-j}
t_i^b(t_i-1)^a(t_i-(1-z))^c
\;\nonumber\\[6pt]
&\times
\prod_{m-j< i_1<i_2\le m}(t_{i_1}-t_{i_2})^g
\prod_{m-j< i\le m}
t_i^b(1-t_i)^a(1-z-t_i)^c
\;\nonumber\\[6pt]
&\times
\prod_{\substack{
1\le i_1\le m-j\\[3pt]
m-j\le i_2\le m}}
(t_{i_1}-t_{i_2})^g\;
dt_1\cdots dt_m,
\end{align*}
where
$$0<t_m<\cdots<t_{m-j+1}<1-z,\quad 1<t_{m-j}<\cdots<t_1$$
\smallskip

\noindent
with the standard orientation, which
is }
$I_j(b,a,c;g;1-z).$ \hfill$\square$

\bigskip

Therefore, combining Theorem 2.3 and Proposition 2.4, 
we reach the following.

\bigskip

\noindent
{\bf Corollary 2.5.}\; {\it For $0\le i,j \le m$, we have
\begin{align*}
&\delta_{ij}
=h_i(e(2c)q^{-1},e(2a)q^{-1}, e(2(b+c))q^{m-1},m;q)\nonumber\\[6pt]
&\times\sum_{0\le x\le m}
\rho(x;a', b', c',m;q)
W_i(x; a', b', c',m;q)
W_j(x; a', b', c',m;q),
\end{align*}

\noindent
where
$$a'=e(2c)q^{-1},b'=e(2a)q^{-1}, c'=e(2(b+c))q^{m-1}.$$}
\medskip

\noindent
{\bf Proof.}\; By Proposition 2.4, we have
\begin{align*}
&I_i(a,b,c;g;z)=\sum_{x=0}^m p_{ix}^{(0,1)}(a,b,c;g)
J_x(a,b,c;g;z)\nonumber\\[6pt]
&=\sum_{x=0}^m p_{ix}^{(0,1)}(a,b,c;g)I_x(b,a,c;g;1-z)\nonumber\\[6pt]
&=\sum_{x=0}^m p_{ix}^{(0,1)}(a,b,c;g)\sum_{j=0}^m 
p_{xj}^{(0,1)}(b,a,c;g)I_j(a,b,c;g;z),
\end{align*}

\noindent
hence

\begin{align*}
\sum_{x=0}^m p_{ix}^{(0,1)}(a,b,c;g) p_{xj}^{(0,1)}(b,a,c;g)=\delta_{ij}.
\end{align*}

\medskip

\noindent
On the other hand, by Theorem 2.3, we have

\begin{align*}
&\sum_{x=0}^m p_{ix}^{(0,1)}(a,b,c,g)p_{xj}^{(0,1)}(b,a,c,g)\\[6pt]
&=
\frac{(e(2a), e(2b);q)_m}{(e(2(a+c)), e(2(b+c));q)_m}
\frac{(e(2(a+b+c))q^{m-1}, q^{-m};q)_i}{(e(2a), q;q)_i}
\nonumber\\[6pt]
&\times
\frac{1-e(2(a+c))q^{2j-1}}{1-e(2(a+c))q^{j-1}}
\frac{(e(2(a+c)), e(2c);q)_j}{(e(2(a+c))q^m, e(-2b)q^{1-m};q)_j}
\nonumber\\[6pt]
&\times 
e((-i-j)b+(2m-i-j)c)\,q^i\nonumber\\[6pt]
&\times\sum_{x=0}^m
\frac{1-e(2(b+c))q^{2x-1}}{1-e(2(b+c))q^{x-1}}
\frac{(e(2(b+c)), e(2c), e(2(a+b+c))q^{m-1}, q^{-m};q)_x}
{(e(2(b+c))q^m, e(-2a)q^{1-m}, e(2b), q ;q)_x}
\nonumber\\[6pt]
&\times
e(-2x(a+c))\,q^x\,
\nonumber\\[6pt]
&\times 
{}_4\varphi_3\hs{q^{-i},\; e(2(a+c))q^{i-1},\; 
q^{-x},\; e(2(c+b))q^{x-1}}
{e(2c),\; q^{-m},\; e(2(a+b+c))q^{m-1}}
{q,\; q}\nonumber\\[6pt]
&\times{}_4\varphi_3\hs{q^{-x},\; e(2(b+c))q^{x-1},\; 
q^{-j},\; e(2(c+a))q^{j-1}}
{e(2c),\; q^{-m},\; e(2(a+b+c))q^{m-1}}
{q,\; q}.
\end{align*}

\noindent
Therefore we obtain the required result. \hfill$\square$

\bigskip

\subsection{The solutions around $0$ in terms of
those around $\infty$}

In this subsection,  we give formulas that connect the fudamental
set of solutions around $0$ with the set of solutions around $\infty.$

For convenience to our purpose, 
we fix a complex variable
$z$ to be real such that $z<0$ and assign the names
$D_{0,j,0, m-j}(t)$ and  $D_{m-j,0,j,0}(t)$ for $0\le j\le m$  
to the domains of the real manifold $T_{\Bbb R}$  by
\begin{align*}
&D_{0,j,0, m-j}(t)=\{(t_1,\ldots,t_m)\mid
z<t_1<\cdots<t_j<0,\; 1<t_{j+1}<\cdots<t_m\},\\[6pt]
&D_{m-j,0,j,0}(t)=\{(t_1,\ldots,t_m)\mid
t_1<\cdots<t_{m-j}<z,\; 0<t_{m-j+1}<\cdots<t_m<1\},
\end{align*}

\noindent
where each orientation is natural one.  Correspondingly,
we define the loaded cycles $C_{0,j,0, m-j}$ and
$C_{m-j,0,j,0}$ to be
$$C_{0,j,0, m-j}=\sum_{\sigma\in S_m}\sigma
\left\{D_{0,j,0, m-j}(t)\otimes u_{D_{0,j,0, m-j}}(t)\right\}$$
and
$$C_{m-j,0,j,0}=\sum_{\sigma\in S_m}\sigma
\left\{D_{m-j,0,j,0}(t)\otimes u_{D_{m-j,0,j,0}}(t)\right\}.$$ 
Then the integrals \begin{align*}
&I_j(a,b,c; g;z)
=\langle\, {\rm reg}\,C_{0,j,0, m-j}, dt_1\ldots dt_m\rangle\\[5pt]
&=m!\int_{D_{0,j,0, m-j}(t)}u_{D_{0,j,0, m-j}}(t)\,dt_1\cdots\,dt_m
\end{align*}
and
\begin{align*}
&K_j(a,b,c; g;z)
=\langle\, {\rm reg}\,C_{m-j, 0, j, 0}, dt_1\ldots dt_m\rangle\\[5pt]
&=m!\int_{D_{m-j, 0, j, 0}(t)}u_{D_{m-j, 0, j, 0}}(t)\,dt_1\cdots\,dt_m
\end{align*}
give fundamental set of solutions around $0$ and 
that around $\infty$, respectively.
Indeed, we have the folowing.
\medskip

\noindent
{\bf Proposition 2.6.} (1)\; {\it For $0\le j\le m,$ we have}

\begin{align*}
&I_j(a,b,c;g;z)\comment{=m!\,
\int_{D_{0,j,0, m-j}}u_{D_{0,j,0, m-j}}(t)\,dt_1\cdots\,dt_m}
\nonumber\\[6pt]
&=m!\;
S_j(c+1, a+1, g/2)\,
S_{m-j}(-a-b-c-(m-1)g-1, b+1, g/2)
\nonumber\\[6pt]
&\times (-z)^{(a+c+1)j+{j\choose 2} g}\;(1+O(z))\quad (z\rightarrow 0).
\end{align*}
\medskip

\noindent
(2)\; {\it For $0\le j\le m,$ we have}
\begin{align*}
&K_j(a,b,c;g;z)\comment{=m!
\int_{D_{j,0, m-j,0}}u_{D_{j,0, m-j, 0 }}(t)\,dt_1\cdots\,dt_m}
\nonumber\\[6pt]
&=m!\;
S_{j}(-a-b-c-(m-1)g-1, b+1, g/2)\,
S_{m-j}(a+1, b+1, g/2)
\nonumber\\[6pt]
&\times 
\left(-z^{-1}\right)^{-(a+b+1)j-mc-\left\{{j\choose 2}+j(m-j)\right\} g}\;
(1+O(z^{-1}))\quad (z\rightarrow \infty)
\end{align*}

\noindent
{\it Here the arguments of $-z$ of $(-z)^{(a+c+1)j+{j\choose 2} g}$
and $-z^{-1}$ of  $\left(-z^{-1}\right)
^{-(a+b+1)j-mc-\left\{{j\choose 2}+j(m-j)\right\} g}$ 
are fixed to be zero on $z<0$, and $S(\alpha, \beta, \gamma)$
is the Selberg integral (2.1).}
\medskip 

\noindent
{\bf Proof.}\;(1)\;
The change of the integration variables 
$t_i\mapsto zt_i$ for $1\le i\le j$ and
$t_i\mapsto t_i^{-1}$ for $j< i\le m$ and the binomial
theorem imply the result.
\comment{
\begin{align*}
&\int_{D_{0, j,0, m-j}}u_{D_{0, j,0, m-j}}(t)\,dt_1\cdots\,dt_m
\nonumber\\[6pt]
&=
(-z)^{(a+c+1)j+{j\choose 2}\,g}
\int
\prod_{1\le i_1<i_2\le j}(t_{i_1}-t_{i_2})^g
\prod_{1\le i\le j}
t_i^a
(1-zt_i)^b
(1-t_i)^c\nonumber\\[6pt]
&\times
\prod_{j< i_1<i_2\le m}(t_{i_1}-t_{i_2})^g
\prod_{j< i\le m}
t_i^{-a-b-c-2-(m-1)g}
(1-zt_i)^c(1-t_i)^b
\;\nonumber\\[6pt]
&\times
\prod_{\substack{
1\le i_1\le j\\[3pt]
j< i_2\le m}}
(1-zt_{i_1}t_{i_2})^g\;
dt_1\cdots dt_m,
\end{align*}
where
$$0<t_j<\cdots<t_1<1,\quad 0<t_m<\cdots<t_{j+1}<1$$

\noindent
with the standard orientation. The binomial theorem
leads to the result required.}

\medskip

\noindent
(2)\; The change of the integration variables 
$t_i\mapsto zt_i^{-1}$ for $1\le i\le j$ 
($t_i$ for $j< i\le m$ is fixed) and the binomial
theorem imply the result. 
\comment{
\begin{align*}
&\int_{D_{j,0, m-j,0}}u_{D_{j,0, m-j,0}}(t)\,dt_1\cdots\,dt_m
\nonumber\\[6pt]
\comment{&=\int_{D_{0,j,0, m-j}}
\prod_{1\le i_1<i_2\le j}(t_{i_2}-t_{i_1})^g
\prod_{1\le i\le j}
(z-t_i)^{\lambda_1}(-t_i)^{\lambda_2}(1-t_i)^{\lambda_3}
\;\nonumber\\[6pt]
&\times
\prod_{j< i_1<i_2\le m}(t_{i_2}-t_{i_1})^g
\prod_{j< i\le m}
(t_i-z)^{\lambda_1}t_i^{\lambda_2}(1-t_i)^{\lambda_3}
\;\nonumber\\[6pt]
&\times
\prod_{\substack{
1\le i_1\le j\\[3pt]
j\le i_2\le m}}
(t_{i_2}-t_{i_1})^g\;
dt_1\cdots dt_m}
&=(-z)^{(a+b+1)j+mc+\{{j\choose 2}+j(m-j)\}\,g}
\nonumber\\[6pt]
&\times\int
\prod_{1\le i_1<i_2\le j}(t_{i_2}-t_{i_1})^g
\prod_{1\le i\le j}
t_i^{-a-b-c-2-(m-1)g}(1-t_i)^b
(1-z^{-1}t_i)^c
\;\nonumber\\[6pt]
&\times
\prod_{j< i_1<i_2\le m}(t_{i_2}-t_{i_1})^g
\prod_{j< i\le m}
(1-z^{-1}t_i)^ct_i^a(1-t_i)^b
\;\nonumber\\[6pt]
&\times
\prod_{\substack{
1\le i_1\le j\\[3pt]
j< i_2\le m}}
(1-z^{-1}t_{i_1}t_{i_2})^g\;
dt_1\cdots dt_m,
\end{align*}
where
$$0<t_1<\cdots<t_j<1,\quad 0<t_{j+1}<\cdots<t_m<1$$
\medskip

\noindent
with the standard orientation. The binomial theorem
leads to the result required.}\hfill$\square$

\medskip

\noindent
When $m=1$, Proposition 2.6 corresponds to
\begin{align}
&I_0(a,b,c;z)\nonumber\\[6pt]
&=B(b+1,-a-b-c-1)\;{}_2F_1
\hs{-c,-a-b-c-1}{-a-c}{z},\\[6pt]
&I_1(a,b,c;z)\nonumber\\[6pt]
&=B(a+1,c+1)\;(-z)^{a+c+1}\,{}_2F_1
\hs{-b,a+1}{a+c+2}{z}
\end{align}
and
\begin{align}
K_0(a,b,c;z)&=B(c+1,-a-b-c-1)\nonumber\\[6pt]
&\times
\;\left(-\frac{1}{z}\right)^{-a-b-c-1}
\;{}_2F_1
\hs{-b,-a-b-c-1}{-a-b}{z^{-1}},\\[6pt]
K_1(a,b,c;z)&=B(a+1,b+1)\nonumber\\[6pt]
&\times
\;\left(-\frac{1}{z}\right)^{-c}\,{}_2F_1
\hs{-c,a+1}{a+b+2}{z^{-1}},
\end{align}
where
$B(a,b)$ denotes the beta function.
\medskip

\noindent
The linear relations between $\{I_0, I_1\}$ and $\{K_0, K_1\}$
are expressed by

\begin{align*}
&I_0(a,b,c;z)=\frac{s(c)}{s(a+b)}K_0(a,b,c;z)
+\frac{-s(a)}{s(a+b)}K_1(a,b,c;z),\nonumber\\[6pt]
&I_1(a,b,c;z)=\frac{-s(a+b+c)}{s(a+b)}K_0(a,b,c;z)
+\frac{-s(b)}{s(a+b)}K_1(a,b,c;z)\nonumber\\[6pt]
\end{align*}

\noindent
and

\medskip
\begin{align*}
&K_0(a,b,c;z)=\frac{s(b)}{s(a+c)}I_0(a,b,c;z)
+\frac{-s(a)}{s(a+c)}I_1(a,b,c;z),\nonumber\\[6pt]
&K_1(a,b,c;z)=\frac{-s(a+b+c)}{s(a+c)}I_0(a,b,c;z)
+\frac{-s(c)}{s(a+c)}I_1(a,b,c;z).
\end{align*}

\noindent
These are the connection formulas between
the solutions around $0$ and those around $\infty$.
\medskip

In general $m$, if we define $p_{ij}^{(0,\infty)}(a,b,c;g)$ 
for $0\le i,j\le m$ by 
\begin{equation*}
I_i(a,b,c;g;z)=\sum_{0\le j\le m} 
p_{ij}^{(0,\infty)}(a,b,c;g)\, K_j(a,b,c;g;z),
\end{equation*}

\noindent
we have the following two expressions.
\medskip

\noindent
{\bf Theorem 2.7.}\; {\it For $0\le i,j\le m,$ we have
\begin{align}
&p_{ij}^{(0,\infty)}(a,b,c;g)\nonumber\\[6pt]
&=(-)^i\sum_{\substack{
0\le k\le m-i\\[3pt]
0\le l\le i\\[3pt]
k+l=j}}
(-)^{k}
\prod_{r=1}^{m-i-k}
\frac{s\bigl(\,c+\frac{i+r-1}{2}g\,\bigr)}
{s\bigl(\,a+b+\left(k+\frac{i+r-1}{2}\right)g\,\bigr)}
\prod_{r=1}^{k}
\frac{s\bigl(a+\frac{i+r-1}{2}g\,\bigr)}
{s\bigl(a+b+\left(k+\frac{i-r-1}{2}\right)g\,\bigr)}
\nonumber\\[6pt]
&\times
\prod_{r=1}^{i-l}
\frac{s\bigl(a+b+c+\frac{m+i+k-r-1}{2}g\,\bigr)
s\bigl(\frac{m-i-k+r}{2}g\,\bigr)}
{s\bigl(a+b+\left(j+\frac{r-1}{2}\right)g\,\bigr)
s\bigl(\frac{r}{2}g\,\bigr)}
\prod_{r=1}^{l}
\frac{s\bigl(b+\frac{k+r-1}{2}g\,\bigr)
s\bigl(\frac{k+r}{2}g\,\bigr)}
{s\bigl(a+b+\left(j-\frac{r+1}{2}\right)g\,\bigr)
s\bigl(\frac{r}{2}g\,\bigr)}
\end{align}

\noindent
and

\begin{align}
&p_{ij}^{(0,\infty)}(a,b,c;g)\nonumber\\[6pt]
&=(-)^i
\sum_{\substack{
0\le k\le i\\[3pt]
0\le l\le m-i\\[3pt]
k+l=j}}
(-)^{l}
\prod_{r=1}^{i-k}
\frac{s\bigl(\,a+b+c+\frac{m+i-r-1}{2}g\,\bigr)}
{s\bigl(\,a+b+\left(k+\frac{m-i+r-1}{2}\right)g\,\bigr)}
\prod_{r=1}^{k}
\frac{s\bigl(\,b+\frac{m-i+r-1}{2}g\,\bigr)}
{s\bigl(\,a+b+\left(k+\frac{m-i-r-1}{2}\right)g\,\bigr)}
\nonumber\\[6pt]
&\times
\prod_{r=1}^{m-i-l}
\frac{s\bigl(\,c+\frac{i-k+r-1}{2}g\,\bigr)
s\bigl(\,\frac{i-k+r}{2}g\,\bigr)}
{s\bigl(\,a+b+\left(j+\frac{r-1}{2}\right)g\,\bigr)
s\bigl(\,\frac{r}{2}g\,\bigr)}
\prod_{r=1}^{l}
\frac{s\bigl(\,a+\frac{k+r-1}{2}g\,\bigr)
s\bigl(\,\frac{k+r}{2}g\,\bigr)}
{s\bigl(\,a+b+\left(j-\frac{r+1}{2}\right)g\,\bigr)
s\bigl(\,\frac{r}{2}g\,\bigr)}.
\end{align}
\medskip

\noindent
where $s(A)=\sin(\pi\,A).$}

\medskip

\noindent
{\bf Proof.}\; Set $\lambda_1=c, \lambda_2=a, \lambda_3=b$ and
$z_1=z, z_2=0, z_3=1$ in Proposition 3.3. Or change $a$ and $c$
in Theorem 2.2. Then we reach the required
result. \hfill$\square$

\bigskip

Similar to Theorem 2.3, we also have the following.

\medskip

\noindent
{\bf Theorem 2.8.}\; {\it For $0\le i, j\le m,$ we have 
\begin{align*}
&p_{ij}^{(0,\infty)}(a,b,c;g)=(-)^{i+j}
\prod_{r=1}^{i}
\frac{s(\frac{m-i+r}{2}g)}
{s(\frac{r}{2}g)}
\nonumber\\[6pt]
&\times
\frac{s(b+a+(j-\frac{1}{2})g)
\prod_{r=1}^{i}s(a+b+c+\frac{m+r-2}{2}g)
\prod_{r=1}^{j}s(a+\frac{r-1}{2}g)
\prod_{r=1}^{m}s(c+\frac{r-1}{2}g)}
{\prod_{r=1}^{m+1}s(b+a+\frac{j+r-2}{2}g)
\prod_{r=1}^{i}s(c+\frac{r-1}{2}g)
\prod_{r=1}^{j}s(c+\frac{m-j+r-1}{2}g)}
\nonumber\\[6pt]
&\times\sum_{l\ge 0}\; \prod_{r=1}^l\;
\frac{s(\frac{-j+r-1}{2}g)s(\frac{-i+r-1}{2}g)
s(b+a+\frac{j+r-2}{2}g)s(a+c+\frac{i+r-2}{2}g)}
{s(\frac{-m+r-1}{2}g)s(a+\frac{r-1}{2}g)
s(a+b+c+\frac{m+r-2}{2}g)
s(\frac{r}{2}g)},
\end{align*}

\noindent
or, equivalently, 
\begin{align*}
&p_{ij}^{(0,\infty)}(a,b,c;g)\\[5pt]
&=\frac{1-e(2(a+b))q^{2j-1}}{1-e(2(a+b))q^{j-1}}
\frac{(e(2(a+b+c))q^{m-1}, q^{-m};q)_i}{(e(2c), q;q)_i}\nonumber\\[6pt]
&\times\frac{(e(2(a+b)), e(2a);q)_j}{(e(2(a+b))q^m, e(-2c)q^{1-m};q)_j}
\frac{(e(2c);q)_m}{(e(2(a+b));q)_m}\nonumber\\[6pt]
&\times
e((m-i-j)a+(m-i)b+(-m-j)c)\,q^i\nonumber\\[6pt]
&\times{}_4\varphi_3\hs{q^{-i},\; e(2(a+c))q^{i-1},\; 
q^{-j},\; e(2(a+b))q^{j-1}}
{e(2a),\; q^{-m},\; e(2(a+b+c))q^{m-1}}
{q,\; q}.
\end{align*}

\noindent
Here 
$$
{}_4\varphi_3\hs{q^{-i},\; e(2(a+c))q^{i-1},\; 
q^{-j},\; e(2(a+b))q^{j-1}}
{e(2a),\; q^{-m},\; e(2(a+b+c))q^{m-1}}
{q,\; q}$$
can be considered as 
$$
W_i(j; e(2a)q^{-1}, e(2c)q^{-1}, e(2(a+b))q^{m-1}, m ;q)
$$
or
$$
W_j(i; e(2a)q^{-1}, e(2b)q^{-1}, e(2(a+c))q^{m-1}, m ;q),
$$

\noindent
where $W_n(x; a', b', c', N ;q)$ denotes the $q$-Racah polynomial
defined by (1.6).

}
\medskip

\noindent
{\bf Proof.} In Proposition 3.6, set $\lambda_1=c, \lambda_2=a, \lambda_3=b$
and $z_1=z, z_2=0, z_3=1$. Then we reach
the required result. \hfill$\square$
\bigskip

On the other hand, it is seen from (2.8-11) that 
$$K_j(a,b,c;z)=(-z)^{a+b+c+1}
I_j(a,c,b;z^{-1})$$

\noindent
for $j=0,1.$  More generally, we have the following.
\medskip

\noindent
{\bf Proposition 2.9.}\; For $0\le j\le m$, 
\begin{equation*}
K_j(a,b,c;g;z)=(-z)^{(a+b+c+1)m+{m\choose 2}g}I_j(a,c,b;g;z^{-1}).
\end{equation*}
\medskip

\noindent
{\bf Proof.}\; The change of integration variables in
$K_j(a,b,c;g;z)$ such as $t_i\mapsto zt_i\;(1\le i\le m)$
implies the required result. \hfill$\square$

\bigskip

\section{Derivation of the connection coefficients}
In this section, let ${\mathcal L}$ be the local system 
determined by the function
\begin{equation*}
u(t)=\prod_{1\le i<j\le m}(t_j-t_i)^g
\prod_{\substack{1\le i\le m\\1\le j\le 3}}
(t_i-z_j)^{\lambda_j}
\end{equation*}

\noindent
on the domain 

$$T=\{t=(t_1,\ldots, t_m)\in {\Bbb C}^m\mid
t_i\neq t_j\; (i\neq j),\; t_i\neq z_1,\,z_2,\,z_3\;\},$$

\noindent
where  $z_1, z_2, z_3$ are fixed to be real and  $z_1<z_2<z_3.$

Each connection formula in the previous section is obtained as a 
special case of the formula in this section. We obtain the formulas in
\S\S 2.1, if we set $z_1=0, z_2=z, z_3=1$ and $\lambda_1=a,
\lambda_2=c, \lambda_3=b.$ Similarly, we obtain the formulas in
\S\S 2.2, if we set $z_1=z, z_2=0, z_3=1$ and $\lambda_1=c,
\lambda_2=a, \lambda_3=b.$ 
\medskip

Set the loaded cycle 
$$C_{i_1j_1i_2j_2}=\sum_{\sigma\in S_m}\sigma
\left\{D_{i_1j_1i_2j_2}(t)\otimes u_{D_{i_1j_1i_2j_2}}(t)\right\},$$
for $i_1,j_1,i_2,j_2\in {\Bbb Z}_{\ge 0}$ with $i_1+j_1+i_2+j_2=m$, 
where $D_{i_1j_1i_2j_2}(t)$ is the domain of $T_{\Bbb R}$  
defined by the inequalities
\begin{align*}
&t_1<t_2<\cdots<t_{i_1}<z_1,\\[6pt]
&z_1<t_{i_1+1}<t_{i_1+2}<\cdots<t_{i_1+j_1}<z_2,\\[6pt]
&z_2<t_{i_1+j_1+1}<t_{i_1+j_1+2}<\cdots<t_{i_1+j_1+i_2}<z_3,\\[6pt]
&z_3<t_{i_1+j_1+i_1+1}<t_{i_1+j_1+i_1+2}<\cdots<t_{i_1+j_1+i_2+j_2}
\end{align*}
with the standard orientation.
\medskip

\noindent
Define the symbols $e(A), \;s(A),\;  
[n]_q, \langle\,A\,\,\rangle_n$ and 
$\lambda_{ijk\cdots l}$ to be
\begin{align*}
&e(A)=e^{\pi\sqrt{-1} A},\quad s(A)=\sin\pi A, \\[6pt]
&[n]_q=1+q+\cdots+q^{n-1}=\frac{1-q^n}{1-q},\\[6pt]
&\langle\,A\,\,\rangle_n=A[n]_q-A^{-1}[n]_{q^{-1}},\\[6pt]
&\lambda_{ijk\cdots l}=\lambda_i+\lambda_j+\lambda_k+\cdots+\lambda_l
\end{align*}
for brevity. Note that
$$\langle\,A\,\,\rangle_n
=\langle\,Aq^{\frac{n-1}{2}}\,\,\rangle_1
\frac{
\langle\,q^{{n}/{2}}\,\rangle_1
}
{
\langle\,q^{{1}/{2}}\,\rangle_1
}.$$
Hence, when $q=e(g)$, we have
\begin{equation*}
\langle\,e(\lambda)\,\,\rangle_n
=
\comment{
\langle\,e(\lambda)q^{\frac{n-1}{2}}\,\,\rangle_1
\frac{
\langle\,q^{{n}/{2}}\,\rangle_1
}
{
\langle\,q^{{1}/{2}}\,\rangle_1
}
=\langle\,e(\lambda+{\frac{n-1}{2}g)}\,\,\rangle_1
\frac{
\langle\,e(\frac{n}{2}g)\,\rangle_1
}
{
\langle\,e(\frac{1}{2}g)\,\rangle_1
}\\[6pt]
&=}
2\sqrt{-1}\;s\left(\lambda+\frac{n-1}{2}g\right)
\frac{s(\frac{n}{2}g)}
{s(\frac{1}{2}g)}.
\end{equation*}
In what follows, we fix $q$ to be $e(g).$
\medskip

\subsection{Connection coefficients}

\medskip

\noindent
{\bf Lemma 3.1.}\; $(1)$ {\it  For integers $i_1, i_2, j_1\ge 0, j_2\ge 1,$
we have}
\begin{align}
C_{i_1 j_1 i_2 j_2}
&=\frac{\langle\,e(\lambda_{1})q^{\frac{j_1}{2}}\,\rangle_{i_1+1}}
{\langle\,e(\lambda_{23})q^{i_2+\frac{j_1}{2}}\,\rangle_{j_2}}\,
C_{i_1+1,j_1,i_2,j_2-1}\nonumber\\[6pt]
&-\frac{\langle\,e(\lambda_{2})q^{\frac{j_1}{2}}\,\rangle_{i_2+1}}
{\langle\,e(\lambda_{23})q^{i_2+\frac{j_1}{2}}\,\rangle_{j_2}}\,
C_{i_1,j_1,i_2+1,j_2-1}.
\end{align}

\noindent
$(2)$ {\it For integers $i_1, i_2, j_2\ge 0, j_1\ge 1,$
we have}
\begin{align}
C_{i_1 j_1 i_2j_2}
=&-\frac{\langle\,e(\lambda_{123})q^{i_2+j_1+\frac{j_2}{2}-1}\,\rangle_{i_1+1}}
{\langle\,e(\lambda_{23})q^{i_2+\frac{j_2}{2}}\,\rangle_{j_1}}\,
C_{i_1+1,j_1-1,i_2,j_2}\nonumber\\[6pt]
&-
\frac{\langle\,e(\lambda_{2})q^{\frac{j_2}{2}}\,\rangle_{i_2+1}}
{\langle\,e(\lambda_{23})q^{i_2+\frac{j_2}{2}}\,\rangle_{j_1}}\,
C_{i_1,j_1-1,i_2+1,j_2}.
\end{align}

\noindent
{\bf Proof.} (1) Fix a point $(t_1,\ldots, t_{m-1})$ of 
$D_{i_1, j_1, i_2, i_2-1}(t_1, \ldots, t_{m-1}),$
where $i_1+j_1+i_2+j_2=m-1$. Then it is seen that
a trivial loop with clockwise direction in the lower
half plane of the $t_m$-plane is homologous to
\begin{align*}
&\sum_{s=1}^{i_1+1}q^{s-1}
D_{i_1+1, j_1, i_2, j_2-1}(t_1,\ldots,t_{s-1}, t_m, t_s,\,\ldots, t_{m-1})
\nonumber\\[6pt]
+&e(\lambda_1)
\sum_{s=i_1+1}^{i_1+j_1+1}q^{s-1}
D_{i_1, j_1+1, i_2, j_2-1}(t_1,\ldots,t_{s-1}, t_m, t_s, \ldots, t_{m-1})
\nonumber\\[6pt]
+&e(\lambda_{12})
\sum_{s=i_1+j_1+1}^{i_1+j_1+i_2+1}q^{s-1}
D_{i_1, j_1, i_2+1, j_2-1}(t_1,\ldots,t_{s-1}, t_m, t_s, \ldots, t_{m-1})
\nonumber\\[6pt]
+&e(\lambda_{123})
\sum_{s=i_1+j_1+i_2+1}^{m}q^{s-1}
D_{i_1, j_1, i_2, j_2}(t_1,\ldots,t_{s-1}, t_m, t_s, \ldots, t_{m-1})
\end{align*}
and a trivial loop with counterclockwise direction in the upper
half plane of the $t_m$-plane is homologous to
\begin{align*}
&\sum_{s=1}^{i_1+1}q^{-s+1}
D_{i_1+1, j_1, i_2, j_2-1}(t_1,\ldots,t_{s-1}, t_m, t_s,\,\ldots, t_{m-1})
\nonumber\\[6pt]
+&e(-\lambda_1)
\sum_{s=i_1+1}^{i_1+j_1+1}q^{-s+1}
D_{i_1, j_1+1, i_2, j_2-1}(t_1,\ldots,t_{s-1}, t_m, t_s, \ldots, t_{m-1})
\nonumber\\[6pt]
+&e(-\lambda_{12})
\sum_{s=i_1+j_1+1}^{i_1+j_1+i_2+1}q^{-s+1}
D_{i_1, j_1, i_2+1, j_2-1}(t_1,\ldots,t_{s-1}, t_m, t_s, \ldots, t_{m-1})
\nonumber\\[6pt]
+&e(-\lambda_{123})
\sum_{s=i_1+j_1+i_2+1}^{m}q^{-s+1}
D_{i_1, j_1, i_2, j_2}(t_1,\ldots,t_{s-1}, t_m, t_s, \ldots, t_{m-1}).
\end{align*}

\noindent
It implies 
\begin{align*}
&\sum_{s=1}^{i_1+1}q^{s-1}
C_{i_1+1, j_1, i_2, j_2-1}
+e(\lambda_1)
\sum_{s=i_1+1}^{i_1+j_1+1}q^{s-1}
C_{i_1, j_1+1, i_2, j_2-1}
\nonumber\\[6pt]
+&e(\lambda_{12})
\sum_{s=i_1+j_1+1}^{i_1+j_1+i_2+1}q^{s-1}
C_{i_1, j_1, i_2+1, j_2-1}
+e(\lambda_{123})
\sum_{s=i_1+j_1+i_2+1}^{m}q^{s-1}
C_{i_1, j_1, i_2, j_2}=0
\end{align*}
and
\begin{align*}
&\sum_{s=1}^{i_1+1}q^{-s+1}
C_{i_1+1, j_1, i_2, j_2-1}
+e(-\lambda_1)
\sum_{s=i_1+1}^{i_1+j_1+1}q^{-s+1}
C_{i_1, j_1+1, i_2, j_2-1}
\nonumber\\[6pt]
+&e(-\lambda_{12})
\sum_{s=i_1+j_1+1}^{i_1+j_1+i_2+1}q^{-s+1}
C_{i_1, j_1, i_2+1, j_2-1}
+e(-\lambda_{123})
\sum_{s=i_1+j_1+i_2+1}^{m}q^{-s+1}
C_{i_1, j_1, i_2, j_2}=0,
\end{align*}
thus,
\comment{
\begin{align*}
&\;D_{i_1+1, j_1, i_2, i_2}(t_m, t_1, \ldots, t_{m-1})
+q\,D_{i_1+1, j_1, i_2, i_2}(t_1, t_m, t_2, \ldots, t_{m-1})+\cdots
\nonumber\\[6pt]
&\qquad\cdots+q^{i_1}\,D_{i_1+1, j_1, i_2, i_2}
(t_1, \ldots, t_{i_1}, t_m,t_{i_1+1},\cdots, t_{m-1})\nonumber\\[5pt]
+&q^{i_1}e(\lambda_1)\,D_{i_1, j_1+1, i_2, i_2}
(t_1, \ldots, t_{i_1}, t_m,t_{i_1+1},\cdots, t_{m-1}) 
+q^{i_1+1}e(\lambda_1)\,D_{i_1, j_1+1, i_2, i_2}
(t_1, \ldots, t_{i_1+1}, t_m,t_{i_1+2},\cdots, t_{m-1})
+\cdots\nonumber\\[6pt]
&\qquad\cdots+q^{i_1+j_1}e(\lambda_1)
\,D_{i_1, j_1+1, i_2, i_2}
(t_1, \ldots, t_{i_1+j_1+1}, t_m,t_{i_1+j_1+2},\cdots, t_{m-1})\nonumber\\[5pt]
+&q^{i_1+j_1}e(\lambda_{12})
\,D_{i_1, j_1+1, i_2, i_2}
(t_1, \ldots, t_{i_1+j_1+1}, t_m,t_{i_1+j_1+2},\cdots, t_{m-1}) 
+q^{i_1+j_1+1}e(\lambda_{12})\overrightarrow{(h_{i_1+j_1+1}, h_{i_1+j_1+2})}
+\cdots\nonumber\\[6pt]
&\qquad\cdots
+q^{i_1+j_1+i_2}e(\lambda_{12})
\overrightarrow{(h_{i_1+j_1+i_2}, z_3)}\nonumber\\[5pt]
+&q^{i_1+j_1+i_2}e(\lambda_{123})\overrightarrow{(z_3, h_{i_1+j_1+i_2+1})} 
+q^{i_1+j_1+i_2+1}e(\lambda_{123})
\overrightarrow{(h_{i_1+j_1+i_2+1}, h_{i_1+j_1+i_2+2})}
+\cdots\nonumber\\[6pt]
&\qquad\cdots
+q^{i_1+j_1+i_2+j_2-1}e(\lambda_{123})
\overrightarrow{(h_{i_1+j_1+i_2+j_2-1}, +\infty)}.
\end{align*}
}
\begin{align}
&[i_1+1]_qC_{i_1+1, j_1, i_2, j_2-1}\nonumber\\[6pt]
&+e(\lambda_{1})q^{i_1}[j_1+1]_qC_{i_1, j_1+1, i_2, j_2-1}\nonumber\\[6pt]
&+e(\lambda_{12})q^{i_1+j_1}[i_2+1]_qC_{i_1, j_1, i_2+1, j_2-1}\nonumber\\[6pt]
&+e(\lambda_{123})q^{i_1+j_1+i_2}[j_2]_qC_{i_1, j_1, i_2, j_2}=0
\end{align}
and
\begin{align}
&[i_1+1]_{q^{-1}}C_{i_1+1, j_1, i_2, j_2-1}\nonumber\\[6pt]
&+e(-\lambda_{1})q^{-i_1}[j_1+1]_{q^{-1}}C_{i_1, j_1+1, i_2, j_2-1}
\nonumber\\[6pt]
&+e(-\lambda_{12})q^{-i_1-j_1}[i_2+1]_{q^{-1}}C_{i_1, j_1, i_2+1, j_2-1}
\nonumber\\[6pt]
&+e(-\lambda_{123})q^{-i_1-j_1-i_2}[j_2]_{q^{-1}}C_{i_1, j_1, i_2, j_2}=0
\end{align}
in the sense of twisted homology.

Therefore, by eliminating the second terms of $(3.3)$ and $(3.4)$,  
we have 
\begin{align*}
&\left(
\frac{[i_1+1]_q}{e(\lambda_{1})q^{i_1}[j_1+1]_q}-
\frac{[i_1+1]_{q^{-1}}}{e(-\lambda_{1})q^{-i_1}[j_1+1]_{q^{-1}}}
\right)
C_{i_1+1, j_1, i_2, j_2-1}\\[6pt]
&+\left(
\frac{e(\lambda_{2})q^{j_1}[i_2+1]_q}{[j_1+1]_q}
-\frac{e(-\lambda_{2})q^{-j_1}[i_2+1]_{q^{-1}}}{[j_1+1]_{q^{-1}}}
\right)
C_{i_1, j_1, i_2+1, j_2-1}\\[6pt]
&+\left(
\frac{e(\lambda_{23})q^{j_1+i_2}[j_2]_q}{[j_1+1]_q}
-\frac{e(-\lambda_{23})q^{-j_1-i_2}[j_2]_{q^{-1}}}{[j_1+1]_{q^{-1}}}
\right)
C_{i_1, j_1, i_2, j_2}=0,
\end{align*}
which is simplified to
\begin{align*}
&\langle\,e(\lambda_{1})q^{\frac{j_1}{2}}\,\rangle_{i_1+1}\,
C_{i_1+1, j_1, i_2, j_2-1}
-\langle\,e(\lambda_{2})q^{\frac{j_1}{2}}\,\rangle_{i_2+1}\,
C_{i_1, j_1, i_2+1, j_2-1}\\[6pt]
&-\langle\,e(\lambda_{23})q^{i_2+\frac{j_1}{2}}\,\rangle_{j_2}\,
C_{i_1, j_1, i_2, j_2}=0.
\end{align*}
This is the required equality.

\medskip

\noindent
(2)\; Similarly, we have
\begin{align*}
&[i_1+1]_q\,C_{i_1+1, j_1-1, i_2, j_2}\\[6pt]
&+e(\lambda_{1})q^{i_1}[j_1]_qC_{i_1, j_1, i_2, j_2}\\[6pt]
&+e(\lambda_{12})q^{i_1+j_1-1}[i_2+1]_qC_{i_1, j_1-1, i_2+1, j_2}\\[6pt]
&+e(\lambda_{123})q^{i_1+j_1+i_2-1}[j_2+1]_qC_{i_1, j_1-1, i_2, j_2}=0
\end{align*}
and
\begin{align*}
&[i_1+1]_{q^{-1}}C_{i_1+1, j_1-1, i_2, j_2-1}\\[6pt]
&+e(-\lambda_{1})q^{-i_1}[j_1]_{q^{-1}}C_{i_1, j_1, i_2, j_2}\\[6pt]
&+e(-\lambda_{12})
q^{-i_1-j_1-1}[i_2+1]_{q^{-1}}C_{i_1, j_1-1, i_2+1, j_2}\\[6pt]
&+e(-\lambda_{123})q^{-i_1-j_1-i_2-1}[j_2+1]_{q^{-1}}
C_{i_1, j_1-1, i_2, j_2+1}=0,
\end{align*}
which imply, by eliminating the last terms, 
\begin{align*}
&\langle\,e(\lambda_{123})q^{i_2+j_1+\frac{j_2}{2}-1}\,\rangle_{i_1+1}\,
C_{i_1+1, j_1-1, i_2, j_2}\\[6pt]
&+\langle\,e(\lambda_{23})q^{i_2+\frac{j_2}{2}}\,\rangle_{j_1}\,
C_{i_1, j_1, i_2, j_2}\\[6pt]
&+\langle\,e(\lambda_{3})q^{\frac{j_2}{2}}\,\rangle_{i_2+1}\,
C_{i_1, j_1-1, i_2+1, j_2}=0.
\end{align*}
It completes the proof of Lemma 3.1. \hfill$\square$
\bigskip

\noindent
{\bf Lemma 3.2.}\;{\it For nonnegative integers $i_1, j_1, i_2, j_2,$
we have the following.}
\medskip

\noindent
$(1)$
\begin{align}
C_{i_1 j_1 i_2 j_2}&=
\sum_{k=0}^{j_2}(-)^k\Biggl\{
\prod_{r=1}^{j_2-k}
\frac{\langle\,e(\lambda_{1})q^{\frac{j_1}{2}}\,\rangle_{i_1+r}
}
{\langle\,e(\lambda_{23})q^{i_2+\frac{j_1}{2}+k}\,\rangle_{r}}
\nonumber\\[6pt]
&\times
\prod_{r=1}^{k}
\frac{\langle\,e(\lambda_{2})q^{\frac{j_1}{2}}\,\rangle_{i_2+r}}
{\langle\,e(\lambda_{23})q^{i_2+\frac{j_1}{2}+k-r}\,\rangle_{r}}
\Biggr\}
\,C_{i_1+j_2-k, j_1, i_2+k, 0}.
\end{align}

\noindent
$(2)$
\begin{align}
C_{i_1 j_1 i_2 j_2}
=&(-)^{j_1}\sum_{k=0}^{j_1}
\Biggl\{\;
\prod_{r=1}^{j_1-k}
\frac{\langle\,e(\lambda_{123})q^{i_2+\frac{j_2}{2}+j_1-r}\,\rangle_{i_1+r}}
{\langle\,e(\lambda_{23})q^{i_2+\frac{j_2}{2}+k}\,\rangle_r}
\nonumber\\[6pt]
&\times
\prod_{r=1}^{k}
\frac{\langle\,e(\lambda_{3})q^{\frac{j_2}{2}}\,\rangle_{i_2+r}}
{\langle\,e(\lambda_{23})q^{i_2+\frac{j_2}{2}+k-r}\,\rangle_{r}}
\;\Biggr\}
\,C_{i_1+j_1-k,0, i_2+k,j_2}.
\end{align}
\medskip

\noindent
{\bf Proof.} 
\;(1) We prove it by induction on $j_2$. 
The equality $(3.5)$ in case $j_2=1$  is equal to  $(3.1)$ 
in case $j_2=1.$ Hence the equality $(3.5)$ in case $j_2=1$ is true.
Next we assume the equality $(3.5)$ for a fixed $j_2.$
It follows from $(3.1)$ by the change of $j_2$ into $j_2+1$ 
that 
\begin{align}
C_{i_1, j_1, i_2, j_2+1}
&=\frac{\langle\,e(\lambda_{1})q^{\frac{j_1}{2}}\,\rangle_{i_1+1}}
{\langle\,e(\lambda_{23})q^{i_2+\frac{j_1}{2}}\,\rangle_{j_2+1}}\,
C_{i_1+1,j_1,i_2,j_2}\nonumber\\[6pt]
&-\frac{\langle\,e(\lambda_{2})q^{\frac{j_1}{2}}\,\rangle_{i_2+1}}
{\langle\,e(\lambda_{23})q^{i_2+\frac{j_1}{2}}\,\rangle_{j_2+1}}\,
C_{i_1,j_1,i_2+1,j_2}.
\end{align}

\noindent
Substitute  $(3.5)$ with the change of $i_1\mapsto i_1+1$ into
$(3.7)$, and substitute $(3.5)$ with the change of 
$i_2\mapsto i_2+1$ into $(3.7)$. Then  we have
\begin{align}
&C_{i_1, j_1, i_2, j_2+1}
=
\frac{\langle\,e(\lambda_{1})q^{\frac{j_1}{2}}\,\rangle_{i_1+1}}
{\langle\,e(\lambda_{23})q^{i_2+\frac{j_1}{2}}\,\rangle_{j_2+1}}
\Biggl\{
\prod_{r=1}^{j_2}
\frac{\langle\,e(\lambda_{1})q^{\frac{j_1}{2}}\,\rangle_{i_1+1+r}}
{\langle\,e(\lambda_{23})q^{i_2+\frac{j_1}{2}}\,\rangle_{r}}\,
C_{i_1+j_2+1, j_1, i_2, 0}\nonumber\\[6pt]
&+\sum_{k=1}^{j_2}(-)^k
\prod_{r=1}^{j_2-k}
\frac{\langle\,e(\lambda_{1})q^{\frac{j_1}{2}}\,\rangle_{i_1+1+r}}
{\langle\,e(\lambda_{23})q^{i_2+\frac{j_1}{2}+k}\,\rangle_{r}}
\prod_{r=1}^k
\frac{\langle\,e(\lambda_{2})q^{\frac{j_1}{2}}\,\rangle_{i_2+r}}
{\langle\,e(\lambda_{23})q^{i_2+\frac{j_1}{2}+k-r}\,\rangle_{r}}\,
C_{i_1+j_2+1-k, j_1, i_2+k, 0}\Biggr\}\nonumber\\[6pt]
&
-\frac{\langle\,e(\lambda_{2})q^{\frac{j_1}{2}}\,\rangle_{i_2+1}}
{\langle\,e(\lambda_{23})q^{i_2+\frac{j_1}{2}}\,\rangle_{j_2+1}}
\Biggl\{
\sum_{k=1}^{j_2}(-)^{k-1}
\prod_{r=1}^{j_2-k+1}
\frac{\langle\,e(\lambda_{1})q^{\frac{j_1}{2}}\,\rangle_{i_1+r}}
{\langle\,e(\lambda_{23})q^{i_2+\frac{j_1}{2}+k}\,\rangle_{r}}
\prod_{r=1}^{k-1}
\frac{\langle\,e(\lambda_{2})q^{\frac{j_1}{2}}\,\rangle_{i_2+1+r}}
{\langle\,e(\lambda_{23})q^{i_2+\frac{j_1}{2}+k-r}\,\rangle_{r}}\,
\nonumber\\[6pt]
&\times
C_{i_1+j_2-k+1, j_1, i_2+k, 0}
+(-)^{j_2}
\prod_{r=1}^{j_2}
\frac{\langle\,e(\lambda_{2})q^{\frac{j_1}{2}}\,\rangle_{i_2+1+r}}
{\langle\,e(\lambda_{23})q^{i_2+\frac{j_1}{2}+j_2+1-r}\,\rangle_{r}}\,
C_{i_1, j_1, i_2+j_2+1, 0}\Biggr\}\nonumber\\[6pt]
&=
\sum_{k=0}^{j_2+1}(-)^k
\prod_{r=1}^{j_2+1-k}\Biggl\{
\frac{\langle\,e(\lambda_{1})q^{\frac{j_1}{2}}\,\rangle_{i_1+r}
}
{\langle\,e(\lambda_{23})q^{i_2+\frac{j_1}{2}+k}\,\rangle_{r}}
\prod_{r=1}^k
\frac{\langle\,e(\lambda_{2})q^{\frac{j_1}{2}}\,\rangle_{i_2+r}}
{\langle\,e(\lambda_{23})q^{i_2+\frac{j_1}{2}+k-r}\,\rangle_{r}}\,
\Biggr\}\nonumber\\[6pt]
&\times\; C_{i_1+j_2+1-k, j_1, i_2+k, 0}.
\end{align}

\noindent
Here the last equality follows from
\begin{equation*}
\langle\,e(\lambda_{23})q^{i_2+\frac{j_1}{2}}\,\rangle_{k}+
\langle\,e(\lambda_{23})q^{i_2+\frac{j_1}{2}+k}\,\rangle_{j_2-k+1}
=\langle\,e(\lambda_{23})q^{i_2+\frac{j_1}{2}}\,\rangle_{j_2+1}.
\end{equation*}
The right most of $(3.8)$ is the right of $(3.5)$ 
with the change of $j_2\mapsto j_2+1$.
Thus we have proved the required equality.
\medskip

\noindent
(2)\; It is similarly proved by induction on $j_1$. 
The equality $(3.6)$ in case $j_1=1$ is equal to $(3.2)$ 
in case $j_1=1.$ Hence the equality $(3.6)$ holds true in case
$j_1=1$.  Next we assume the equality $(3.6)$ for a fixed $j_1.$
It follows from $(3.2)$ by the change of $j_1\mapsto j_1+1$ that
\begin{align}
C_{i_1, j_1+1, i_2, j_2}
=&-\frac{\langle\,e(\lambda_{123})q^{i_2+j_1+\frac{j_2}{2}}\,\rangle_{i_1+1}}
{\langle\,e(\lambda_{23})q^{i_2+\frac{j_2}{2}}\,\rangle_{j_1+1}}\,
C_{i_1+1,j_1,i_2,j_2}\nonumber\\[6pt]
&-
\frac{\langle\,e(\lambda_{2})q^{\frac{j_2}{2}}\,\rangle_{i_2+1}}
{\langle\,e(\lambda_{23})q^{i_2+\frac{j_2}{2}}\,\rangle_{j_1+1}}\,
C_{i_1,j_1,i_2+1,j_2}.
\end{align}
Substitute $(3.6)$ with the change of $i_1\mapsto i_1+1$ into 
$(3.9),$ and substitute $(3.6)$ with the change of $i_2\mapsto i_2+1$
into $(3.9).$ Then we have

\begin{align}
&C_{i_1, j_1+1, i_2, j_2}
=(-)^{j_1+1}
\frac{\langle\,e(\lambda_{123})q^{i_2+j_1+\frac{j_2}{2}}\,\rangle_{i_1+1}}
{\langle\,e(\lambda_{23})q^{i_2+\frac{j_2}{2}}\,\rangle_{j_1+1}}
\nonumber\\[6pt]
&\times\,
\left\{
\prod_{r=1}^{j_1}
\frac{\langle\,e(\lambda_{123})q^{i_2+\frac{j_2}{2}+j_1-r}\,\rangle_{i_1+1+r}}
{\langle\,e(\lambda_{23})q^{i_2+\frac{j_2}{2}}\,\rangle_r}
\,C_{i_1+j_1+1,0, i_2,j_2}\right.\nonumber\\[6pt]
&+\sum_{k=1}^{j_1}
\prod_{r=1}^{j_1-k}
\frac{\langle\,e(\lambda_{123})q^{i_2+\frac{j_2}{2}+j_1-r}\,\rangle_{i_1+1+r}}
{\langle\,e(\lambda_{23})q^{i_2+\frac{j_2}{2}+k}\,\rangle_r}
\nonumber\\[6pt]
&\times
\prod_{r=1}^k
\frac{\langle\,e(\lambda_{3})q^{\frac{j_2}{2}}\,\rangle_{i_2+r}}
{\langle\,e(\lambda_{23})q^{i_2+\frac{j_2}{2}+k-r}\,\rangle_{r}}
\,C_{i_1+j_1+1-k,0, i_2+k,j_2}\Biggr\}\nonumber\\[6pt]
&
+(-)^{j_1+1}
\frac{\langle\,e(\lambda_{2})q^{\frac{j_2}{2}}\,\rangle_{i_2+1}}
{\langle\,e(\lambda_{23})q^{i_2+\frac{j_2}{2}}\,\rangle_{j_1+1}}\,
\nonumber\\[6pt]
&\times
\Biggl\{
\sum_{k=1}^{j_1}
\prod_{r=1}^{j_1+1-k}
\frac{\langle\,e(\lambda_{123})q^{i_2+\frac{j_2}{2}+j_1+1-r}\,\rangle_{i_1+r}}
{\langle\,e(\lambda_{23})q^{i_2+\frac{j_2}{2}+k}\,\rangle_r}
\nonumber\\[6pt]
&\times
\prod_{r=1}^{k-1}
\frac{\langle\,e(\lambda_{3})q^{\frac{j_2}{2}}\,\rangle_{i_2+1+r}}
{\langle\,e(\lambda_{23})q^{i_2+\frac{j_2}{2}+k-r}\,\rangle_{r}}
\,C_{i_1+j_1+1-k,0, i_2+k,j_2}\nonumber\\[6pt]
&
+
\prod_{r=1}^{j_1}
\frac{\langle\,e(\lambda_{3})q^{\frac{j_2}{2}}\,\rangle_{i_2+1+r}}
{\langle\,e(\lambda_{23})q^{i_2+\frac{j_2}{2}+j_1+1-r}\,\rangle_{r}}
\,C_{i_1,0, i_2+j_1+1,j_2}\Biggr\}\nonumber\\[6pt]
&=
(-1)^{j_1+1}\sum_{k=0}^{j_1+1}
\prod_{r=1}^{j_1+1-k}
\frac{\langle\,e(\lambda_{123})q^{i_2+\frac{j_2}{2}+j_1+1-r}\,\rangle_{i_1+r}}
{\langle\,e(\lambda_{23})q^{i_2+\frac{j_2}{2}+k}\,\rangle_{r}}
\nonumber\\[6pt]
&\times
\prod_{r=1}^{k}
\frac{\langle\,e(\lambda_{3})q^{\frac{j_2}{2}}\,\rangle_{i_2+r}}
{\langle\,e(\lambda_{23})q^{i_2+\frac{j_2}{2}+k-r}\,\rangle_{r}}
\, C_{i_1+j_1+1-k, 0, i_2+k, j_2}.
\end{align}
The right most of $(3.10)$ is the right of $(3.6)$ with the change
of $j_1\mapsto j_1+1.$ It competes the proof.
\hfill$\square$
\medskip

By using the equalities in Lemma 3.2, we obtain the 
expressions of the  connection coefficients $p_{ij}$ defined by
\begin{equation}
C_{0,i,0,m-i}=\sum_{j=0}^mp_{ij}\,C_{m-j,0,j,0}
\end{equation}
\medskip
for $0\le i\le m.$
\medskip

\noindent
{\bf Proposition 3.3.} {\it For $0\le i,j\le m$, we have the following.}

\smallskip

\noindent
$(1)$\;

\begin{align}
p_{ij}&=(-)^i\sum_{\substack{
0\le k\le m-i\\[3pt]
0\le l\le i\\[3pt]
k+l=j}}
(-)^{k}
\prod_{r=1}^{m-i-k}
\frac{s\bigl(\,\lambda_{1}+\frac{i+r-1}{2}g\,\bigr)}
{s\bigl(\,\lambda_{23}+\left(k+\frac{i+r-1}{2}\right)g\,\bigr)}
\prod_{r=1}^{k}
\frac{s\bigl(\lambda_{2}+\frac{i+r-1}{2}g\,\bigr)}
{s\bigl(\lambda_{23}+\left(k+\frac{i-r-1}{2}\right)g\,\bigr)}
\nonumber\\[6pt]
&\times
\prod_{r=1}^{i-l}
\frac{s\bigl(\lambda_{123}+\frac{m+i+k-r-1}{2}g\,\bigr)
s\bigl(\frac{m-i-k+r}{2}g\,\bigr)}
{s\bigl(\lambda_{23}+\left(j+\frac{r-1}{2}\right)g\,\bigr)
s\bigl(\frac{r}{2}g\,\bigr)}
\prod_{r=1}^{l}
\frac{s\bigl(\lambda_{3}+\frac{k+r-1}{2}g\,\bigr)
s\bigl(\frac{k+r}{2}g\,\bigr)}
{s\bigl(\lambda_{23}+\left(j-\frac{r+1}{2}\right)g\,\bigr)
s\bigl(\frac{r}{2}g\,\bigr)}.
\end{align}

\noindent
$(2)$\;

\begin{align}
p_{ij}
&=(-)^i
\sum_{\substack{
0\le k\le i\\[3pt]
0\le l\le m-i\\[3pt]
k+l=j}}
(-)^{l}
\prod_{r=1}^{i-k}
\frac{s\bigl(\,\lambda_{123}+\frac{m+i-r-1}{2}g\,\bigr)}
{s\bigl(\,\lambda_{23}+\left(k+\frac{m-i+r-1}{2}\right)g\,\bigr)}
\prod_{r=1}^{k}
\frac{s\bigl(\,\lambda_{3}+\frac{m-i+r-1}{2}g\,\bigr)}
{s\bigl(\,\lambda_{23}+\left(k+\frac{m-i-r-1}{2}\right)g\,\bigr)}
\nonumber\\[6pt]
&\times
\prod_{r=1}^{m-i-l}
\frac{s\bigl(\,\lambda_{1}+\frac{i-k+r-1}{2}g\,\bigr)
s\bigl(\,\frac{i-k+r}{2}g\,\bigr)}
{s\bigl(\,\lambda_{23}+\left(j+\frac{r-1}{2}\right)g\,\bigr)
s\bigl(\,\frac{r}{2}g\,\bigr)}
\prod_{r=1}^{l}
\frac{s\bigl(\,\lambda_{2}+\frac{k+r-1}{2}g\,\bigr)
s\bigl(\,\frac{k+r}{2}g\,\bigr)}
{s\bigl(\,\lambda_{23}+\left(j-\frac{r+1}{2}\right)g\,\bigr)
s\bigl(\,\frac{r}{2}g\,\bigr)}.
\end{align}

\medskip

\noindent
{\bf Proof.} $(1)$\; When $i_1=i_2=0,\, j_1=i,\, j_2=m-i$, (1) of Lemma 3.2
implies

\begin{align}
&C_{0,i,0,m-i}\nonumber\\[6pt] 
&=\sum_{k=0}^{m-i}(-)^k
\Biggl\{\prod_{r=1}^{m-i-k}
\frac{\langle\,e(\lambda_{1})q^{\frac{i}{2}}\,\rangle_{r}}
{\langle\,e(\lambda_{23})q^{\frac{i}{2}+k}\,\rangle_{r}}
\prod_{r=1}^{k}
\frac{\langle\,e(\lambda_{2})q^{\frac{i}{2}}\,\rangle_{r}}
{\langle\,e(\lambda_{23})q^{\frac{i}{2}+k-r}\,\rangle_{r}}\Biggr\}\;
C_{m-i-k,i,k,0}.
\end{align}
\medskip

\noindent
When $i_1=m-i-k,\, i_2=k,\,  j_1=i,\,  j_2=0$, (2) of Lemma 3.2 implies

\begin{align}
&C_{m-i-k,i,k,0}\nonumber\\[6pt] 
&=(-)^{i}\sum_{l=0}^i
\Biggl\{\prod_{r=1}^{i-l}
\frac{\langle\,e(\lambda_{123})q^{k+i-r}\,\rangle_{m-i-k+r}}
{\langle\,e(\lambda_{23})q^{j}\,\rangle_{r}}
\prod_{r=1}^{l}
\frac{\langle\,e(\lambda_{3})\,\rangle_{k+r}}
{\langle\,e(\lambda_{23})q^{j-r}\,\rangle_{r}}\Biggr\}
\;C_{m-k-l,0,k+l,0}.
\end{align}
\medskip

\noindent
Substituting (3.14) into (3.15) leads to the relation
\medskip
\begin{align}
p_{ij}&=(-)^i
\sum_{\substack{
0\le k\le m-i\\[3pt]
0\le l\le i\\[3pt]
k+l=j}}
(-)^{k}
\prod_{r=1}^{m-i-k}
\frac{\langle\,e(\lambda_{1})q^{\frac{i}{2}}\,\rangle_{r}}
{\langle\,e(\lambda_{23})q^{\frac{i}{2}+k}\,\rangle_{r}}
\prod_{r=1}^{k}
\frac{\langle\,e(\lambda_{2})q^{\frac{i}{2}}\,\rangle_{r}}
{\langle\,e(\lambda_{23})q^{\frac{i}{2}+k-r}\,\rangle_{r}}
\nonumber\\[6pt]
&\times
\prod_{r=1}^{i-l}
\frac{\langle\,e(\lambda_{123})q^{k+i-r}\,\rangle_{m-i-k+r}}
{\langle\,e(\lambda_{23})q^{j}\,\rangle_{r}}
\prod_{r=1}^{l}
\frac{\langle\,e(\lambda_{2})\,\rangle_{k+r}}
{\langle\,e(\lambda_{23})q^{j-r}\,\rangle_{r}}\nonumber\\[6pt]
&=(-)^i
\sum_{\substack{
0\le k\le m-i\\[3pt]
0\le l\le i\\[3pt]
k+l=j}}
(-)^{k}
\prod_{r=1}^{m-i-k}
\frac{\langle\,e(\lambda_{1})q^{\frac{i+r-1}{2}}\,\rangle_{1}}
{\langle\,e(\lambda_{23})q^{k+\frac{i+r-1}{2}}\,\rangle_{1}}
\prod_{r=1}^{k}
\frac{\langle\,e(\lambda_{2})q^{\frac{i+r-1}{2}}\,\rangle_{1}}
{\langle\,e(\lambda_{23})q^{k+\frac{i-r-1}{2}}\,\rangle_{1}}
\nonumber\\[6pt]
&\times
\prod_{r=1}^{i-l}
\frac{\langle\,e(\lambda_{123})q^{\frac{m+i+k-r-1}{2}}\,\rangle_{1}
\langle\,q^{\frac{m-i-k+r}{2}}\,\rangle_{1}}
{\langle\,e(\lambda_{23})q^{j+\frac{r-1}{2}}\,\rangle_{1}
\langle\,q^{\frac{r}{2}}\,\rangle_{1}}
\prod_{r=1}^{l}
\frac{\langle\,e(\lambda_{3})q^{\frac{k+r-1}{2}}\,\rangle_{1}
\langle\,q^{\frac{k+r}{2}}\,\rangle_{1}}
{\langle\,e(\lambda_{23})q^{j-\frac{r+1}{2}}\,\rangle_{1}
\langle\,q^{\frac{r}{2}}\,\rangle_{1}},
\end{align}

\noindent
which implies the required result (3.12).

\medskip

\noindent
$(2)$\; When $i_1=i_2=0,\, j_1=i,\, j_2=m-i$, (2) of Lemma 3.2 implies

\begin{align}
&C_{0,i,0,m-i}\nonumber\\[6pt] 
&=(-)^i 
\sum_{k=0}^{i}
\Biggl\{\prod_{r=1}^{i-k}
\frac{\langle\,e(\lambda_{123})q^{\frac{m+i}{2}-r}\,\rangle_{r}}
{\langle\,e(\lambda_{23})q^{\frac{m-i}{2}+k}\,\rangle_{r}}
\prod_{r=1}^{k}
\frac{\langle\,e(\lambda_{3})q^{\frac{m-i}{2}}\,\rangle_{r}}
{\langle\,e(\lambda_{23})q^{\frac{m-i}{2}+k-r}\,\rangle_{r}}
\Biggl\}\;
C_{i-k,0,k,m-i}.
\end{align}
\medskip

\noindent
When $i_1=i-k,\, i_2=k,\, j_1=0,\, j_2=m-i$, (1) of Lemma 3.2 implies

\begin{align}
&C_{i-k,0,k,m-i}\nonumber\\[6pt] 
&=\sum_{l=0}^{m-i}(-)^l\Biggl\{
\prod_{r=1}^{m-i-l}
\frac{\langle\,e(\lambda_{1})\,\rangle_{i-k+r}}
{\langle\,e(\lambda_{23})q^{j}\,\rangle_{r}}
\prod_{r=1}^{l}
\frac{\langle\,e(\lambda_{2})\,\rangle_{k+r}}
{\langle\,e(\lambda_{23})q^{j-r}\,\rangle_{r}}
\Biggr\}\;
C_{m-k-l,0,k+l,0}.
\end{align}
\medskip

\noindent
Substituting (3.18) into (3.17) leads to the relation
\medskip

\begin{align}
p_{ij}&=(-)^i
\sum_{\substack{
0\le k\le i\\[3pt]
0\le l\le m-i\\[3pt]
k+l=j}}
(-)^{l}
\prod_{r=1}^{i-k}
\frac{\langle\,e(\lambda_{123})q^{\frac{m+i}{2}-r}\,\rangle_{r}}
{\langle\,e(\lambda_{23})q^{\frac{m-i}{2}+k}\,\rangle_{r}}
\prod_{r=1}^{k}
\frac{\langle\,e(\lambda_{3})q^{\frac{m-i}{2}}\,\rangle_{r}}
{\langle\,e(\lambda_{23})q^{\frac{m-i}{2}+k-r}\,\rangle_{r}}
\nonumber\\[6pt]
&\times
\prod_{r=1}^{m-i-l}
\frac{\langle\,e(\lambda_{1})\,\rangle_{i-k+r}}
{\langle\,e(\lambda_{23})q^{j}\,\rangle_{r}}
\prod_{r=1}^{l}
\frac{\langle\,e(\lambda_{2})\,\rangle_{k+r}}
{\langle\,e(\lambda_{23})q^{j-r}\,\rangle_{r}}
\nonumber\\[6pt]
&=(-)^i
\sum_{\substack{
0\le k\le i\\[3pt]
0\le l\le m-i\\[3pt]
k+l=j}}
(-)^{l}
\prod_{r=1}^{i-k}
\frac{\langle\,e(\lambda_{123})q^{\frac{m+i-r-1}{2}}\,\rangle_{1}}
{\langle\,e(\lambda_{23})q^{k+\frac{m-i+r-1}{2}}\,\rangle_{1}}
\prod_{r=1}^{k}
\frac{\langle\,e(\lambda_{3})q^{\frac{m-i+r-1}{2}}\,\rangle_{1}}
{\langle\,e(\lambda_{23})q^{k+\frac{m-i-r-1}{2}}\,\rangle_{1}}
\nonumber\\[6pt]
&\times
\prod_{r=1}^{m-i-l}
\frac{\langle\,e(\lambda_{1})q^{\frac{i-k+r-1}{2}}\,\rangle_1
\langle\,q^{\frac{i-k+r}{2}}\,\rangle_1}
{\langle\,e(\lambda_{23})q^{j+\frac{r-1}{2}}\,\rangle_1
\langle\,q^{\frac{r}{2}}\,\rangle_1}
\prod_{r=1}^{l}
\frac{\langle\,e(\lambda_{2})q^{\frac{k+r-1}{2}}\,\rangle_1
\langle\,q^{\frac{k+r}{2}}\,\rangle_1}
{\langle\,e(\lambda_{23})q^{j-\frac{r+1}{2}}\,\rangle_1
\langle\,q^{\frac{r}{2}}\,\rangle_1},
\end{align}
\medskip

\noindent
which implies (3.13). 
\medskip

It completes the proof.\hfill$\square$

\medskip

\noindent
In the next subsection, we shall give an expression of 
$p_{ij}$ in terms of the basic hypergeometric polynomial 
${}_8\varphi_7$ and that in terms of ${}_4\varphi_3$. 

\bigskip

\subsection{Connection coefficients in terms of the
basic hypergeometric polynomials}

\noindent
{\bf Proposition 3.4.} (1)\; {\it For} $0\le i+j\le m,$ 
{\it we have}

\begin{align*} 
p_{ij}&=
(-)^{i+j}
\prod_{r=1}^{m-i-j}
\frac{\langle\,e(\lambda_{1})q^{\frac{i+r-1}{2}}\,\rangle_{1}}
{\langle\,e(\lambda_{23})q^{j+\frac{i+r-1}{2}}\,\rangle_{1}}
\prod_{r=1}^{j}
\frac{\langle\,e(\lambda_{2})q^{\frac{i+r-1}{2}}\,\rangle_{1}}
{\langle\,e(\lambda_{23})q^{\frac{i+j+r-2}{2}}\,\rangle_{1}}
\nonumber\\[6pt]
&\times
\prod_{r=1}^{i}
\frac{\langle\,e(\lambda_{123})q^{\frac{m+j+r-2}{2}}\,\rangle_{1}
\langle\,q^{\frac{m-i-j+r}{2}}\,\rangle_{1}}
{\langle\,e(\lambda_{23})q^{j+\frac{r-1}{2}}\,\rangle_{1}
\langle\,q^{\frac{r}{2}}\,\rangle_{1}}\nonumber\\[6pt]
&\times\;
{}_8\varphi_7
\Biggl(
\begin{array}{ccc}
e(-\lambda_{23})q^{-\frac{i}{2}-j+\frac{3}{2}},&
-e(-\lambda_{23})q^{-\frac{i}{2}-j+\frac{3}{2}},&
e(2\lambda_{1})q^{m-j},\\[6pt]
e(-\lambda_{23})q^{-\frac{i}{2}-j+\frac{1}{2}},&
-e(-\lambda_{23})q^{-\frac{i}{2}-j+\frac{1}{2}},&
e(-2\lambda_{2})q^{1-i-j},
\end{array}\nonumber\\[8pt]
&
\begin{array}{ccccc}
e(-2\lambda_{23})q^{1-m-j},&
e(-2\lambda_{23})q^{-2j+1-i},&
q^{-i}, &
\,e(-2\lambda_{3})q^{1-j},&
q^{-j}\\[6pt]
e(-2\lambda_{23})q^{2-i-j},&
e(-2\lambda_{123})q^{2-m-i-j},&
q^{m-i-j+1},&
e(-2\lambda_{23})q^{2-2j}
\end{array}\\[8pt]
&\qquad
q;\quad e(-2\lambda_{12})q^{2-i}\quad \Biggr).
\end{align*}

\noindent
(2)\; {\it For } $2m\ge i+j\ge m,$ {\it we have}

\begin{align*}
p_{ij}&=
(-)^{m}
\prod_{r=1}^{m-i}
\frac{\langle\,e(\lambda_{2})q^{\frac{i+r-1}{2}}\,\rangle_{1}}
{\langle\,e(\lambda_{23})q^{\frac{m+r-2}{2}}\,\rangle_{1}}
\prod_{r=1}^{m-j}
\frac{
\langle\,e(\lambda_{123})q^{\frac{m+j+r-2}{2}}\,\rangle_{1}}
{\langle\,e(\lambda_{23})q^{j+\frac{r-1}{2}}\,\rangle_{1}}
\nonumber\\[6pt]
&\times
\prod_{r=1}^{i+j-m}
\frac{\langle\,e(\lambda_{3})q^{\frac{m-i+r-1}{2}}\,\rangle_{1}
\langle\,q^{\frac{m-i+r}{2}}\,\rangle_{1}}
{\langle\,e(\lambda_{23})q^{\frac{m-i+j+r-2}{2}}\,\rangle_{1}
\langle\,q^{\frac{r}{2}}\,\rangle_{1}}
\nonumber\\[6pt]
&\times{}_8\varphi_7
\Biggl(
\begin{array}{lll}
e(-\lambda_{23})q^{\frac{i+3}{2}-m},&
-e(-\lambda_{23})q^{\frac{i+3}{2}-m},&
e(-2\lambda_{23})q^{1+i-2m},\\[6pt]
e(-\lambda_{23})q^{\frac{i+1}{2}-m},&
-e(-\lambda_{23})q^{\frac{i+1}{2}-m},&
e(-2\lambda_{23})q^{2-m},
\end{array}\nonumber\\[8pt]
&
\begin{array}{lllll}
e(-2\lambda_{23})q^{1-m-j},&
e(2\lambda_{1})q^{i},&
\,e(-2\lambda_{3})q^{1+i-m},&
q^{j-m}, &
q^{i-m}
\\[6pt]
e(-2\lambda_{23})q^{2-m+i-j},&
e(-2\lambda_{123})q^{2-2m},&
e(-2\lambda_{2})q^{1-m},&
q^{1+i+j-m}
\end{array}\nonumber\\[8pt]
&;\,
q,\;\quad e(-2\lambda_{12})q^{2-i}\quad \Biggr).
\end{align*}
\medskip

\noindent
{\bf Proof.}\;$(1)$\; When $0\le i+j\le m,$ (1) of Proposition 3.3 shows
\begin{align}
&p_{ij}
=(-)^{i+j}
\sum_{0\le l\le \min\{i, j\}}
(-)^{l}
\prod_{r=1}^{m-i-j+l}
\frac{\langle\,e(\lambda_{1})q^{\frac{i+r-1}{2}}\,\rangle_{1}}
{\langle\,e(\lambda_{23})q^{j-l+\frac{i+r-1}{2}}\,\rangle_{1}}
\prod_{r=1}^{j-l}
\frac{\langle\,e(\lambda_{2})q^{\frac{i+r-1}{2}}\,\rangle_{1}}
{\langle\,e(\lambda_{23})q^{j-l+\frac{i-r-1}{2}}\,\rangle_{1}}
\nonumber\\[6pt]
&\times
\prod_{r=1}^{i-l}
\frac{\langle\,e(\lambda_{123})q^{\frac{m+i+j-l-r-1}{2}}\,\rangle_{1}
\langle\,q^{\frac{m-i-j+l+r}{2}}\,\rangle_{1}}
{\langle\,e(\lambda_{23})q^{j+\frac{r-1}{2}}\,\rangle_{1}
\langle\,q^{\frac{r}{2}}\,\rangle_{1}}
\;
\prod_{r=1}^l
\frac{\langle\,e(\lambda_{3})q^{\frac{j-l+r-1}{2}}\,\rangle_{1}
\langle\,q^{\frac{j-l+r}{2}}\,\rangle_{1}}
{\langle\,e(\lambda_{23})q^{j-\frac{r+1}{2}}\,\rangle_{1}
\langle\,q^{\frac{r}{2}}\,\rangle_{1}}.
\end{align}

\noindent
It is seen that

\begin{align}
&\prod_{r=1}^{m-i-j+l}
\frac{\langle\,e(\lambda_{1})q^{\frac{i+r-1}{2}}\,\rangle_{1}}
{\langle\,e(\lambda_{23})q^{j-l+\frac{i+r-1}{2}}\,\rangle_{1}}
\prod_{r=1}^{j-l}
\frac{\langle\,e(\lambda_{2})q^{\frac{i+r-1}{2}}\,\rangle_{1}}
{\langle\,e(\lambda_{23})q^{j-l+\frac{i-r-1}{2}}\,\rangle_{1}}
\nonumber\\[6pt]
&=\prod_{r=1}^{l}
\frac{\langle\,e(\lambda_{23})q^{\frac{m+j-r}{2}}\,\rangle_{1}
\langle\,e(\lambda_{1})q^{\frac{m-j+r-1}{2}}\,\rangle_{1}}
{\langle\,e(\lambda_{23})q^{\frac{i+j-r-1}{2}}\,\rangle_{1}
\langle\,e(\lambda_{2})q^{\frac{i+j-r}{2}}\,\rangle_{1}}
\nonumber\\[6pt]
&\times
\frac{\langle\,e(\lambda_{23})q^{\frac{i-1}{2}+j-l}\,\rangle_{1}}
{\langle\,e(\lambda_{23})q^{j+\frac{i-1}{2}}\,\rangle_{1}}
\prod_{r=1}^{m-i-j}
\frac{\langle\,e(\lambda_{1})q^{\frac{i+r-1}{2}}\,\rangle_{1}}
{\langle\,e(\lambda_{23})q^{j+\frac{i+r-1}{2}}\,\rangle_{1}}
\prod_{r=1}^{j}
\frac{\langle\,e(\lambda_{2})q^{\frac{i+r-1}{2}}\,\rangle_{1}}
{\langle\,e(\lambda_{23})q^{\frac{i+j+r-2}{2}}\,\rangle_{1}}
\nonumber\\[6pt]
&=(-)^l\prod_{r=1}^{l}
\frac{\langle\,e(-\lambda_{23})q^{\frac{-m-j+r}{2}}\,\rangle_{1}
\langle\,e(\lambda_{1})q^{\frac{m-j+r-1}{2}}\,\rangle_{1}}
{\langle\,e(-\lambda_{23})q^{\frac{-i-j+r+1}{2}}\,\rangle_{1}
\langle\,e(-\lambda_{2})q^{\frac{-i-j+r}{2}}\,\rangle_{1}}
\nonumber\\[6pt]
&\times\;\;
\prod_{r=1}^{l}
\frac{\langle\,e(-\lambda_{23})q^{-j-\frac{i-1}{2}+r}\,\rangle_{1}}
{\langle\,e(-\lambda_{23})q^{-j-\frac{i+1}{2}+r}\,\rangle_{1}}
\nonumber\\[6pt]
&\times
\prod_{r=1}^{m-i-j}
\frac{\langle\,e(\lambda_{1})q^{\frac{i+r-1}{2}}\,\rangle_{1}}
{\langle\,e(\lambda_{23})q^{j+\frac{i+r-1}{2}}\,\rangle_{1}}
\prod_{r=1}^{j}
\frac{\langle\,e(\lambda_{2})q^{\frac{i+r-1}{2}}\,\rangle_{1}}
{\langle\,e(\lambda_{23})q^{\frac{i+j+r-2}{2}}\,\rangle_{1}}
\end{align}
\medskip

\noindent
and

\begin{align}
&
\prod_{r=1}^{i-l}
\frac{\langle\,e(\lambda_{123})q^{\frac{m+i+j-l-r-1}{2}}\,\rangle_{1}
\langle\,q^{\frac{m-i-j+l+r}{2}}\,\rangle_{1}}
{\langle\,e(\lambda_{23})q^{j+\frac{r-1}{2}}\,\rangle_{1}
\langle\,q^{\frac{r}{2}}\,\rangle_{1}}
\nonumber\\[6pt]
&=
\prod_{r=1}^{i}
\frac{\langle\,e(\lambda_{123})q^{\frac{m+j+r-2}{2}}\,\rangle_{1}
\langle\,q^{\frac{m-i-j+r}{2}}\,\rangle_{1}}
{\langle\,e(\lambda_{23})q^{j+\frac{r-1}{2}}\,\rangle_{1}
\langle\,q^{\frac{r}{2}}\,\rangle_{1}}
\nonumber\\[6pt]
&\times
\prod_{r=1}^{l}
\frac
{\langle\,e(\lambda_{23})q^{j+\frac{i-r}{2}}\,\rangle_{1}
\langle\,q^{\frac{i+1-r}{2}}\,\rangle_{1}}
{\langle\,e(\lambda_{123})q^{\frac{m+i+j-1-r}{2}}\,\rangle_{1}
\langle\,q^{\frac{m-i-j+r}{2}}\,\rangle_{1}}
\nonumber\\[6pt]
&=
\prod_{r=1}^{i}
\frac{\langle\,e(\lambda_{123})q^{\frac{m+j+r-2}{2}}\,\rangle_{1}
\langle\,q^{\frac{m-i-j+r}{2}}\,\rangle_{1}}
{\langle\,e(\lambda_{23})q^{j+\frac{r-1}{2}}\,\rangle_{1}
\langle\,q^{\frac{r}{2}}\,\rangle_{1}}
\nonumber\\[6pt]
&\times
(-)^l
\prod_{r=1}^{l}
\frac
{\langle\,e(-\lambda_{23})q^{-j+\frac{r-i}{2}}\,\rangle_{1}
\langle\,q^{\frac{r-1-i}{2}}\,\rangle_{1}}
{\langle\,e(-\lambda_{123})q^{\frac{1+r-m-i-j}{2}}\,\rangle_{1}
\langle\,q^{\frac{m-i-j+r}{2}}\,\rangle_{1}}
\end{align}
\medskip

\noindent
and

\begin{align}
&\prod_{r=1}^l
\frac{\langle\,e(\lambda_{3})q^{\frac{j-l+r-1}{2}}\,\rangle_{1}
\langle\,q^{\frac{j-l+r}{2}}\,\rangle_{1}}
{\langle\,e(\lambda_{23})q^{j-\frac{r+1}{2}}\,\rangle_{1}
\langle\,q^{\frac{r}{2}}\,\rangle_{1}}
=
\prod_{r=1}^l
\frac{\langle\,e(\lambda_{3})q^{\frac{j-r}{2}}\,\rangle_{1}
\langle\,q^{\frac{j-r+1}{2}}\,\rangle_{1}}
{\langle\,e(\lambda_{23})q^{j-\frac{r+1}{2}}\,\rangle_{1}
\langle\,q^{\frac{r}{2}}\,\rangle_{1}}
\nonumber\\[6pt]
&=(-)^l
\prod_{r=1}^l
\frac{\langle\,e(-\lambda_{3})q^{\frac{r-j}{2}}\,\rangle_{1}
\langle\,q^{\frac{r-1-j}{2}}\,\rangle_{1}}
{\langle\,e(-\lambda_{23})q^{-j+\frac{r+1}{2}}\,\rangle_{1}
\langle\,q^{\frac{r}{2}}\,\rangle_{1}}.
\end{align}
\medskip

\noindent
Therefore, substituting (3.21-23) into (3.20) implies

\begin{align}
p_{ij}
&=(-)^{i+j}
\prod_{r=1}^{m-i-j}
\frac{\langle\,e(\lambda_{1})q^{\frac{i+r-1}{2}}\,\rangle_{1}}
{\langle\,e(\lambda_{23})q^{j+\frac{i+r-1}{2}}\,\rangle_{1}}
\prod_{r=1}^{j}
\frac{\langle\,e(\lambda_{2})q^{\frac{i+r-1}{2}}\,\rangle_{1}}
{\langle\,e(\lambda_{23})q^{\frac{i+j+r-2}{2}}\,\rangle_{1}}
\nonumber\\[6pt]
&\times
\prod_{r=1}^{i}
\frac{\langle\,e(\lambda_{123})q^{\frac{m+j+r-2}{2}}\,\rangle_{1}
\langle\,q^{\frac{m-i-j+r}{2}}\,\rangle_{1}}
{\langle\,e(\lambda_{23})q^{j+\frac{r-1}{2}}\,\rangle_{1}
\langle\,q^{\frac{r}{2}}\,\rangle_{1}}\nonumber\\[6pt]
&\times\sum_{0\le l\le \min\{i, j\}}
\prod_{r=1}^{l}
\frac{\langle\,e(-\lambda_{23})q^{-j-\frac{i-1}{2}+r}\,\rangle_{1}}
{\langle\,e(-\lambda_{23})q^{-j-\frac{i+1}{2}+r}\,\rangle_{1}}
\frac{\langle\,e(-\lambda_{23})q^{\frac{-m-j+r}{2}}\,\rangle_{1}
\langle\,e(\lambda_{1})q^{\frac{m-j+r-1}{2}}\,\rangle_{1}}
{\langle\,e(-\lambda_{23})q^{\frac{-i-j+r+1}{2}}\,\rangle_{1}
\langle\,e(-\lambda_{2})q^{\frac{-i-j+r}{2}}\,\rangle_{1}}
\nonumber\\[6pt]
&\times
\prod_{r=1}^{l}
\frac
{\langle\,e(-\lambda_{23})q^{-j+\frac{r-i}{2}}\,\rangle_{1}
\langle\,q^{\frac{r-1-i}{2}}\,\rangle_{1}}
{\langle\,e(-\lambda_{123})q^{\frac{1+r-m-i-j}{2}}\,\rangle_{1}
\langle\,q^{\frac{m-i-j+r}{2}}\,\rangle_{1}}
\frac{\langle\,e(-\lambda_{3})q^{\frac{r-j}{2}}\,\rangle_{1}
\langle\,q^{\frac{r-1-j}{2}}\,\rangle_{1}}
{\langle\,e(-\lambda_{23})q^{-j+\frac{r+1}{2}}\,\rangle_{1}
\langle\,q^{\frac{r}{2}}\,\rangle_{1}}\nonumber\\[6pt]
&=(-)^{i+j}
\prod_{r=1}^{m-i-j}
\frac{\langle\,e(\lambda_{1})q^{\frac{i+r-1}{2}}\,\rangle_{1}}
{\langle\,e(\lambda_{23})q^{j+\frac{i+r-1}{2}}\,\rangle_{1}}
\prod_{r=1}^{j}
\frac{\langle\,e(\lambda_{2})q^{\frac{i+r-1}{2}}\,\rangle_{1}}
{\langle\,e(\lambda_{23})q^{\frac{i+j+r-2}{2}}\,\rangle_{1}}
\nonumber\\[6pt]
&\times
\prod_{r=1}^{i}
\frac{\langle\,e(\lambda_{123})q^{\frac{m+j+r-2}{2}}\,\rangle_{1}
\langle\,q^{\frac{m-i-j+r}{2}}\,\rangle_{1}}
{\langle\,e(\lambda_{23})q^{j+\frac{r-1}{2}}\,\rangle_{1}
\langle\,q^{\frac{r}{2}}\,\rangle_{1}}\nonumber\\[6pt]
&\times
{}_8\varphi_7
\Biggl(
\begin{array}{ccc}
e(-\lambda_{23})q^{-\frac{i}{2}-j+\frac{3}{2}},&
-e(-\lambda_{23})q^{-\frac{i}{2}-j+\frac{3}{2}},&
e(2\lambda_{1})q^{m-j},\\[6pt]
e(-\lambda_{23})q^{-\frac{i}{2}-j+\frac{1}{2}},&
-e(-\lambda_{23})q^{-\frac{i}{2}-j+\frac{1}{2}},&
e(-2\lambda_{2})q^{1-i-j},
\end{array}\nonumber\\[8pt]
&
\begin{array}{ccccc}
e(-2\lambda_{23})q^{1-m-j},&
e(-2\lambda_{23})q^{-2j+1-i},&
q^{-i}, &
\,e(-2\lambda_{3})q^{1-j},&
q^{-j}\\[6pt]
e(-2\lambda_{23})q^{2-i-j},&
e(-2\lambda_{123})q^{2-m-i-j},&
q^{m-i-j+1},&
e(-2\lambda_{23})q^{2-2j}
\end{array}\nonumber\\[8pt]
&;\,q,\quad e(-2\lambda_{12})q^{2-i}\quad \Biggr).
\end{align}
\medskip

\noindent
Here we have used the identities

\begin{equation}
\prod_{r=1}^l
\frac{\langle\,A q^{\frac{r}{2}}\,\rangle_{1}}
{\langle\,B q^{\frac{r}{2}}\,\rangle_{1}}
=\left(\frac{B}{A}\right)^l
\prod_{r=1}^l
\frac{1-A^2\,q^r}{1-B^2\,q^r}
=\left(\frac{B}{A}\right)^l
\frac{(A^2\,q;q)_l}{(B^2\,q;q)_l}
\end{equation}
and
\begin{equation}
\prod_{r=1}^l
\frac{\langle\,A q^{r}\,\rangle_{1}}
{\langle\,B q^{r}\,\rangle_{1}}
=\left(\frac{B}{A}\right)^l
\frac{(A^2\,q^2;q^2)_l}{(B^2\,q^2;q^2)_l}
=\left(\frac{B}{A}\right)^l
\frac{(A\,q;q)_l(-A\,q;q)_l}{(B\,q;q)_l(-B\,q;q)_l}.
\end{equation}

\medskip

\noindent
$(2)$ \;When $2m\ge i+j\ge m,$ (1) of Proposition 3.3 with 
the change of the running index $k\mapsto m-i-k$ shows that
\begin{align}
p_{ij}=
&(-)^{m}\sum_{
0\le k\le \min\{m-i, m-j\}}
(-)^{k}\prod_{r=1}^k
\frac{\langle\,e(\lambda_{1})q^{\frac{i+r-1}{2}}\,\rangle_{1}}
{\langle\,e(\lambda_{23})q^{m-k+\frac{-i+r-1}{2}}\,\rangle_{1}}
\prod_{r=1}^{m-i-k}
\frac{\langle\,e(\lambda_{2})q^{\frac{i+r-1}{2}}\,\rangle_{1}}
{\langle\,e(\lambda_{23})q^{m-k-\frac{i+r+1}{2}}\,\rangle_{1}}
\nonumber\\[6pt]
&\times
\prod_{r=1}^{m-j-k}
\frac{\langle\,e(\lambda_{123})q^{m-\frac{k+r+1}{2}}\,\rangle_{1}}
{\langle\,e(\lambda_{23})q^{j+\frac{r-1}{2}}\,\rangle_{1}}
\frac{\langle\,q^{\frac{k+r}{2}}\,\rangle_{1}}
{\langle\,q^{\frac{r}{2}}\,\rangle_{1}}
\prod_{r=1}^{i+j-m+k}
\frac{\langle\,e(\lambda_{3})q^{\frac{m-i-k+r-1}{2}}\,\rangle_{1}
\langle\,q^{\frac{m-i-k+r}{2}}\,\rangle_{1}}
{\langle\,e(\lambda_{23})q^{j-\frac{r+1}{2}}\,\rangle_{1}
\langle\,q^{\frac{r}{2}}\,\rangle_{1}}.
\end{align}

\medskip

\noindent
It is seen that

\begin{align}
&\prod_{r=1}^k
\frac{\langle\,e(\lambda_{1})q^{\frac{i+r-1}{2}}\,\rangle_{1}}
{\langle\,e(\lambda_{23})q^{m-k+\frac{-i+r-1}{2}}\,\rangle_{1}}
\prod_{r=1}^{m-i-k}
\frac{\langle\,e(\lambda_{2})q^{\frac{i+r-1}{2}}\,\rangle_{1}}
{\langle\,e(\lambda_{23})q^{m-k-\frac{i+r+1}{2}}\,\rangle_{1}}
\nonumber\\[6pt]
&=
\frac{
\langle\,e(\lambda_{23})q^{m-k-\frac{i+1}{2}}\,\rangle_{1}
\prod_{r=1}^k
\langle\,e(\lambda_{1})q^{\frac{i+r-1}{2}}\,\rangle_{1}
\prod_{r=1}^{m-i-k}
\langle\,e(\lambda_{2})q^{\frac{i+r-1}{2}}\,\rangle_{1}}
{\prod_{r=1}^{m-i+1}
\langle\,e(\lambda_{23})q^{\frac{m-k+r-2}{2}}\,\rangle_{1}}
\nonumber\\[6pt]
&=
\prod_{r=1}^{k}
\frac{
\langle\,e(\lambda_{23})q^{m-\frac{i+r}{2}}\,\rangle_{1}
\langle\,e(\lambda_{1})q^{\frac{i+r-1}{2}}\,\rangle_{1}}
{\langle\,e(\lambda_{2})q^{\frac{m-r}{2}}\,\rangle_{1}
\langle\,e(\lambda_{23})q^{\frac{m-r-1}{2}}\,\rangle_{1}}
\nonumber\\[6pt]
&\times
\frac{\langle\,e(\lambda_{23})q^{m-k-\frac{i+1}{2}}\,\rangle_{1}}
{\langle\,e(\lambda_{23})q^{m-\frac{i+1}{2}}\,\rangle_{1}}
\prod_{r=1}^{m-i}
\frac{\langle\,e(\lambda_{2})q^{\frac{i+r-1}{2}}\,\rangle_{1}}
{\langle\,e(\lambda_{23})q^{\frac{m+r-2}{2}}\,\rangle_{1}}
\nonumber\\[6pt]
&=(-)^k
\prod_{r=1}^{k}
\frac{
\langle\,e(-\lambda_{23})q^{-m+\frac{r+i}{2}}\,\rangle_{1}
\langle\,e(\lambda_{1})q^{\frac{i+r-1}{2}}\,\rangle_{1}}
{\langle\,e(-\lambda_{23})q^{\frac{-m+r+1}{2}}\,\rangle_{1}
\langle\,e(-\lambda_{2})q^{\frac{r-m}{2}}\,\rangle_{1}}
\nonumber\\[6pt]
&\times
\prod_{r=1}^{k}
\frac{\langle\,e(-\lambda_{23})q^{\frac{i+1}{2}-m+r}\,\rangle_{1}}
{\langle\,e(-\lambda_{23})q^{\frac{i-1}{2}-m+r}\,\rangle_{1}}
\prod_{r=1}^{m-i}
\frac{\langle\,e(\lambda_{2})q^{\frac{i+r-1}{2}}\,\rangle_{1}}
{\langle\,e(\lambda_{23})q^{\frac{m+r-2}{2}}\,\rangle_{1}}
\end{align}
\medskip

\noindent
and

\medskip

\begin{align}
&\prod_{r=1}^{m-j-k}
\frac{\langle\,e(\lambda_{123})q^{m-\frac{k+r+1}{2}}\,\rangle_{1}}
{\langle\,e(\lambda_{23})q^{j+\frac{r-1}{2}}\,\rangle_{1}}
\frac{\langle\,q^{\frac{k+r}{2}}\,\rangle_{1}}
{\langle\,q^{\frac{r}{2}}\,\rangle_{1}}
\nonumber\\[6pt]
&=
\prod_{r=1}^{m-j}
\frac{
\langle\,e(\lambda_{123})q^{\frac{m+j+r-2}{2}}\,\rangle_{1}}
{\langle\,e(\lambda_{23})q^{j+\frac{r-1}{2}}\,\rangle_{1}}
\prod_{r=1}^{k}
\frac
{\langle\,e(\lambda_{23})q^{\frac{m+j-r}{2}}\,\rangle_{1}
\langle\,q^{\frac{m-j-r+1}{2}}\,\rangle_{1}}
{\langle\,e(\lambda_{123})q^{m-\frac{r+1}{2}}\,\rangle_{1}
\langle\,q^{\frac{r}{2}}\,\rangle_{1}}
\nonumber\\[6pt]
&=
(-)^k
\prod_{r=1}^{m-j}
\frac{
\langle\,e(\lambda_{123})q^{\frac{m+j+r-2}{2}}\,\rangle_{1}}
{\langle\,e(\lambda_{23})q^{j+\frac{r-1}{2}}\,\rangle_{1}}
\prod_{r=1}^{k}
\frac
{\langle\,e(-\lambda_{23})q^{\frac{r-m-j}{2}}\,\rangle_{1}
\langle\,q^{\frac{j-m+r-1}{2}}\,\rangle_{1}}
{\langle\,e(-\lambda_{123})q^{\frac{r+1}{2}-m}\,\rangle_{1}
\langle\,q^{\frac{r}{2}}\,\rangle_{1}}
\end{align}
\medskip

\noindent
and

\medskip

\begin{align}
&\prod_{r=1}^{i+j-m+k}
\frac{\langle\,e(\lambda_{3})q^{\frac{m-i-k+r-1}{2}}\,\rangle_{1}
\langle\,q^{\frac{m-i-k+r}{2}}\,\rangle_{1}}
{\langle\,e(\lambda_{23})q^{j-\frac{r+1}{2}}\,\rangle_{1}
\langle\,q^{\frac{r}{2}}\,\rangle_{1}}
\nonumber\\[6pt]
&=
\prod_{r=1}^{k}
\frac{\langle\,e(\lambda_{3})q^{\frac{m-i-r}{2}}\,\rangle_{1}
\langle\,q^{\frac{m-i+1-r}{2}}\,\rangle_{1}}
{\langle\,e(\lambda_{23})q^{\frac{m-i+j-1-r}{2}}\,\rangle_{1}
\langle\,q^{\frac{i+j-m+r}{2}}\,\rangle_{1}}
\prod_{r=1}^{i+j-m}
\frac{\langle\,e(\lambda_{3})q^{\frac{m-i+r-1}{2}}\,\rangle_{1}
\langle\,q^{\frac{m-i+r}{2}}\,\rangle_{1}}
{\langle\,e(\lambda_{23})q^{\frac{m-i+j+r-2}{2}}\,\rangle_{1}
\langle\,q^{\frac{r}{2}}\,\rangle_{1}}\nonumber\\[6pt]
&=(-)^k
\prod_{r=1}^{k}
\frac{\langle\,e(-\lambda_{3})q^{\frac{r-m+i}{2}}\,\rangle_{1}
\langle\,q^{\frac{i-m+r-1}{2}}\,\rangle_{1}}
{\langle\,e(-\lambda_{23})q^{\frac{i-j-m+r+1}{2}}\,\rangle_{1}
\langle\,q^{\frac{i+j-m+r}{2}}\,\rangle_{1}}
\prod_{r=1}^{i+j-m}
\frac{\langle\,e(\lambda_{3})q^{\frac{m-i+r-1}{2}}\,\rangle_{1}
\langle\,q^{\frac{m-i+r}{2}}\,\rangle_{1}}
{\langle\,e(\lambda_{23})q^{\frac{m-i+j+r-2}{2}}\,\rangle_{1}
\langle\,q^{\frac{r}{2}}\,\rangle_{1}}.\nonumber\\
\end{align}

\medskip

\noindent
Therefore, substituting (3.28-30) into (3.27) shows

\begin{align}
p_{ij}=
&(-)^{m}
\prod_{r=1}^{m-i}
\frac{\langle\,e(\lambda_{2})q^{\frac{i+r-1}{2}}\,\rangle_{1}}
{\langle\,e(\lambda_{23})q^{\frac{m+r-2}{2}}\,\rangle_{1}}
\prod_{r=1}^{m-j}
\frac{
\langle\,e(\lambda_{123})q^{\frac{m+j+r-2}{2}}\,\rangle_{1}}
{\langle\,e(\lambda_{23})q^{j+\frac{r-1}{2}}\,\rangle_{1}}
\nonumber\\[6pt]
&\times
\prod_{r=1}^{i+j-m}
\frac{\langle\,e(\lambda_{3})q^{\frac{m-i+r-1}{2}}\,\rangle_{1}
\langle\,q^{\frac{m-i+r}{2}}\,\rangle_{1}}
{\langle\,e(\lambda_{23})q^{\frac{m-i+j+r-2}{2}}\,\rangle_{1}
\langle\,q^{\frac{r}{2}}\,\rangle_{1}}
\nonumber\\[6pt]
&\times
\sum_{
0\le k\le \min\{m-i, m-j\}}
\prod_{r=1}^{k}
\frac{\langle\,e(-\lambda_{23})q^{\frac{i+1}{2}-m+r}\,\rangle_{1}}
{\langle\,e(-\lambda_{23})q^{\frac{i-1}{2}-m+r}\,\rangle_{1}}
\frac{
\langle\,e(-\lambda_{23})q^{-m+\frac{r+i}{2}}\,\rangle_{1}
\langle\,e(\lambda_{1})q^{\frac{i+r-1}{2}}\,\rangle_{1}}
{\langle\,e(-\lambda_{23})q^{\frac{-m+r+1}{2}}\,\rangle_{1}
\langle\,e(-\lambda_{2})q^{\frac{r-m}{2}}\,\rangle_{1}}
\nonumber\\[6pt]
&
\prod_{r=1}^{k}
\frac
{\langle\,e(-\lambda_{23})q^{\frac{r-m-j}{2}}\,\rangle_{1}}
{\langle\,e(-\lambda_{123})q^{\frac{r+1}{2}-m}\,\rangle_{1}}
\frac{\langle\,q^{\frac{j-m+r-1}{2}}\,\rangle_{1}}
{\langle\,q^{\frac{r}{2}}\,\rangle_{1}}
\frac{\langle\,e(-\lambda_{3})q^{\frac{r-m+i}{2}}\,\rangle_{1}
\langle\,q^{\frac{i-m+r-1}{2}}\,\rangle_{1}}
{\langle\,e(-\lambda_{23})q^{\frac{i-j-m+r+1}{2}}\,\rangle_{1}
\langle\,q^{\frac{i+j-m+r}{2}}\,\rangle_{1}}
\nonumber\\[6pt]
&=(-)^{m}
\prod_{r=1}^{m-i}
\frac{\langle\,e(\lambda_{2})q^{\frac{i+r-1}{2}}\,\rangle_{1}}
{\langle\,e(\lambda_{23})q^{\frac{m+r-2}{2}}\,\rangle_{1}}
\prod_{r=1}^{m-j}
\frac{
\langle\,e(\lambda_{123})q^{\frac{m+j+r-2}{2}}\,\rangle_{1}}
{\langle\,e(\lambda_{23})q^{j+\frac{r-1}{2}}\,\rangle_{1}}
\nonumber\\[6pt]
&\times
\prod_{r=1}^{i+j-m}
\frac{\langle\,e(\lambda_{3})q^{\frac{m-i+r-1}{2}}\,\rangle_{1}
\langle\,q^{\frac{m-i+r}{2}}\,\rangle_{1}}
{\langle\,e(\lambda_{23})q^{\frac{m-i+j+r-2}{2}}\,\rangle_{1}
\langle\,q^{\frac{r}{2}}\,\rangle_{1}}
\nonumber\\[6pt]
&\times{}_8\varphi_7
\Biggl(
\begin{array}{lll}
e(-\lambda_{23})q^{\frac{i+3}{2}-m},&
-e(-\lambda_{23})q^{\frac{i+3}{2}-m},&
e(-2\lambda_{23})q^{1+i-2m},\\[6pt]
e(-\lambda_{23})q^{\frac{i+1}{2}-m},&
-e(-\lambda_{23})q^{\frac{i+1}{2}-m},&
e(-2\lambda_{23})q^{2-m},
\end{array}\nonumber\\[8pt]
&
\begin{array}{lllll}
e(-2\lambda_{23})q^{1-m-j},&
e(2\lambda_{1})q^{i},&
\,e(-2\lambda_{3})q^{1+i-m},&
q^{j-m}, &
q^{i-m}
\\[6pt]
e(-2\lambda_{23})q^{2-m+i-j},&
e(-2\lambda_{123})q^{2-2m},&
e(-2\lambda_{2})q^{1-m},&
q^{1+i+j-m}
\end{array}\nonumber\\[8pt]
&;\,
q,\;\quad e(-2\lambda_{12})q^{2-i}\quad \Biggr).
\end{align}
\medskip

\noindent
Here we have used the identity (3.25) and (3.26).
\hfill$\square$
\bigskip

In the next step, we rewrite the connection coefficients
$p_{ij}$ in terms of ${}_4\varphi_3.$


\medskip

\noindent
{\bf Proposition 3.5.}\; (1) {\it In case $0\le i+j\le m,$ we have }
\begin{align}
&p_{ij}=(-)^{i+j}
\prod_{r=1}^{i}
\frac{\langle\,q^{\frac{m-i+r}{2}}\,\rangle_{1}}
{\langle\,q^{\frac{r}{2}}\,\rangle_{1}}\nonumber\\[6pt]
&\times
\frac{\langle\,e(\lambda_{23})q^{j-\frac{1}{2}}\,\rangle_{1}
\prod_{r=1}^{i}\langle\,e(\lambda_{123})q^{\frac{m+j+r-2}{2}}\,\rangle_{1}
\prod_{r=1}^{j}\langle\,e(\lambda_{2})q^{\frac{i+r-1}{2}}\,\rangle_{1}
\prod_{r=1}^{m-i-j}\langle\,e(\lambda_{1})q^{\frac{i+r-1}{2}}\,\rangle_{1}}
{\prod_{r=1}^{m+1}\langle\,e(\lambda_{23})q^{\frac{j+r-2}{2}}\,\rangle_{1}}
\nonumber\\[6pt]
&\times{}_4\varphi_3
\hs{q^{-j},  q^{-i}, e(-2\lambda_{23})q^{1-j-m}, e(-2\lambda_{12})q^{1-i-m}}
{q^{-m}, e(-2\lambda_{2})q^{1-i-j}, e(-2\lambda_{123})q^{2-i-j-m} }
{q, q}\\[6pt]
&=(-)^{i+j}
\prod_{r=1}^{i}
\frac{\langle\,q^{\frac{m-i+r}{2}}\,\rangle_{1}}
{\langle\,q^{\frac{r}{2}}\,\rangle_{1}}\nonumber\\[6pt]
&\times
\frac{\langle\,e(\lambda_{23})q^{j-\frac{1}{2}}\,\rangle_{1}
\prod_{r=1}^{i}\langle\,e(\lambda_{123})q^{\frac{m+r-2}{2}}\,\rangle_{1}
\prod_{r=1}^{j}\langle\,e(\lambda_{2})q^{\frac{r-1}{2}}\,\rangle_{1}
\prod_{r=1}^{m-i-j}\langle\,e(\lambda_{1})q^{\frac{i+r-1}{2}}\,\rangle_{1}}
{\prod_{r=1}^{m+1}\langle\,e(\lambda_{23})q^{\frac{j+r-2}{2}}\,\rangle_{1}}
\nonumber\\[6pt]
&\times
\;{}_4\varphi_3
\hs{q^{-j},  q^{-i}, e(2\lambda_{23})q^{j-1}, e(2\lambda_{12})q^{i-1}}
{q^{-m}, e(2\lambda_{2}), e(2\lambda_{123})q^{m-1}}
{q, q}.
\end{align}
\medskip

\noindent
(2)\; {\it In case $m\le i+j\le 2m,$ we have}

\begin{align}
&p_{ij}=(-)^m
\prod_{r=1}^{i}
\frac{\langle\,q^{\frac{m-i+r}{2}}\,\rangle_{1}}
{\langle\,q^{\frac{r}{2}}\,\rangle_{1}}\nonumber\\[6pt]
&\times
\frac{\langle\,e(\lambda_{23})q^{j-\frac{1}{2}}\,\rangle_{1}
\prod_{r=1}^{m-j}\langle\,e(\lambda_{123})q^{\frac{m+j+r-2}{2}}\,\rangle_{1}
\prod_{r=1}^{i+j-m}\langle\,e(\lambda_{3})q^{\frac{m-i+r-1}{2}}\,\rangle_{1}
\prod_{r=1}^{m-i}\langle\,e(\lambda_{2})q^{\frac{i+r-1}{2}}\,\rangle_{1}}
{\prod_{r=1}^{m+1}\langle\,e(\lambda_{23})q^{\frac{j+r-2}{2}}\,\rangle_{1}}
\nonumber\\[6pt]
&\times{}_4\varphi_3
\hs{e(-2\lambda_{12})q^{1-i-m},  e(-2\lambda_{23})q^{1-j-m},
q^{j-m},  q^{i-m} }
{e(-2\lambda_{123})q^{2-2m}, e(-2\lambda_{2})q^{1-m}, q^{-m} }
{q, q}\\[6pt]
&=(-)^{i+j}
\prod_{r=1}^{i}
\frac{\langle\,q^{\frac{m-i+r}{2}}\,\rangle_{1}}
{\langle\,q^{\frac{r}{2}}\,\rangle_{1}}\nonumber\\[6pt]
&\times
\frac{\langle\,e(\lambda_{23})q^{j-\frac{1}{2}}\,\rangle_{1}
\prod_{r=1}^{i}\langle\,e(\lambda_{123})q^{\frac{m+r-2}{2}}\,\rangle_{1}
\prod_{r=1}^{j}\langle\,e(\lambda_{2})q^{\frac{r-1}{2}}\,\rangle_{1}}
{\prod_{r=1}^{m+1}\langle\,e(\lambda_{23})q^{\frac{j+r-2}{2}}\,\rangle_{1}
\prod_{r=1}^{i+j-m}\langle\,e(\lambda_{1})q^{\frac{m-j+r-1}{2}}\,\rangle_{1}}
\nonumber\\[6pt]
&\times{}_4\varphi_3
\hs{q^{-i},  e(2\lambda_{23})q^{j-1}, q^{-j},   e(2\lambda_{12})q^{i-1}}
{q^{-m}, e(2\lambda_{2}), e(2\lambda_{123})q^{m-1} }
{q, q}.
\end{align}
\medskip

\noindent
{\bf Proof.}\; (1) Watson's transformation formula 
for a terminating very-well poised ${}_8\varphi_7$ series
\cite{GR} is
\begin{align}
&{}_8\varphi_7\hs{a, qa^{1/2}, -qa^{1/2}, b, c, d, e, q^{-n}}
{a^{1/2}, -a^{1/2}, aq/b, aq/c, aq/d, aq/e, aq^{n+1}}
{q, \frac{a^2q^{n+2}}{bcde}}\nonumber\\[6pt]
&=\frac{(aq, aq/de ;q)_n}{(aq/d, aq/e ;q)_n}
{}_4\varphi_3\hs{aq/bc,  d, e, q^{-n}}
{aq/b, aq/c, deq^{-n}/a}
{q, q}.
\end{align}

\noindent
The substitution 
\begin{align*}
&n=i, \quad a=e(-2\lambda_{23})q^{-2j+1-i}, \quad
b=e(2\lambda_{1})q^{m-j},\\[6pt]
&c=e(-2\lambda_{3})q^{1-j}, \quad
d=e(-2\lambda_{23})q^{1-m-j},\quad e=q^{-j}
\end{align*}
into Watson's formula (3.36) leads to
\begin{align}
&{}_8\varphi_7
\Biggl(
\begin{array}{ccc}
e(-\lambda_{23})q^{-\frac{i}{2}-j+\frac{3}{2}},&
-e(-\lambda_{23})q^{-\frac{i}{2}-j+\frac{3}{2}},&
e(2\lambda_{1})q^{m-j},\\[6pt]
e(-\lambda_{23})q^{-\frac{i}{2}-j+\frac{1}{2}},&
-e(-\lambda_{23})q^{-\frac{i}{2}-j+\frac{1}{2}},&
e(-2\lambda_{2})q^{1-i-j},
\end{array}\nonumber\\[8pt]
&
\begin{array}{ccccc}
e(-2\lambda_{23})q^{1-m-j},&
e(-2\lambda_{23})q^{-2j+1-i},&
q^{-i}, &
\,e(-2\lambda_{3})q^{1-j},&
q^{-j}\\[6pt]
e(-2\lambda_{23})q^{2-i-j},&
e(-2\lambda_{123})q^{2-m-i-j},&
q^{m-i-j+1},&
e(-2\lambda_{23})q^{2-2j}
\end{array}\nonumber\\[8pt]
&\quad
;\; q,\; e(-2\lambda_{23})\,q^{2-i}\quad \Biggr)\nonumber\\[6pt]
&=\frac{(e(-2\lambda_{23})q^{2-2j-i}, q^{1+m-i} ;q)_i}
{(q^{1+m-i-j}, e(-2\lambda_{23})q^{2-i-j};q)_i}\,
\nonumber\\[6pt]
&\times{}_4\varphi_3
\hs{q^{-j},  q^{-i}, e(-2\lambda_{23})q^{1-j-m}, e(-2\lambda_{12})q^{1-i-m}}
{q^{-m}, e(-2\lambda_{2})q^{1-i-j}, e(-2\lambda_{123})q^{2-i-j-m} }
{q, q},\nonumber
\end{align}

\noindent
which is equal to
\begin{align}
&\prod_{r=1}^{i}
\frac{\langle\,e(\lambda_{23})q^{j+\frac{r-2}{2}}\,\rangle_{1}
\langle\,q^{\frac{m-i+r}{2}}\,\rangle_{1}}
{\langle\,e(\lambda_{23})q^{\frac{j+r-2}{2}}\,\rangle_{1}
\langle\,q^{\frac{m-i-j+r}{2}}\,\rangle_{1}}
\nonumber\\[6pt]
&\times{}_4\varphi_3
\hs{q^{-j},  q^{-i}, e(-2\lambda_{23})q^{1-j-m}, e(-2\lambda_{12})q^{1-i-m}}
{q^{-m}, e(-2\lambda_{2})q^{1-i-j}, e(-2\lambda_{123})q^{2-i-j-m} }
{q, q}.
\end{align}

\noindent
Hence (3.37) and (2) of Proposition 3.4 implies (3.32).
\medskip

Sears's transformation formula for a terminating series \cite{GR} is
\begin{align}
&{}_4\varphi_3\hs{q^{-n},  a_1, a_2, a_3}
{b_1, b_2, b_3}
{q, q}\nonumber\\[6pt]
&=\frac{(b_2/a_1, b_3/a_1 ;q)_n}{(b_2,b_3;q)_n}\,a_1^n\;
{}_4\varphi_3\hs{q^{-n},  a_1, b_1/a_2, b_1/a_3}
{b_1, a_1q^{1-n}/b_2, a_1q^{1-n}/b_3}
{q, q}.
\end{align}

\medskip

\noindent
The substitution of
\begin{align*}
&n=j,\;a_1=q^{-i},\; a_2=e(-2\lambda_{23})q^{1-j-m}, \;
a_3=e(-2\lambda_{12})q^{1-i-m}, \nonumber\\[6pt]
&b_1=q^{-m},\;
b_2=e(-2\lambda_{2})q^{1-i-j},\; b_3=e(-2\lambda_{123})q^{2-i-j-m}
\end{align*}
into the Sears' formula (3.38) implies
\begin{align}
&{}_4\varphi_3
\hs{q^{-j},  q^{-i}, e(-2\lambda_{23})q^{1-j-m}, e(-2\lambda_{12})q^{1-i-m}}
{q^{-m}, e(-2\lambda_{2})q^{1-i-j}, e(-2\lambda_{123})q^{2-i-j-m} }
{q, q}\nonumber\\[6pt]
&=\frac{(e(-2\lambda_{2})q^{1-j}, e(-2\lambda_{123})q^{2-j-m} ;q)_j}
{(e(-2\lambda_{2})q^{1-i-j}, e(-2\lambda_{123})q^{2-i-j-m};q)_j}\,q^{-ij}
\nonumber\\[6pt]
&\times
\;{}_4\varphi_3
\hs{q^{-j},  q^{-i}, e(2\lambda_{23})q^{j-1}, e(2\lambda_{12})q^{i-1}}
{q^{-m}, e(2\lambda_{2}), e(2\lambda_{123})q^{m-1}}
{q, q}\nonumber\\[6pt]
&=
\prod_{r=1}^{j}
\frac{\langle\,e(\lambda_{2})q^{\frac{r-1}{2}}\,\rangle_{1}
\langle\,e(\lambda_{123})q^{\frac{m+r-2}{2}}\,\rangle_{1}}
{\langle\,e(\lambda_{2})q^{\frac{i+r-1}{2}}\,\rangle_{1}
\langle\,e(\lambda_{123})q^{\frac{m+i+r-2}{2}}\,\rangle_{1}}
\nonumber\\[6pt]
&\times
\;{}_4\varphi_3
\hs{q^{-j},  q^{-i}, e(2\lambda_{23})q^{j-1}, e(2\lambda_{12})q^{i-1}}
{q^{-m}, e(2\lambda_{2}), e(2\lambda_{123})q^{m-1}}
{q, q}.
\end{align}

\medskip

\noindent
Hence, combining (3.37), (3.39) with the expression (1) of Proposition 3.4
implies (3.33). 
\medskip

\noindent
(2)\; The substitution
\begin{align*}
&n=m-i, \quad a=e(-2\lambda_{23})q^{-2m+1+i}, \quad
b=e(2\lambda_{1})q^{i},\\[6pt]
&c=e(-2\lambda_{3})q^{1+i-m},\quad
d=e(-2\lambda_{23})q^{1-m-j}, \quad e=q^{j-m}
\end{align*}
into Watson's formula (3.36) implies

\begin{align}
&{}_8\varphi_7
\Biggl(
\begin{array}{lll}
e(-2\lambda_{23})q^{1+i-2m},&
e(-\lambda_{23})q^{\frac{i+3}{2}-m},&
-e(-\lambda_{23})q^{\frac{i+3}{2}-m},\\[6pt]
e(-\lambda_{23})q^{\frac{i+1}{2}-m},&
-e(-\lambda_{23})q^{\frac{i+1}{2}-m},&
e(-2\lambda_{123})q^{2-2m},
\end{array}\nonumber\\[8pt]
&
\begin{array}{lllll}
e(2\lambda_{1})q^{i},&
\,e(-2\lambda_{3})q^{1+i-m},&
e(-2\lambda_{23})q^{1-m-j},&
q^{j-m}, &
q^{i-m}
\\[6pt]
e(-2\lambda_{2})q^{1-m},&
q^{1+i+j-m},&
e(-2\lambda_{23})q^{2-m+i-j},&
e(-2\lambda_{23})q^{2-m}
\end{array}\nonumber\\[8pt]
&;\,
q,\;\quad e(-2\lambda_{12})q^{2-i}\quad \Biggr)\nonumber\\[6pt]
&=\frac{(e(-2\lambda_{23})q^{2+i-2m}, q^{1+i} ;q)_{m-i}}
{(q^{1+i+j-m},e(-2\lambda_{23})q^{2+i-j-m};q)_{m-i}}
\nonumber\\[6pt]
&\times{}_4\varphi_3
\hs{e(-2\lambda_{12})q^{1-i-m},  e(-2\lambda_{23})q^{1-j-m},
q^{j-m},  q^{i-m} }
{e(-2\lambda_{123})q^{2-2m}, e(-2\lambda_{2})q^{1-m}, q^{-m} }
{q, q}\nonumber\\[6pt]
&=
\prod_{r=1}^{m-i}
\frac{\langle\,e(\lambda_{23})q^{\frac{m+r-2}{2}}\,\rangle_{1}
\langle\,q^{\frac{i+r}{2}}\,\rangle_{1}}
{\langle\,e(\lambda_{23})q^{\frac{j+r-2}{2}}\,\rangle_{1}
\langle\,q^{\frac{i+j-m+r}{2}}\,\rangle_{1}}
\nonumber\\[6pt]
&\times{}_4\varphi_3
\hs{e(-2\lambda_{12})q^{1-i-m},  e(-2\lambda_{23})q^{1-j-m},
q^{j-m},  q^{i-m} }
{e(-2\lambda_{123})q^{2-2m}, e(-2\lambda_{2})q^{1-m}, q^{-m} }
{q, q}.
\end{align}

\noindent
Hence (3.40) and (2) of Propostion 3.4 implies (3.34).

\medskip

The substitution
\begin{align*}
&n=m-j, \quad a_1=e(-2\lambda_{12})q^{1-i-m}, \quad
a_2=q^{i-m}, \quad a_3=e(-2\lambda_{23})q^{1-m-j}\\[6pt]
&
b_1=q^{-m},\quad b_2=e(-2\lambda_{123})q^{2-2m},
\quad b_3=e(-2\lambda_{2})q^{1-m}
\end{align*}
into Sears's formula (3.38) implies

\begin{align}
&{}_4\varphi_3
\hs{q^{j-m}, e(-2\lambda_{12})q^{1-i-m},  q^{i-m}, e(-2\lambda_{23})q^{1-j-m} }
{q^{-m}, e(-2\lambda_{123})q^{2-2m}, e(-2\lambda_{2})q^{1-m} }
{q, q}\nonumber\\[6pt]
&=\frac{(e(-2\lambda_{3})q^{1-m+i}, e(2\lambda_{1})q^{i} ;q)_{m-j}}
{(e(-2\lambda_{123})q^{2-2m}, e(-2\lambda_{2})q^{1-m};q)_{m-j}}\,
\left(e(-2\lambda_{12})\,q^{1-i-m}\right)^{m-j}
\nonumber\\[6pt]
&\times{}_4\varphi_3
\hs{q^{j-m}, e(-2\lambda_{12})q^{1-i-m}, q^{-i},  e(2\lambda_{23})q^{j-1}}
{q^{-m}, e(2\lambda_{3})q^{j-i}, e(-2\lambda_{1})q^{1-m-i+j} }
{q, q}
\nonumber\\[6pt]
&=(-)^{m-j}
\prod_{r=1}^{m-j}
\frac{\langle\,e(\lambda_{3})q^{\frac{j-i+r-1}{2}}\,\rangle_{1}
\langle\,e(\lambda_{1})q^{\frac{i+r-1}{2}}\,\rangle_{1}}
{\langle\,e(\lambda_{123})q^{\frac{m+j+r-2}{2}}\,\rangle_{1}
\langle\,e(\lambda_{2})q^{\frac{j+r-1}{2}}\,\rangle_{1}}
\nonumber\\[6pt]
&\times{}_4\varphi_3
\hs{q^{j-m}, e(-2\lambda_{12})q^{1-i-m}, q^{-i},  e(2\lambda_{23})q^{j-1}}
{q^{-m}, e(2\lambda_{3})q^{j-i}, e(-2\lambda_{1})q^{1-m-i+j} }
{q, q}.
\end{align}

\noindent
The substitution
\begin{align*}
&n=i, \quad a_1=e(2\lambda_{23})q^{j-1}, \quad
a_2=q^{j-m}, \quad a_3=e(-2\lambda_{12})q^{1-i-m}\\[6pt]
&
b_1=q^{-m},\quad b_2=e(2\lambda_{3})q^{j-i},
\quad b_3=e(-2\lambda_{1})q^{1-m+j-i}
\end{align*}
into Sears's formula (3.38) implies

\begin{align}
&{}_4\varphi_3
\hs{q^{-i}, e(2\lambda_{23})q^{j-1}, q^{j-m}, e(-2\lambda_{12})q^{1-i-m}}
{q^{-m}, e(2\lambda_{3})q^{j-i}, e(-2\lambda_{1})q^{1-m-i+j} }
{q, q}\nonumber\\[6pt]
&=\frac{(e(-2\lambda_{2})q^{1-i}, e(-2\lambda_{123})q^{2-m-i} ;q)_{i}}
{(e(2\lambda_{3})q^{j-i}, e(-2\lambda_{1})q^{1-m+j-i};q)_{i}}\,
\left(e(2\lambda_{23})q^{j-1}\right)^{i}
\nonumber\\[6pt]
&\times{}_4\varphi_3
\hs{q^{-i},  e(2\lambda_{23})q^{j-1}, q^{-j},   e(2\lambda_{12})q^{i-1}}
{q^{-m}, e(2\lambda_{2}), e(2\lambda_{123})q^{m-1} }
{q, q}\nonumber\\[6pt]
&=(-)^i
\prod_{r=1}^{i}
\frac{\langle\,e(\lambda_{2})q^{\frac{r-1}{2}}\,\rangle_{1}
\langle\,e(\lambda_{123})q^{\frac{m+r-2}{2}}\,\rangle_{1}}
{\langle\,e(\lambda_{3})q^{\frac{j-i+r-1}{2}}\,\rangle_{1}
\langle\,e(\lambda_{1})q^{\frac{m-j+r-1}{2}}\,\rangle_{1}}
\nonumber\\[6pt]
&\times{}_4\varphi_3
\hs{q^{-i},  e(2\lambda_{23})q^{j-1}, q^{-j},   e(2\lambda_{12})q^{i-1}}
{q^{-m}, e(2\lambda_{2}), e(2\lambda_{123})q^{m-1} }
{q, q}.
\end{align}
\medskip

\noindent
Hence, combining (3.40), (3.41), (3.42) with the expression (2) 
of Proposition 3.4 implies (3.35). 
It completes the proof.
\hfill$\square$

\bigskip
The equality
\begin{equation*}
\frac{
\prod_{r=1}^{m-i}\langle\,e(\lambda_{1})q^{\frac{i+r-1}{2}}\,\rangle_{1}}
{\prod_{r=1}^{j}\langle\,e(\lambda_{1})q^{\frac{m-j+r-1}{2}}\,\rangle_{1}}
=
\begin{cases}
&\prod_{r=1}^{m-i-j}
\langle\,e(\lambda_{1})q^{\frac{i+r-1}{2}}\,\rangle_{1}
\quad \mbox{for}\quad i+j\le m,\\[6pt]
&\dfrac{1}
{\prod_{r=1}^{i+j-m}
\langle\,e(\lambda_{1})q^{\frac{m-j+r-1}{2}}\,\rangle_{1}}
\quad \mbox{for} \quad i+j\ge m
\end{cases}
\end{equation*}

\noindent
rewrites (3.33) and (3.35) into the following.
\medskip

\noindent
{\bf Proposition 3.6.}\; {\it For $0\le i,j \le m,$ we have}
\begin{align*}
&p_{ij}=(-)^{i+j}
\prod_{r=1}^{i}
\frac{\langle\,q^{\frac{m-i+r}{2}}\,\rangle_{1}}
{\langle\,q^{\frac{r}{2}}\,\rangle_{1}}\nonumber\\[6pt]
&\times
\frac{\langle\,e(\lambda_{23})q^{j-\frac{1}{2}}\,\rangle_{1}
\prod_{r=1}^{i}\langle\,e(\lambda_{123})q^{\frac{m+r-2}{2}}\,\rangle_{1}
\prod_{r=1}^{j}\langle\,e(\lambda_{2})q^{\frac{r-1}{2}}\,\rangle_{1}
\prod_{r=1}^{m}\langle\,e(\lambda_{1})q^{\frac{r-1}{2}}\,\rangle_{1}}
{\prod_{r=1}^{m+1}\langle\,e(\lambda_{23})q^{\frac{j+r-2}{2}}\,\rangle_{1}\prod_{r=1}^{i}\langle\,e(\lambda_{1})q^{\frac{r-1}{2}}\,\rangle_{1}
\prod_{r=1}^{j}\langle\,e(\lambda_{1})q^{\frac{m-j+r-1}{2}}\,\rangle_{1}}
\nonumber\\[6pt]
&\times{}_4\varphi_3\hs{q^{-i},\; e(2\lambda_{12})q^{i-1},\; 
q^{-j},\; e(2\lambda_{23})q^{j-1}}
{e(2\lambda_{2}),\; q^{-m},\; e(2\lambda_{123})q^{m-1}}
{q,\; q}\\[6pt]
&=\frac{1-e(2\lambda_{23})q^{2j-1}}{1-e(2\lambda_{23})q^{j-1}}
\frac{(e(2\lambda_{123})q^{m-1};q)_i}{(e(2\lambda_{1});q)_i}
\frac{(e(2\lambda_{1});q)_m}{(e(2\lambda_{23})q^j;q)_m}
\frac{(e(2\lambda_{2});q)_j}{(e(-2\lambda_{1})q^{1-m};q)_j}
\nonumber\\[6pt]
&\times
\frac{(q^{-m};q)_i}{(q;q)_i}
e((-m-j)\lambda_{1}+(m-i-j)\lambda_{2}
+(m-i)\lambda_{3})q^i\nonumber\\[6pt]
&\times{}_4\varphi_3\hs{q^{-i},\; e(2\lambda_{12})q^{i-1},\; 
q^{-j},\; e(2\lambda_{23})q^{j-1}}
{e(2\lambda_{2}),\; q^{-m},\; e(2\lambda_{123})q^{m-1}}
{q,\; q}.
\end{align*}

\bigskip

\section{Invariant Hermitian form of non-diagonal type}
A Hermitian form which is invariant with respect to 
the monodromy group is called the {\it monodromy-invariant 
Hermitian form} or, simply, the {\it invariant Hermitian form}.
One of the method for constructing an invariant Hermitian form is
that by means of the intersection number of twisted cycles,
as is explained in Subsection 1.2.

The invariant Herimitian form 
$$F(z,\overline{z})=\sum_{i,j}a_{ij}I_i(z)\overline{I_j(z)},$$
where $\{I_j(z)\}$ is a  fundamental set of solutions 
around a singularity, 
is called {\it diagonal} if $a_{ij}=c_i\delta_{ij}$,
where $c_i$ is a multiplicative constant, and is 
called {\it nondiagonal} if not the case. 
The example displayed in (1.9) is an invariant Hermitian
form of diagonal type. 
 
In this section, as an application of our connection formulas,
we provide some examples of the invariant Hermitian form
of nondiagonal type. To obtain them, we consider the special case
of the exponents $a, b, c, g$: for $\rho\in {\Bbb Z}_{\ge 0},$ 
\begin{equation}
m=2\rho,\quad a=b=c=-\dfrac{\rho}{2\rho+1},
\quad g=\dfrac{1}{2\rho+1}.
\end{equation}
In this case, the characteristic exponents at $0, 1$ and $\infty$
are  
\begin{equation}
e_j^{(0)}=e_j^{(1)}=\frac{j(j+1)}{2(2\rho+1)},\quad 
e_j^{(\infty)}=\frac{2m\rho+(1-2m)j-j^2}{2(2\rho+1)}
\end{equation}
for $0\le j\le m,$ and it is seen that some of two differences of them
are integer; indeed, $e_{2\rho-i}^{(0)}-e_{i}^{(0)}=\rho-i$ for
$0\le i\le m.$
As a result of such degeneration of characteristic exponents,
in the solution space of the differential equation of rank $m+1$
a submodule emerges.
This submodue induces the required
Hermitian form, and the connection matrix in Theorem 2.2
is used to find it.
\medskip

In what follows, under the condition (4.1), 
$T_z$ and ${\mathcal L}_z$ are taken as in Section 2, hence  
$$q=e(g)=e\left(\frac{1}{2\rho+1}\right), \quad q^{2\rho+1}=-1
\quad\text{and}\quad e(a)=e(b)=e(c)=q^{-\rho}.$$

For a related work in  conformal field theory, 
we refer the reader to \cite{Kat} and \cite{CIZ}, where
the $\widehat{sl_2}$ modular-invariant partition functions are 
classified; our functions below  correspond to
the $D_{2\rho+1}$ type invariants in the ADE-classification. 
\medskip

\bigskip

In case $\rho=1$, (4.2) means
$$e_0^{(0)}=e_0^{(1)}=0,\; e_1^{(0)}=e_1^{(1)}=1/3,\;
e_2^{(0)}=e_2^{(1)}=1,$$ and Theorem 2.2 imlies  
\begin{equation*}
\left[
\begin{array}{c}
I_0\\[6pt] I_1\\[6pt] I_2
\end{array}
\right]
=
\left[
\begin{array}{ccc}
s(1/6)&-1&s(1/6)\\[8pt]
-s(1/6)&0&s(1/6)\\[8pt]
s(1/6)&1&s(1/6)
\end{array}\right]
\left[
\begin{array}{c}
J_0\\[6pt] J_1\\[6pt] J_2
\end{array}
\right],
\end{equation*}

\noindent
where $s(A)=\sin(\pi A).$ 
Thus, it is easily seen that
\begin{equation*}
\left[
\begin{array}{c}
I_0+I_2\\[6pt] I_1
\end{array}
\right]
=
\left[
\begin{array}{cc}
2s(1/6)&2s(1/6)\\[8pt]
-s(1/6)&s(1/6)
\end{array}\right]
\left[
\begin{array}{c}
J_0\\[6pt] J_2
\end{array}
\right],
\end{equation*}
which leads to the submodule
generated by $I_0+I_2$ and $I_1$. 
Moreover, it is seen that
$
I_0+I_2=2\,s(1/6) (J_0+J_2),
$
which leads to the submodule of rank one
generated by $I_0+I_2$.

\medskip

In case $\rho=2$, (4.2) means
\begin{align*}
&e_0^{(0)}=e_0^{(1)}=0, \quad e_1^{(0)}=e_1^{(1)}=1/5,\quad
e_2^{(0)}=e_2^{(1)}=3/5,\\[6pt] 
&e_3^{(0)}=e_3^{(1)}=6/5,\quad
e_4^{(0)}=e_4^{(1)}=2,
\end{align*}
and Theorem 2.2 implies

\begin{equation*}
\left[
\begin{array}{c}\\
I_0\\[10pt] \vdots\\[6pt]\vdots\\[6pt]\vdots\\[10pt] I_4\\[6pt] 
\end{array}
\right]
=
\left[
\begin{array}{ccccc}
\dfrac{\sqrt{5}-1}{4}&\dfrac{-\sqrt{5}-1}{4}&1&\dfrac{-\sqrt{5}-1}{4}
&\dfrac{\sqrt{5}-1}{4}\\[8pt]
\dfrac{1-\sqrt{5}}{4}&\dfrac{1}{2}&0&
-\dfrac{1}{2}&\dfrac{\sqrt{5}-1}{4}\\[8pt]
\dfrac{\sqrt{5}-1}{4}&0&\dfrac{1-\sqrt{5}}{2}&0&\dfrac{\sqrt{5}-1}{4}\\[8pt]
\dfrac{1-\sqrt{5}}{4}&-\dfrac{1}{2}&0&\dfrac{1}{2}&
\dfrac{\sqrt{5}-1}{4}\\[8pt]
\dfrac{\sqrt{5}-1}{4}&\dfrac{\sqrt{5}+1}{4}&1&
\dfrac{\sqrt{5}+1}{4}&\dfrac{\sqrt{5}-1}{4}
\end{array}\right]
\left[
\begin{array}{c}\\
J_0\\[10pt] \vdots\\[6pt]\vdots\\[6pt]\vdots\\[10pt] J_4\\[6pt] 
\end{array}
\right].
\end{equation*}

\noindent
This induces
\begin{equation*}
\left[
\begin{array}{c}
I_0+I_4\\[6pt]I_1+I_3\\[6pt] I_2
\end{array}
\right]
=
\left[
\begin{array}{ccc}
\dfrac{\sqrt{5}-1}{2}&2&\dfrac{\sqrt{5}-1}{2}\\[8pt]
\dfrac{1-\sqrt{5}}{2}&0&\dfrac{\sqrt{5}-1}{2}\\[8pt]
\dfrac{\sqrt{5}-1}{2}&\dfrac{1-\sqrt{5}}{2}&\dfrac{\sqrt{5}-1}{2}
\end{array}\right]
\left[
\begin{array}{c}
J_0\\[6pt] J_2\\[6pt]J_4
\end{array}
\right],
\end{equation*}
which leads to the submodule of rank $3$ generated by
$I_0+I_4, I_1+I_3, I_2$, and
\begin{equation*}
\left[
\begin{array}{c}
I_0+I_4\\[12pt] 
I_2
\end{array}
\right]
=
\left[
\begin{array}{ccc}
\dfrac{\sqrt{5}-1}{2}&2\\[8pt]
\dfrac{\sqrt{5}-1}{4}&\dfrac{1-\sqrt{5}}{2}
\end{array}\right]
\left[
\begin{array}{c}
J_0+J_4\\[12pt]  
J_2
\end{array}
\right],
\end{equation*}
which leads to the submodule of rank $2$ generated by
$I_0+I_4, I_2$.

\bigskip

Generally, we have the following.
\medskip

\noindent
{\bf Proposition 4.1.}\; {\it For $0\le i\le \rho$, we have}
\begin{equation*}
\begin{cases}
\;I_i+I_{2\rho-i}&=\sum_{0\le j\le \rho} 2p_{i,2j} J_{2j}
\qquad  (0\le i< \rho),\\[6pt]
\;I_{\rho}&=\sum_{0\le j\le \rho} 2p_{\rho,2j} J_{2j}.\\[6pt]
\end{cases}
\end{equation*}
\medskip

\noindent
{\bf Proof.} It follows from (2) of Lemma 4.5 below.
\hfill$\square$

\comment{
$$\begin{cases}
\;I_0+I_m&=2p_{00} J_0\quad+2p_{02} J_2\quad+\cdots+
2p_{0,m-2} J_{m-2}\quad+2p_{0,m} J_{m},\\[6pt]
\;I_1+I_{m-1}&=2p_{10} J_0\quad+2p_{12} J_2\quad+\cdots+
2p_{1,m-2} J_{m-2}\quad+2p_{1,m} J_{m},\\[6pt]
\quad\cdots&\qquad \cdots\qquad \cdots\qquad \cdots\qquad \cdots \\[6pt]
\;I_{\rho-1}+I_{\rho+1}&=2p_{\rho-1,0}J_0+2p_{\rho-1,2} J_2+\cdots+
2p_{\rho-1,m-2} J_{m-2}+2p_{\rho-1,m} J_{m},\\[6pt]
\;I_{\rho}&=2p_{\rho,0} J_0\quad+2p_{\rho,2} J_2\quad+\cdots+
2p_{\rho,m-2} J_{m-2}\quad+2p_{\rho,m} J_{m},\\[6pt]
\end{cases}$$
} 
\medskip

\noindent
{\bf Proposition 4.2.}\; {\it If $\rho$ is even, we have}
\begin{equation*}
\begin{cases}
\;I_{2i}+I_{2\rho-2i}&=\sum_{0\le j< \rho} 2p_{2i,2j} 
(J_{2j}+J_{2\rho-2j})+2p_{2i,\rho} J_{\rho}
\qquad  (0\le i\le \frac{\rho}{2}-1),\\[6pt]
\;I_{\rho}&=\sum_{0\le j\le \rho} 2p_{\rho,2j} 
(J_{2j}+J_{2\rho-2j})+2p_{\rho,\rho} J_{\rho}.\\[6pt]
\end{cases}
\end{equation*}
{\it If $\rho$ is odd, we have} 
\begin{equation*}
I_{2i}+I_{2\rho-2i}=\sum_{0\le j< \rho} 2p_{2i,2j} 
(J_{2j}+J_{2\rho-2j})
\qquad  (0\le i\le \frac{\rho-1}{2}).
\end{equation*}
\medskip

\noindent
{\bf Proof.} It follows from (1) with (2) of Lemma 4.5 below.
\hfill$\square$

\medskip
Therefore, we reach the following:
\medskip

\noindent
{\bf Theorem 4.3.} {\it The Hermitian form given by

$$F(z,\overline{z})=\sum_{0\le i\le \rho-1}
\frac{1}{(C_i+C_{2\rho-i})^2}|I_i+I_{2\rho-i}|^2
+\frac{1}{C_{\rho}^2}|I_\rho|^2$$

\noindent
is the monodromy-invariant Hermitian form. Here
\begin{align*}
&(C_i+C_{2\rho-i})^2=(C_i+C_{2\rho-i})\bullet (C_i+C_{2\rho-i})
\nonumber\\[6pt]
&=C_i\bullet C_i+C_{2\rho-i}\bullet C_{2\rho-i}=
2 C_i\bullet C_i=2C_i^2
\end{align*}
for $0\le i\le \rho-1$, and
$$C_i^2=m!\left(\frac{\sqrt{-1}}{2}\right)^m
\prod_{j=1}^i
\frac{s\left(\frac{-2\rho+j-2}{2(2\rho+1)}\right)
s\left(\frac{1}{2(2\rho+1)}\right)}
{s\left(\frac{-2\rho+j-1}{2(2\rho+1)}\right)^2
s\left(\frac{j}{2(2\rho+1)}\right)}
\prod_{j=1}^{2\rho-i}
\frac{s\left(\frac{-2\rho+j-2}{2(2\rho+1)}\right)
s\left(\frac{1}{2(2\rho+1)}\right)}
{s\left(\frac{-2\rho+j-1}{2(2\rho+1)}\right)^2
s\left(\frac{j}{2(2\rho+1)}\right)}
$$
for $0\le i\le \rho.$}

\bigskip

\noindent
{\bf Theorem 4.4.} {\it The Hermitian form given by

$$F(z,\overline{z})=
\begin{cases}
&\sum_{i=0}^{\frac{\rho}{2}-1}
\dfrac{1}{(C_{2i}+C_{2\rho-2i})^2}|I_{2i}+I_{2\rho-2i}|^2
+\dfrac{1}{C_{\rho}^2}|I_\rho|^2,\qquad 
(\rho : \text{even})\\[10pt]
&\sum_{i=0}^{\frac{\rho-1}{2}}
\dfrac{1}{(C_{2i}+C_{2\rho-2i})^2}|I_{2i}+I_{2\rho-2i}|^2,
\qquad (\rho : \text{odd})
\end{cases}
$$

\noindent
is the monodromy-invariant Hermitian form. Here
$$
(C_{2i}+C_{2\rho-2i})^2=2C_{2i}^2,
$$
for $0\le i< \rho/2$, and
$$C_i^2=m!\left(\frac{\sqrt{-1}}{2}\right)^m
\prod_{j=1}^i
\frac{s\left(\frac{-2\rho+j-2}{2(2\rho+1)}\right)
s\left(\frac{1}{2(2\rho+1)}\right)}
{s\left(\frac{-2\rho+j-1}{2(2\rho+1)}\right)^2
s\left(\frac{j}{2(2\rho+1)}\right)}
\prod_{j=1}^{2\rho-i}
\frac{s\left(\frac{-2\rho+j-2}{2(2\rho+1)}\right)
s\left(\frac{1}{2(2\rho+1)}\right)}
{s\left(\frac{-2\rho+j-1}{2(2\rho+1)}\right)^2
s\left(\frac{j}{2(2\rho+1)}\right)}
$$
for $0\le i\le \rho.$}

\medskip

\noindent
{\bf Lamma 4.5.} {\it We have }
\bigskip

\noindent
(1) $p_{ij}=(-)^i\,p_{i,m-j}\,,$
\qquad
(2)  $p_{ij}=(-)^j\,p_{m-i,j}\,.$
\bigskip

\noindent
{\bf Proof.}  $(1)$\; Under the condition (4.1),  
the expression (2.6) reduces to 

\begin{align}
p_{ij}=(-)^i
&\sum_{\substack{
0\le k\le m-i\\[3pt]
0\le l\le i\\[3pt]
k+l=j}}
(-)^{k}
\prod_{r=1}^{m-i-k}
\frac{\langle\,q^{-\rho+\frac{i+r-1}{2}}\,\rangle_{1}}
{\langle\,q^{-2\rho+k+\frac{i+r-1}{2}}\,\rangle_{1}}
\prod_{r=1}^{k}
\frac{\langle\,q^{-\rho+\frac{i+r-1}{2}}\,\rangle_{1}}
{\langle\,q^{-2\rho+k+\frac{i-r-1}{2}}\,\rangle_{1}}
\nonumber\\[6pt]
&\times
\prod_{r=1}^{i-l}
\frac{\langle\,q^{-2\rho+\frac{k+i-r-1}{2}}\,\rangle_{1}
\langle\,q^{\rho+\frac{-i-k+r}{2}}\,\rangle_{1}}
{\langle\,q^{-2\rho+j+\frac{r-1}{2}}\,\rangle_{1}
\langle\,q^{\frac{r}{2}}\,\rangle_{1}}
\prod_{r=1}^{l}
\frac{\langle\,q^{-\rho+\frac{k+r-1}{2}}\,\rangle_{1}
\langle\,q^{\frac{k+r}{2}}\,\rangle_{1}}
{\langle\,q^{-2\rho+j+\frac{-r-1}{2}}\,\rangle_{1}
\langle\,q^{\frac{r}{2}}\,\rangle_{1}}.
\end{align}

\medskip

\noindent
Hence we have

\begin{align}
p_{i,m-j}&=(-)^i
\sum_{\substack{
0\le k\le m-i\\[3pt]
0\le l\le i\\[3pt]
k+l=m-j}}
(-)^{k}
\prod_{r=1}^{m-i-k}
\frac{\langle\,q^{-\rho+\frac{i+r-1}{2}}\,\rangle_{1}}
{\langle\,q^{-2\rho+k+\frac{i+r-1}{2}}\,\rangle_{1}}
\prod_{r=1}^{k}
\frac{\langle\,q^{-\rho+\frac{i+r-1}{2}}\,\rangle_{1}}
{\langle\,q^{-2\rho+k+\frac{i-r-1}{2}}\,\rangle_{1}}
\nonumber\\[6pt]
&\times
\prod_{r=1}^{i-l}
\frac{\langle\,q^{-2\rho+\frac{k+i-r-1}{2}}\,\rangle_{1}
\langle\,q^{\rho+\frac{-i-k+r}{2}}\,\rangle_{1}}
{\langle\,q^{-j+\frac{r-1}{2}}\,\rangle_{1}
\langle\,q^{\frac{r}{2}}\,\rangle_{1}}
\prod_{r=1}^{l}
\frac{\langle\,q^{-\rho+\frac{k+r-1}{2}}\,\rangle_{1}
\langle\,q^{\frac{k+r}{2}}\,\rangle_{1}}
{\langle\,q^{-j+\frac{-r-1}{2}}\,\rangle_{1}
\langle\,q^{\frac{r}{2}}\,\rangle_{1}}\nonumber\\[6pt]
&=\sum_{\substack{
0\le k\le m-i\\[3pt]
0\le l\le i\\[3pt]
k+l=j}}
(-)^{k}
\prod_{r=1}^{k}
\frac{\langle\,q^{-\rho+\frac{i+r-1}{2}}\,\rangle_{1}}
{\langle\,q^{-k+\frac{-i+r-1}{2}}\,\rangle_{1}}
\prod_{r=1}^{m-i-k}
\frac{\langle\,q^{-\rho+\frac{i+r-1}{2}}\,\rangle_{1}}
{\langle\,q^{-k+\frac{-i-r-1}{2}}\,\rangle_{1}}
\nonumber\\[6pt]
&\times
\prod_{r=1}^{l}
\frac{\langle\,q^{-\rho+\frac{-k-r-1}{2}}\,\rangle_{1}
\langle\,q^{\frac{r+k}{2}}\,\rangle_{1}}
{\langle\,q^{-j+\frac{r-1}{2}}\,\rangle_{1}
\langle\,q^{\frac{r}{2}}\,\rangle_{1}}
\prod_{r=1}^{i-l}
\frac{\langle\,q^{\frac{-i-k+r-1}{2}}\,\rangle_{1}
\langle\,q^{\rho+\frac{-i-k+r}{2}}\,\rangle_{1}}
{\langle\,q^{-j+\frac{-r-1}{2}}\,\rangle_{1}
\langle\,q^{\frac{r}{2}}\,\rangle_{1}}
\nonumber\\[6pt]
&=\sum_{\substack{
0\le k\le m-i\\[3pt]
0\le l\le i\\[3pt]
k+l=j}}
(-)^{k}
\prod_{r=1}^{m-i-k}
\frac{\langle\,q^{-\rho+\frac{i+r-1}{2}}\,\rangle_{1}}
{\langle\,q^{-k+\frac{-i-r-1}{2}}\,\rangle_{1}}
\prod_{r=1}^{k}
\frac{\langle\,q^{-\rho+\frac{i+r-1}{2}}\,\rangle_{1}}
{\langle\,q^{-k+\frac{-i+r-1}{2}}\,\rangle_{1}}
\nonumber\\[6pt]
&\times
\prod_{r=1}^{i-l}
\frac{\langle\,q^{\frac{-i-k+r-1}{2}}\,\rangle_{1}
\langle\,q^{\rho+\frac{-i-k+r}{2}}\,\rangle_{1}}
{\langle\,q^{-j+\frac{-r-1}{2}}\,\rangle_{1}
\langle\,q^{\frac{r}{2}}\,\rangle_{1}}
\prod_{r=1}^{l}
\frac{\langle\,q^{-\rho+\frac{-k-r-1}{2}}\,\rangle_{1}
\langle\,q^{\frac{r+k}{2}}\,\rangle_{1}}
{\langle\,q^{-j+\frac{r-1}{2}}\,\rangle_{1}
\langle\,q^{\frac{r}{2}}\,\rangle_{1}}.
\end{align}
\medskip

\noindent
Here the second equality follows from
the change of the running indeces
$k\rightarrow m-i-k$ and $l\rightarrow i-l$.
\medskip

\noindent
Comparing (4.3) with (4.4) leads to 
$$p_{ij}=(-)^i\,p_{i,m-j},$$ 
because 
\begin{equation}
\langle\,q^{\alpha}\,\rangle_{1}=
\langle\,q^{\beta}\,\rangle_{1}, 
\quad \text{if}\quad \alpha+\beta= \pm(2\rho+1).
\end{equation}
\bigskip

\noindent
(2)\; Under the condition (4.1), the expression (2.7)
reduces to

\begin{align}
p_{ij}=(-)^i
&\sum_{\substack{
0\le k\le i\\[3pt]
0\le l\le m-i\\[3pt]
k+l=j}}
(-)^{l}
\prod_{r=1}^{i-k}
\frac{\langle\,q^{-2\rho+\frac{i-r-1}{2}}\,\rangle_{1}}
{\langle\,q^{-\rho+k+\frac{-i+r-1}{2}}\,\rangle_{1}}
\prod_{r=1}^{k}
\frac{\langle\,q^{\frac{-i+r-1}{2}}\,\rangle_{1}}
{\langle\,q^{-\rho+k+\frac{-i-r-1}{2}}\,\rangle_{1}}
\nonumber\\[6pt]
&\times
\prod_{r=1}^{m-i-l}
\frac{\langle\,q^{-\rho+\frac{i-k+r-1}{2}}\,\rangle_{1}
\langle\,q^{\frac{i-r+k}{2}}\,\rangle_{1}}
{\langle\,q^{-2\rho+j+\frac{r-1}{2}}\,\rangle_{1}
\langle\,q^{\frac{r}{2}}\,\rangle_{1}}
\prod_{r=1}^{l}
\frac{\langle\,q^{-\rho+\frac{k+r-1}{2}}\,\rangle_{1}
\langle\,q^{\frac{k+r}{2}}\,\rangle_{1}}
{\langle\,q^{-2\rho+j+\frac{-r-1}{2}}\,\rangle_{1}
\langle\,q^{\frac{r}{2}}\,\rangle_{1}}.
\end{align}

\medskip

\noindent
Hence we have

\begin{align}
p_{m-i,j}=(-)^{i+j}
&\sum_{\substack{
0\le k\le m-i\\[3pt]
0\le l\le i\\[3pt]
k+l=j}}
(-)^{k}
\prod_{r=1}^{m-i-k}
\frac{\langle\,q^{-\rho+\frac{-i-r-1}{2}}\,\rangle_{1}}
{\langle\,q^{-2\rho+k+\frac{i+r-1}{2}}\,\rangle_{1}}
\prod_{r=1}^{k}
\frac{\langle\,q^{-\rho+\frac{i+r-1}{2}}\,\rangle_{1}}
{\langle\,q^{-2\rho+k+\frac{i-r-1}{2}}\,\rangle_{1}}
\nonumber\\[6pt]
&\times
\prod_{r=1}^{i-l}
\frac{\langle\,q^{\frac{-i-k+r-1}{2}}\,\rangle_{1}
\langle\,q^{\rho+\frac{-i-r+k}{2}}\,\rangle_{1}}
{\langle\,q^{-2\rho+j+\frac{r-1}{2}}\,\rangle_{1}
\langle\,q^{\frac{r}{2}}\,\rangle_{1}}
\prod_{r=1}^{l}
\frac{\langle\,q^{-\rho+\frac{k+r-1}{2}}\,\rangle_{1}
\langle\,q^{\frac{k+r}{2}}\,\rangle_{1}}
{\langle\,q^{-2\rho+j+\frac{-r-1}{2}}\,\rangle_{1}
\langle\,q^{\frac{r}{2}}\,\rangle_{1}}.
\end{align}

\medskip

\noindent
Therefore, by comparing (4.6) and (4.7) with noting (4.5), 
it is seen that
$$p_{m-i, j}=(-)^j\,p_{i,j}.$$
It completes the proof. \hfill$\square$

\bigskip

\section{Appendix}

\noindent
Under the condition (4.1), Theorem 2.2 implies that,
for $0\le j\le m=2\rho$ ,
\begin{align*}
&p_{0j}=(-)^js\left(\frac{2j+1}{2(2\rho+1)}\right),
\qquad
p_{2\rho,j}=s\left(\frac{2j+1}{2(2\rho+1)}\right),
\\[6pt]
&p_{1j}=(-)^{j+1}
\frac{s\left(\frac{1}{2(2\rho+1)}\right)
s\left(\frac{2j+1}{2(2\rho+1)}\right)}
{s\left(\frac{1}{2\rho+1}\right)},
\qquad
p_{2\rho-1,j}=-
\frac{s\left(\frac{1}{2(2\rho+1)}\right)
s\left(\frac{2j+1}{2(2\rho+1)}\right)}
{s\left(\frac{1}{2\rho+1}\right)},\nonumber\\[6pt]
&p_{2,j}=
(-)^js\left(\frac{2j+1}{2(2\rho+1)}\right)
\left\{1-
\frac{s\left(\frac{2}{2(2\rho+1)}\right)
s\left(\frac{3}{2(2\rho+1)}\right)
s\left(\frac{j}{2(2\rho+1)}\right)
s\left(\frac{j+1}{2(2\rho+1)}\right)}
{s\left(\frac{1}{2(2\rho+1)}\right)^2
s\left(\frac{\rho}{2\rho+1}\right)
s\left(\frac{\rho+1}{2\rho+1}\right)}\right.\\[6pt]
&\left.
+\frac{s\left(\frac{3}{2(2\rho+1)}\right)
s\left(\frac{4}{2(2\rho+1)}\right)
s\left(\frac{j-1}{2(2\rho+1)}\right)
s\left(\frac{j}{2(2\rho+1)}\right)
s\left(\frac{j+1}{2(2\rho+1)}\right)
s\left(\frac{j+2}{2(2\rho+1)}\right)}
{s\left(\frac{1}{2\rho+1}\right)
s\left(\frac{\rho}{2\rho+1}\right)^2
s\left(\frac{1}{2(2\rho+1)}\right)
s\left(\frac{2\rho-1}{2(2\rho+1)}\right)^2}\;
\right\},\nonumber\\[6pt]
&p_{m-2,j}=
s\left(\frac{2j+1}{2(2\rho+1)}\right)
\left\{1-
\frac{s\left(\frac{2}{2(2\rho+1)}\right)
s\left(\frac{3}{2(2\rho+1)}\right)
s\left(\frac{j}{2(2\rho+1)}\right)
s\left(\frac{j+1}{2(2\rho+1)}\right)}
{s\left(\frac{1}{2(2\rho+1)}\right)^2
s\left(\frac{\rho}{2\rho+1}\right)
s\left(\frac{\rho+1}{2\rho+1}\right)}\right.\\[6pt]
&\left.
+\frac{s\left(\frac{3}{2(2\rho+1)}\right)
s\left(\frac{4}{2(2\rho+1)}\right)
s\left(\frac{j-1}{2(2\rho+1)}\right)
s\left(\frac{j}{2(2\rho+1)}\right)
s\left(\frac{j+1}{2(2\rho+1)}\right)
s\left(\frac{j+2}{2(2\rho+1)}\right)}
{s\left(\frac{1}{2\rho+1}\right)
s\left(\frac{\rho}{2\rho+1}\right)^2
s\left(\frac{1}{2(2\rho+1)}\right)
s\left(\frac{2\rho-1}{2(2\rho+1)}\right)^2}\;
\right\};
\end{align*}
\noindent
and, for $0\le i\le m=2\rho$,
\begin{align*}
&p_{i,0}=(-)^is\left(\frac{1}{2(2\rho+1)}\right),
\qquad
p_{i,2\rho}=s\left(\frac{1}{2(2\rho+1)}\right),
\\[6pt]
&p_{i,1}=(-)^{i+1}
s\left(\frac{3}{2(2\rho+1)}\right)
\frac{c\left(\frac{2i+1}{2(2\rho+1)}\right)}
{c\left(\frac{1}{2(2\rho+1)}\right)},
\quad
p_{i,2\rho-1}=-
s\left(\frac{3}{2(2\rho+1)}\right)
\frac{c\left(\frac{2i+1}{2(2\rho+1)}\right)}
{c\left(\frac{1}{2(2\rho+1)}\right)}.
\end{align*}

\newpage

\bigskip
\begin{flushleft}
Katsuhisa Mimachi\\
Department of Mathematics\\
Tokyo Institute of Technology\\
Oh-okayama, Meguro-ku, Tokyo 152-8551\\
Japan\\[5pt]
mimachi@math.titech.ac.jp

\comment{
\bigskip

Yasuhiko Yamada\\
Department of Mathematics\\
Kobe University\\
Rokko-dai, Nada-ku, Kobe 657-8501\\
Japan\\[5pt]
yamaday@math.kobe-u.ac.jp
}

\end{flushleft}
\end{document}